%% file: cen.tex
\documentclass[usenatbib]{mnras}

\usepackage[utf8]{inputenc}
\usepackage[pdftex]{graphicx}
\usepackage{mathptmx}
\usepackage{fixltx2e}
\usepackage{color}
\usepackage{amssymb}

\input{defn}

\begin{document}

\title[A very deep view of the Centaurus cluster]
{A very deep \emph{Chandra} view of metals, sloshing and feedback in
  the Centaurus cluster of galaxies}

\author
[J.~S. Sanders et al.]
{
  \begin{minipage}[b]{\linewidth}
    \begin{flushleft}
      J.~S.~Sanders$^1$,
      A.~C.~Fabian$^2$,
      G.~B.~Taylor$^3$,
      H.~R.~Russell$^2$,
      K.~M.~Blundell$^4$,
      R.~E.~A.~Canning$^{5,6}$,
      J.~Hlavacek-Larrondo$^{7,5,6}$,
      S.~A.~Walker$^2$
      and C.~K.~Grimes$^3$
    \end{flushleft}
  \end{minipage}
  \\
  $^1$ Max-Planck-Institut für extraterrestrische Physik,
  Giessenbachstrasse 1, 85748 Garching, Germany\\
  $^2$ Institute of Astronomy, Madingley Road, Cambridge CB3 0FT\\
  $^3$ Department of Physics and Astronomy, University of New Mexico, Albuquerque, NM 87131, USA\\
  $^4$ University of Oxford, Astrophysics, Keble Road, Oxford OX1 3RH\\
  $^5$ Kavli Institute for Particle Astrophysics and Cosmology, Stanford University, 452 Lomita Mall, Stanford, CA 94305, USA\\
  $^6$ Department of Physics, Stanford University, 382 Via Pueblo Mall, Stanford, CA 94305, USA\\
  $^7$ Département de Physique, Université de Montréal, C.P. 6128, Succ. Centre-Ville, Montreal, Quebec H3C 3J7, Canada\\
\
}
\maketitle

\begin{abstract}
  We examine deep \emph{Chandra} X-ray observations
  of the Centaurus cluster of galaxies, Abell 3526. Applying a
  gradient magnitude filter reveals a wealth of structure, from
  filamentary soft emission on 100pc (0.5 arcsec) scales close to the
  nucleus to features 10s of kpc in size at larger radii. The cluster
  contains multiple high-metallicity regions with sharp
  edges. Relative to an azimuthal average, the deviations of
  metallicity and surface brightness are correlated, and the
  temperature is inversely correlated, as expected if the larger scale
  asymmetries in the cluster are dominated by sloshing motions. Around
  the western cold front are a series of $\sim 7$~kpc `notches',
  suggestive of Kelvin-Helmholtz instabilities. The cold front width varies from
  4 kpc down to close to the electron mean free path. Inside the front
  are multiple metallicity blobs on scales of 5--10~kpc, which could
  have been uplifted by AGN activity, also explaining the central
  metallicity drop and flat inner metallicity profile. Close to the
  nucleus are multiple shocks, including a 1.9-kpc-radius inner
  shell-like structure and a weak 1.1--1.4 Mach number shock around
  the central cavities. Within a 10~kpc radius are 9 depressions in
  surface brightness, several of which appear to be associated with
  radio emission. The shocks and cavities imply that the nucleus has
  been repeatedly active on $5-10$~Myr timescales, indicating a tight
  balance between heating and cooling. We confirm the presence of a
  series of linear quasi-periodic structures. If they are sound waves,
  the $\sim 5$~kpc spacing implies a period of 6~Myr, similar to the
  ages of the shocks and cavities. Alternatively, these structures may
  be Kelvin-Helmholtz instabilities, their associated turbulence or
  amplified magnetic field layers.
\end{abstract}


\section{Introduction}
The Centaurus cluster, Abell 3526, is one of the X-ray brightest and
nearest galaxy clusters in the sky, with a flux of $2.7 \times
10^{-10} \ergpcmsqps$ in the 0.1 to 2.4 keV band, corresponding to a
bolometric luminosity of $1.1\times 10^{44} \ergps$
\citep{Reiprich02}. The cluster has two sub-components, a main cluster
Cen~30, with a redshift of 0.0104 and a subcluster, Cen~45, 15 arcmin
($\sim 200$ kpc) to the east with a velocity $1500 \kmps$ higher
\citep{LuceyCurrieDickens86a}.

The first \emph{Chandra} observations \citep{SandersCent02} of the
cluster revealed a plume-like soft X-ray structure extending from the
nucleus and multiphase X-ray material on small spatial scales. Deeper
\emph{Chandra} observations \citep{Fabian05} showed that the central
region contains multiple inner cavities, where the radio plasma
displaces the X-ray emitting material \citep{Taylor02}. The plume was
resolved into wispy filaments of soft, cool, X-ray emitting gas. The data
also showed two semicircular edges in surface brightness (SB) to the
east and west, where the temperature also changes, likely to be cold
fronts \citep{MarkevitchCFShock07},  discontinuities
  where the temperature and density change but the gas is in pressure
  equilibrium. Comparison with simulations indicates that the gas is
  sloshing about in the potential well.

\emph{XMM-Newton} Reflection Grating Spectrometer (RGS) observations
\citep{SandersRGS08} of the cluster core revealed a spectrum with
emission from Fe \textsc{xvii} to \textsc{xxiv}, but not
O~\textsc{vii}, showing there is X-ray emitting material down to
temperatures between 0.3 to 0.45 keV. Above 2 keV the spectra are
consistent with $40\Msunpyr$ of radiative cooling, but at 1 keV this
reduces to $4\Msunpyr$ and less than $0.8 \Msunpyr$ appears to be
cooling below 0.4 keV. From the outskirts to the centre, the
temperature of the intracluster medium (ICM) decreases by a factor of
$\sim 10$. The mean radiative cooling time of the ICM drops to $\sim
10^7$~yr in the centre.

The soft X-ray emission in the core of the cluster is correlated with
atomic and molecular material at much lower temperatures. The central
galaxy, NGC 4696, contains a dust lane \citep{Shobbrook63} and has
bright filamentary central H$\alpha$ emission \citep{Fabian82}, both
of which have similar morphologies \citep{Sparks89}. These components
are themselves correlated with the cool X-ray emitting filaments
\citep{Crawford05}, and appear to be drawn up by buoyant bubbles
rising in the ICM \citep{Crawford05}. This model is supported by deep
integral-field spectroscopy \citep{CanningIFU11}, who find no evidence
for fast shock excitation and whose spectra support the particle
heating excitation model of \cite{Ferland09}.

Using \emph{Spitzer}, \cite{Johnstone07} found evidence for
pure-rotational lines from molecular hydrogen, indicating an
excitation temperature of $300-400$~K. The flux in the 0-0 S(1)
molecular hydrogen line correlates well with the strength of the
optical lines. Molecular material cooler than $400$~K dominates the
mass of the outer filaments. With \emph{Herschel}, \cite{Mittal11}
found two of the strongest cooling lines in the
interstellar medium, [C\,\textsc{ii}] and [O\,\textsc{i}]. The
[C\,\textsc{ii}] emission has a similar morphology and velocity
structure compared to the H$\alpha$ emission, suggesting that the soft
X-rays, optical lines and far-infrared have the same energy source. In
addition \emph{Herschel} found $1.6 \times 10^{6} \Msun$ of dust at
19~K, although the FIR luminosity suggests a low $0.13\Msunpyr$ star
formation rate.

\cite{Canning10611} detected optical coronal emission from $10^6$ K
gas in the core of the cluster. The lowest temperature probe
[Fe\,\textsc{x}] $\lambda 6374$ is twice as bright as
expected. \cite{Chatzikos15} suggest that the coronal spectrum is not
indicative of cooling material, but instead comes from a conductive or
mixing interface between the X-ray plume and optical filaments.

The \emph{Chandra} observations confirmed that the cluster has a high
($1.5-2\Zsun$) metallicity flat core with a sharp edge, previously
seen using \emph{ROSAT} and \emph{ASCA} data \citep{AllenFabian94,
  Fukazawa94, Ikebe99,Allen01}. The boxy metallicity profile requires
a large effective diffusion coefficient which drops very rapidly with
radius \citep{Graham06}. The metallicity ratios in the core of the
cluster are roughly consistent with solar values
\citep{SandersEnrich06}, indicating enrichment by both Type Ia and II
supernovae, although O and Mg are $\sim 40$ per cent less abundant
\citep{Sakuma11}. The metallicity profile and ratios show that the
centre of the cluster appears to have not suffered major disruption
over the past 8 Gyr or longer.

In the centre of the cluster there is a drop in the observed
metallicity \citep{SandersCent02}, which appears not to be caused by
resonance scattering \citep{SandersReson06} or inadequate spectral
modelling \citep{Fabian05}. The decrement is consistent with the
deposition of metals onto grains which are incorporated into dusty
filaments \citep{Panagoulia15}.

The Cen\,45 subcluster has a temperature excess surrounding it,
interpreted as having been heated by its interaction with Cen\,30
\citep{Churazov99}. This excess was confirmed using \emph{XMM} and can
be explained by a simple shock heating model \citep{WalkerCen13}. A
pressure jump is also seen in the direction of the merger and the
merging subcluster appears to have retained its metals. No bulk
velocities greater than $1400 \kmps$ in the ICM were detected in the
cluster \citep{Ota07}.

The extended component of the central radio source in the cluster,
PKS\,1246--410, appears to be interacting with and displacing the ICM
\citep{Taylor02,RudnickBlundell03}, most obviously seen by the
presence of two central cavities in the ICM. The most extended radio
emission appears to go beyond to regions of low thermal pressure,
indicating there may be further X-ray cavities there
\citep{Crawford05}. The optical nucleus is
  double \citep{Laine03}. The radio nucleus is associated with compact
  low-luminosity X-ray emission, but is offset from the
  X-ray-brightest region of the cluster \citep{Taylor06}. If the
nucleus were accreting at the Bondi rate, it would overproduce by 3.5
orders of magnitude its observed X-ray radiative luminosity.
Very long baseline interferometry (VLBI) observations show
a broad one-sided jet \citep{Taylor06}. The cluster shows Faraday
rotation measures, indicating magnetic fields of 25 $\mu$G on 1~kpc scales.
The thermal pressure appears dominant over the magnetic
pressure. The line-emitting gas, soft X-ray material, regions with an
excess of rotation-measure and depolarised regions appear to be
spatially correlated \citep{RudnickBlundell03,Taylor07}.

\cite{SandersSound08} found X-ray SB ripples in
several directions within the inner 40 kpc, postulating that they are
sound waves in the ICM generated by the activity of the central
nucleus, similar to those found in Perseus \citep{FabianPer06} and
Abell 2052 \citep{Blanton11}. Such sound waves could transport energy
from the nucleus to provide the distributed heating required to
prevent the excepted high cooling rates.

In this paper we present new results from 760 ks of
  \emph{Chandra} observations, including 480 ks of new data. We assume
  the galaxy cluster lies at a redshift of 0.0104
  \citep{LuceyCurrieDickens86a}. Using $H_0=70 \kmpspMpc$, 1 arcsec on
  the sky corresponds to 0.213 kpc. Images are aligned with north to
  the top and east to the left. For details of the data reduction, see
  Appendix \ref{sect:dataanalysis}. We discuss the larger scale
  structure in Section \ref{sect:larger}, including in detail sloshing
  gas in the potential well (\S\ref{sect:sloshing}), the western cold
  front (\S\ref{sect:cfront}), the linear structures inside the cold
  front (\S\ref{sect:sbstructure}) and the high metallicity blobs
  (\S\ref{sect:Zstructure}). In Section \ref{sect:central} we examine
  the inner region around the nucleus in detail, including the inner
  shock (\S\ref{sect:innershock}), the interaction of the radio source
  and X-ray plasma (\S\ref{sect:radio_icm}), the cavities
  (\S\ref{sect:cavities}), the shock surrounding the cavities
  (\S\ref{sect:centcavity}), the association between soft X-rays and
  dust (\S\ref{sect:optical}), the origin of the plume
  (\S\ref{sect:plumeorigin}) and the rotation measures and magnetic
  fields (\S\ref{sect:rm}). We discuss the inner multiphase structure
  in Section \ref{sect:thermprof} (also Appendix
  \ref{sect:thermprofscentre}), the central abundance drop in Section
  \ref{sect:elemabund} and conclude in Section
  \ref{sect:conclusions}.

\section{Larger scale structure}
\label{sect:larger}
\subsection{X-ray images}
\label{sect:images}

\begin{figure*}
  \includegraphics[width=0.83\textwidth]{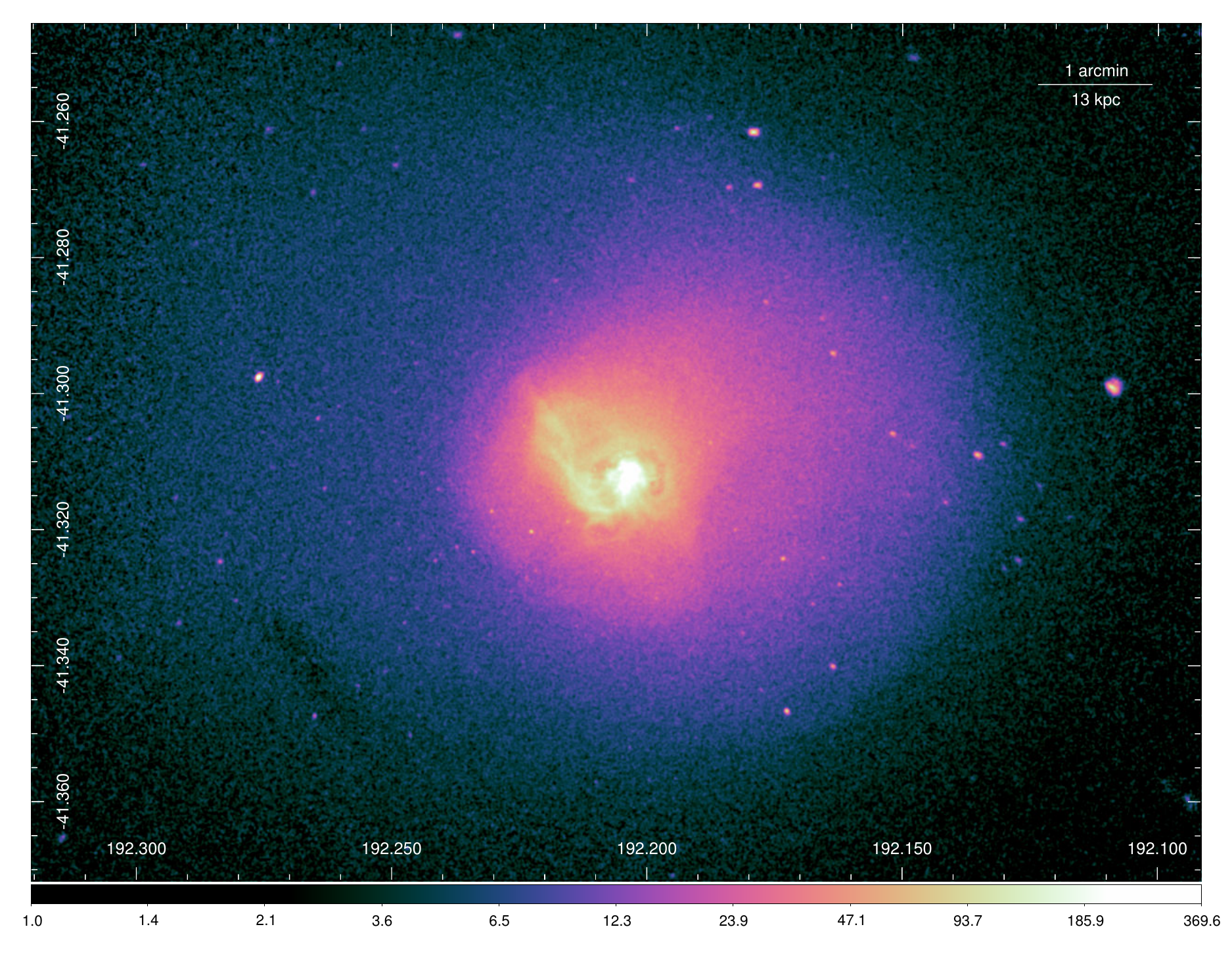}
  \includegraphics[width=0.83\textwidth]{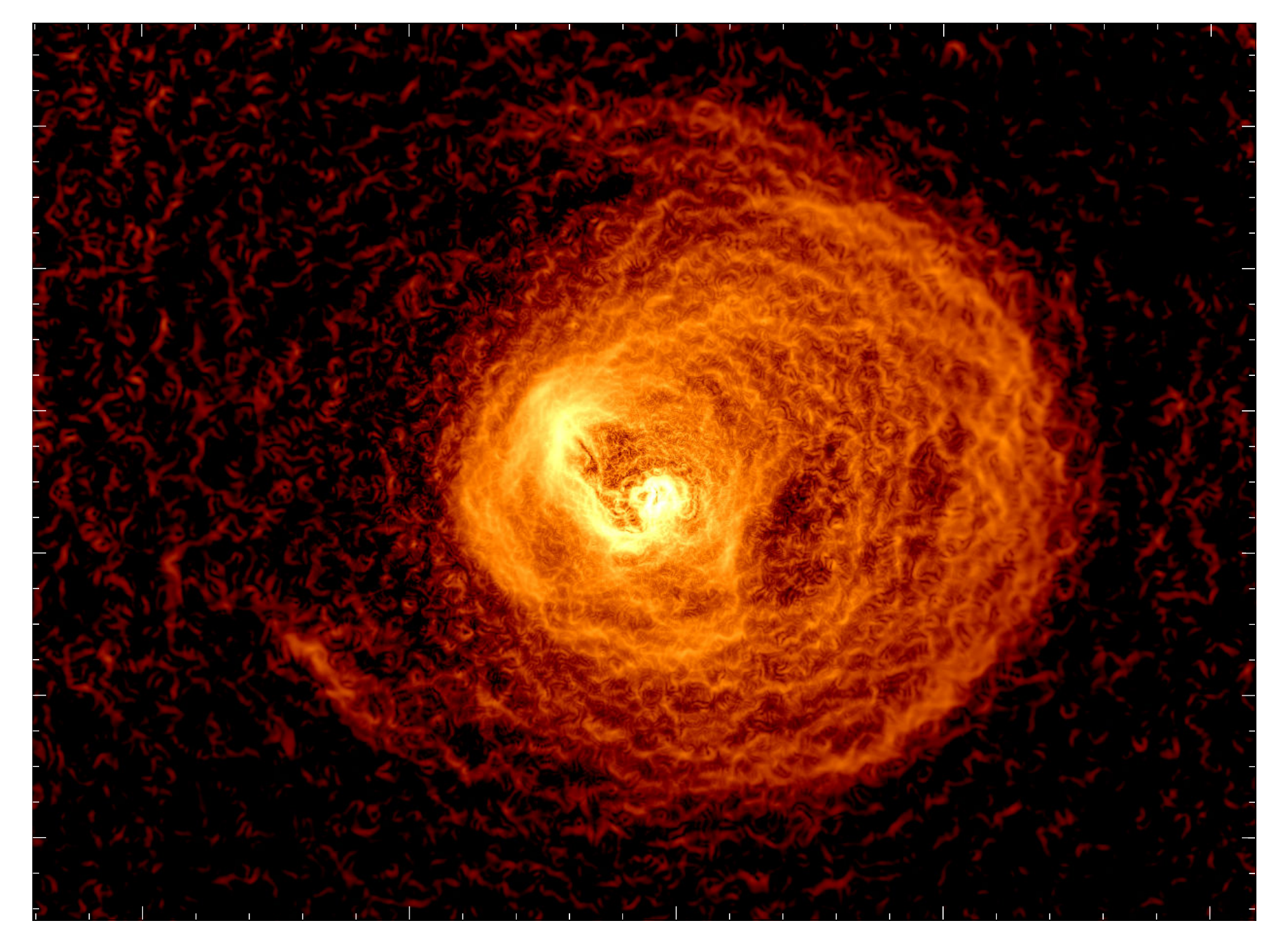}
  \caption{(Top panel) X-ray image of the central $130 \times 100$~kpc
    of the cluster. The 0.5 to 7 keV image is background-subtracted
    and exposure-corrected.  To the south-east can
      be seen the X-ray shadow of a disc galaxy. The colour bar shows
    the SB, scaled to show the central number of counts per 0.492
    arcsec pixel. (Bottom panel) The same region after applying GGM
    gradient filters with different scales and summing with radial
    scaling. Note that the scale is not consistent
      across the image and does not accurately reflect the magnitude
      of jumps. It emphasises the regions with largest SB gradients
      after point source removal.}
  \label{fig:centre}
\end{figure*}

\begin{figure*}
  \includegraphics[width=\textwidth]{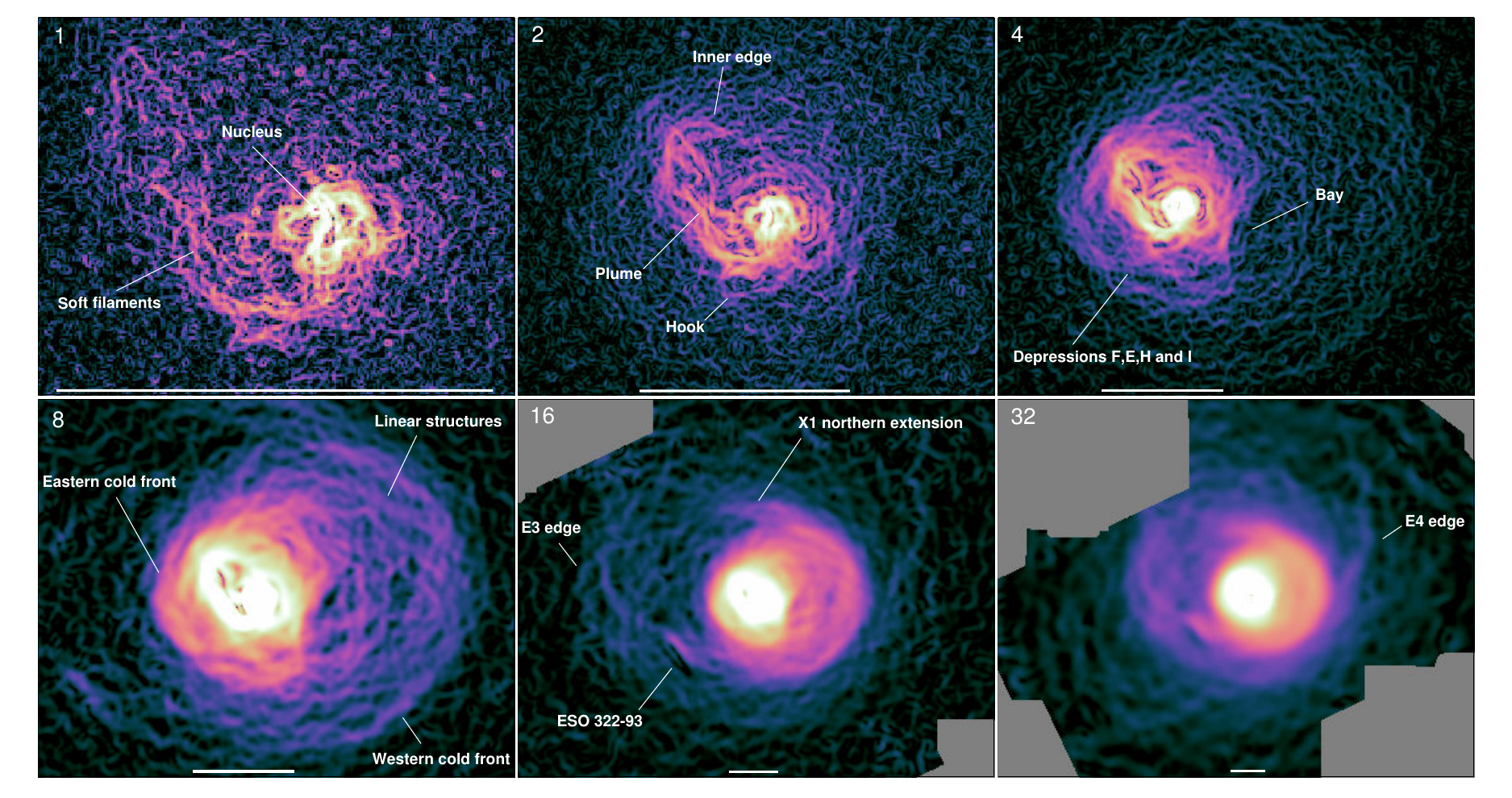}
  \caption{GGM gradient-filtered images using scales
    from 1 to 32 pixels (1 pixel is 0.492 arcsec). The white bar has a
    length of 20 kpc ($\sim 94$ arcsec). Labelled features are
    referred to in the text.}
  \label{fig:filtscale}
\end{figure*}

Fig.~\ref{fig:centre} (top panel) is an X-ray image of the central
$130\times 100$~kpc of the cluster. There are $8.1 \times 10^6$ counts
in the dataset inside a 5 arcmin radius.  Prominent in the image is the
central feedback region associated with NGC 4696, which we examine in
depth in Section \ref{sect:central}. There are SB edges, previously
identified as cold fronts \citep{SandersCent02,Fabian05}, 1.5 arcmin
to the east and 3.3 arcmin to the west. We examine the western cold
front in detail in Section \ref{sect:cfront}. The depression 3.3
arcmin to the south-east is due to an edge-on disc galaxy,
ESO\,322-93, in the main Cen~30 cluster which absorbs some soft
cluster X-ray emission \citep{Fabian05}.

To better examine the features in the image we applied a gradient
filter to highlight sharp and flat regions.
  Gradient filtering has previously been used to examine cluster
  simulations \citep{Roediger13}. The Gaussian gradient magnitude
(GGM) filter is similar to the Sobel filter, calculating the magnitude
of the image gradient assuming Gaussian derivatives. By varying the
Gaussian width, $\sigma$, the filter is sensitive to gradients on
different scales. GGM filtering has the advantage of being a simple
convolution applied to the data and is unlikely to introduce
artifacts.

We applied the GGM filter from \textsc{scipy}
(\url{http://www.scipy.org/}) to the exposure-corrected
background-subtracted image using $\sigma=1$, 2, 4, 8, 16 and 32 detector
pixels, yielding the images shown in Fig.~\ref{fig:filtscale}. Regions
with large gradients, such as edges, are lighter, while flatter SB
areas are darker. Point sources were first removed from the input
image to avoid their contamination of larger structures. They
were initially identified using \textsc{wavdetect} but manually edited
to exclude false detections when it became confused by the bright central
X-ray structures. The values of pixels in the removed source regions were
replaced by random selections of those in the 1-pixel-wide areas
immediately surrounding each point source.

The ability of the filter to detect features is a
  function of the SB in the region of interest and the magnitude of
  the gradient. In the cluster outskirts where the number of counts
  per pixel is low the small length scale filters cannot detect
  gradients unless they are very large. This is the case with normal
  X-ray images, where the detectability of structures depends on their size,
  SB and contrast with their surroundings.
  The noise in the images, however, is seen as
  azimuthal structures. As the largest SB gradient in a cluster is
  radial, Poisson fluctuations in the X-ray image will amplify or
  weaken this radial signal. Therefore noise in the gradient image lies
  perpendicular to the gradient in the original image.

To show the gradients on different scales in a single image
(Fig.~\ref{fig:centre} bottom panel), we added the GGM-filtered images
with radial scaling factors. These radial scalings were applied to
increase the contribution of small scales to the centre (measured from
the central radio nucleus) and reduce the contribution of large scales
there. We experimented with scaling factors, finding linear summations
of the different filtered images do not introduce features which are
not present in the individual filtered images. We note that the image
does not accurately show the relative magnitude of features but
instead highlights those features which are there. The image shows a
number of edges and flat structures around the nucleus and inside the
western cold front (see Section \ref{sect:sbstructure}). Further
discussion of gradient filtering and its application to other X-ray
cluster datasets will be made in a forthcoming paper.

\subsection{Temperature and metallicity maps}
\label{sect:maps}
\begin{figure*}
  \includegraphics[width=0.77\textwidth]{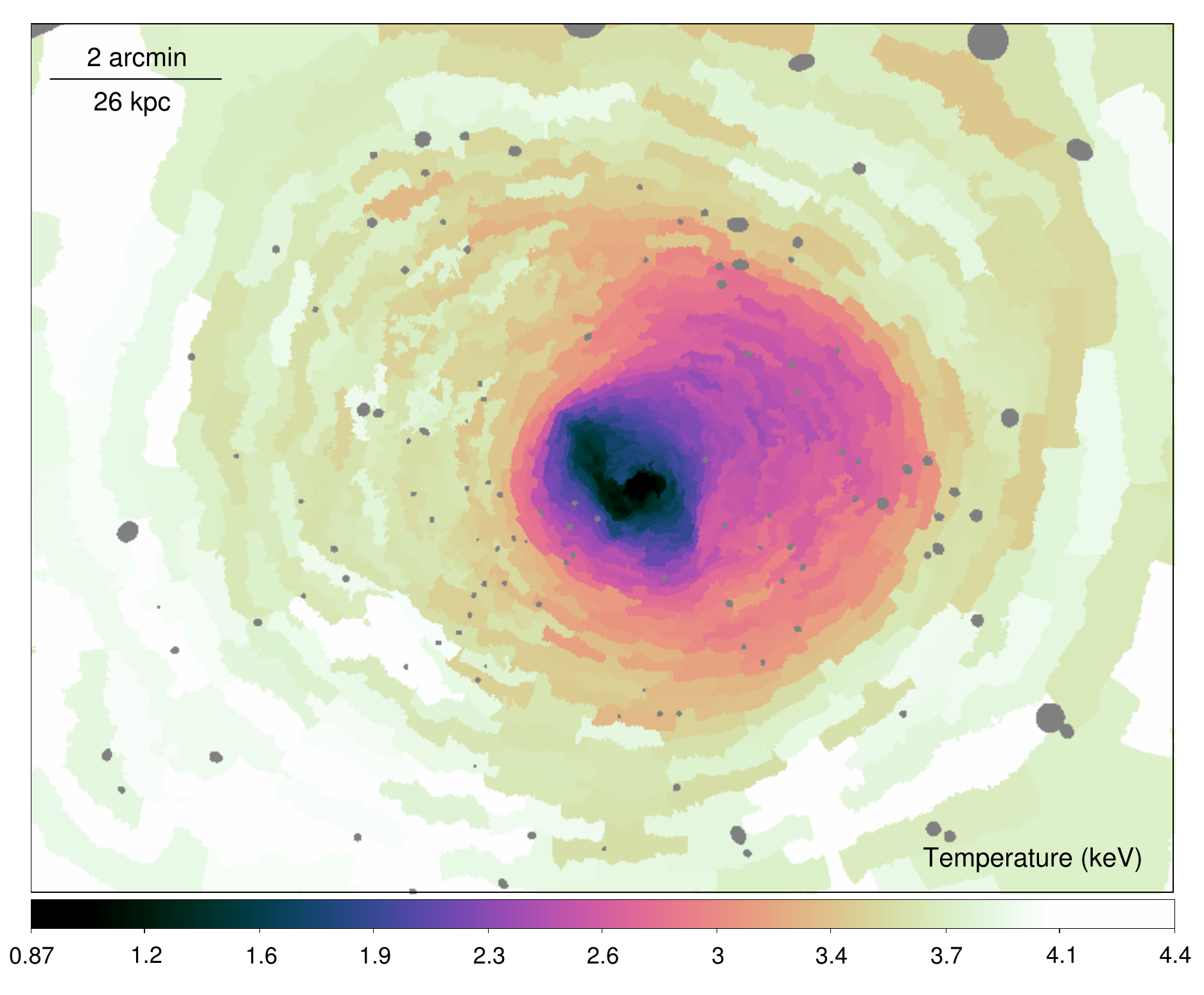}
  \includegraphics[width=0.77\textwidth]{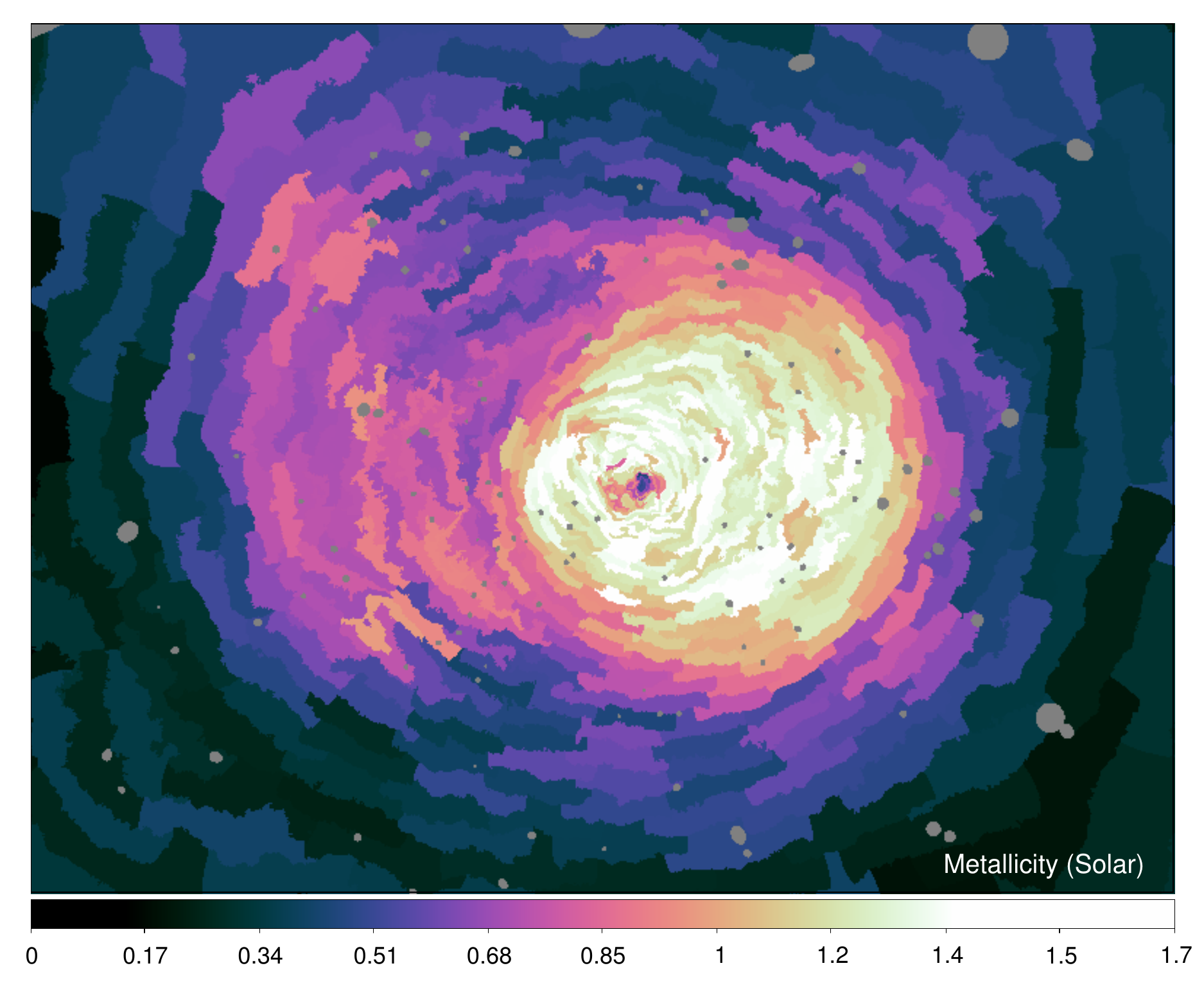}
  \caption{Temperature and metallicity maps of the cluster using
    regions with $S/N=100$. The temperature map shows the
    single-component temperature. The metallicity map shows the results from
    single- or two-component fits, depending on which is
    preferred. Grey regions indicate excluded point sources.
    Uncertainties in the temperatures vary from 0.6 per cent in the
    centre to 2 per cent to the east of the western edge and 3.5 per
    cent at large radius. Metallicity uncertainties are typically from
    $0.05-0.10\Zsun$, excluding the very centre.}
  \label{fig:wide_T_Z}
\end{figure*}

Fig.~\ref{fig:wide_T_Z} shows maps of the temperature and metallicity,
obtained by spectral fitting (Appendix \ref{sect:specmapping}).  The
cool central plume and cooler regions inside the western and eastern
SB edges can be clearly seen. From the core is another plume-like
cooler structure extending north-westwards towards the western
edge. Outside the edges the temperature is relatively flat, although
there may be a cooler region towards the north-east.

The metallicity map contains a great deal of structure. There is a
previously reported high metallicity region inside the western cold front edge,
where the ICM is cooler \citep{SandersCent02,Fabian05}. The abundance
distribution, however, is not steadily rising, but is rather flat with
fluctuations. There are strong variations in metallicity (see Section
\ref{sect:Zstructure}) with a characteristic length scale of around 20
arcsec ($\sim 5\kpc$; see Section \ref{sect:zfluctdetails}). On the
eastern side of the cluster, outside the eastern SB edge
(Fig.~\ref{fig:centre}), is another region of higher metallicity gas
with a tail-like appearance.

\subsection{Sloshing}
\label{sect:sloshing}
\begin{figure}
  \centering
  \includegraphics[width=0.87\columnwidth]{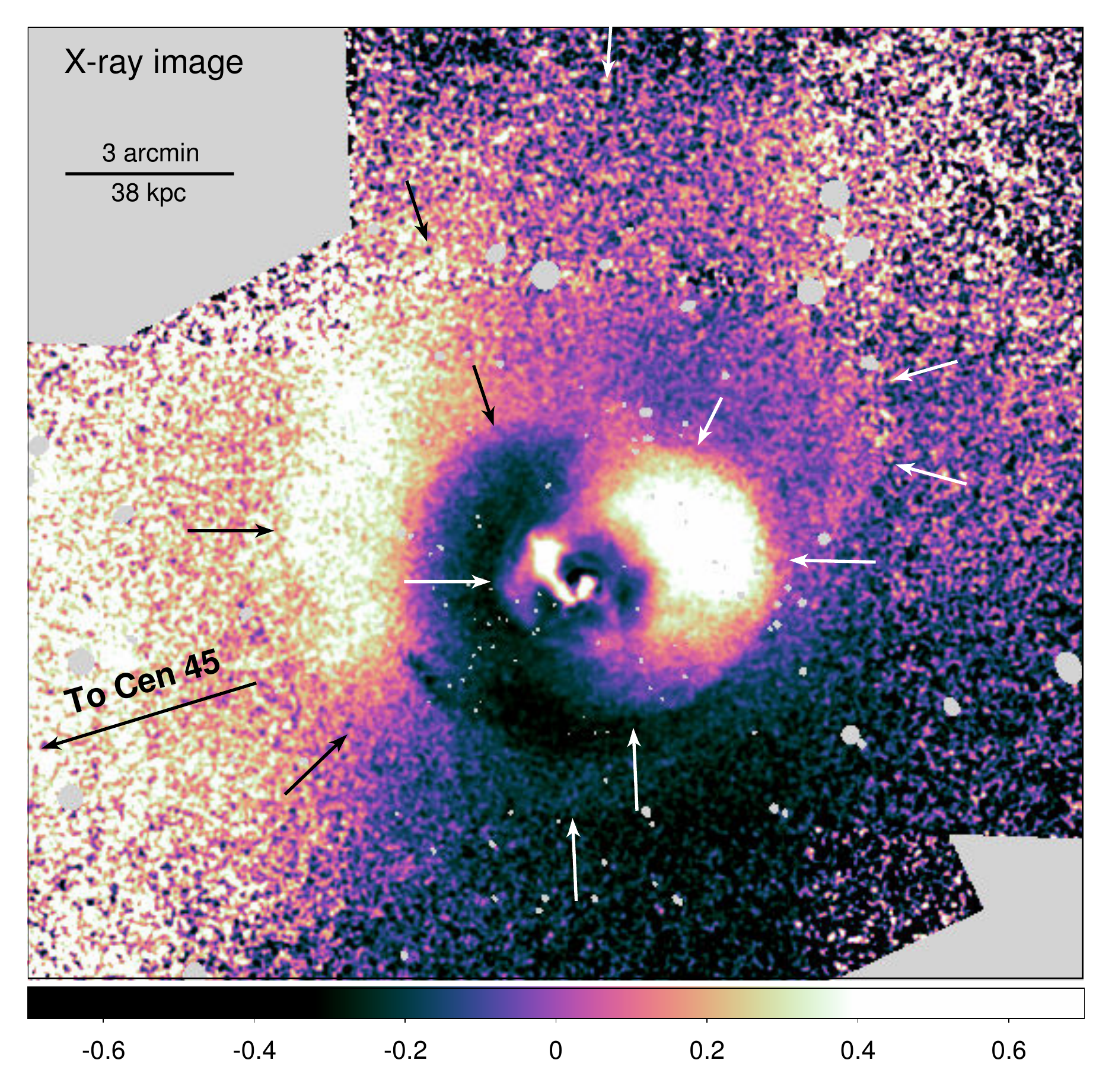}
  \includegraphics[width=0.87\columnwidth]{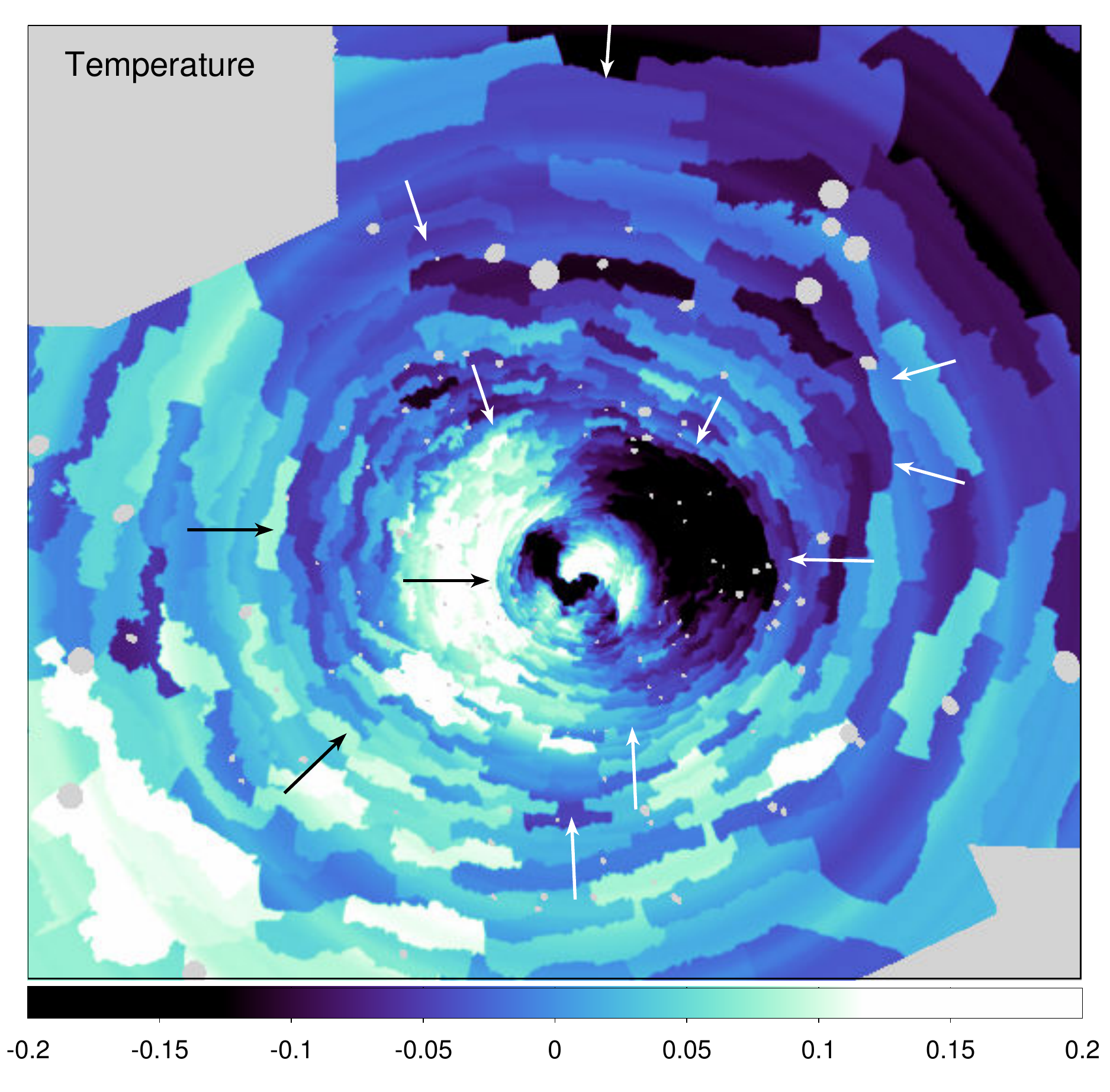}
  \includegraphics[width=0.87\columnwidth]{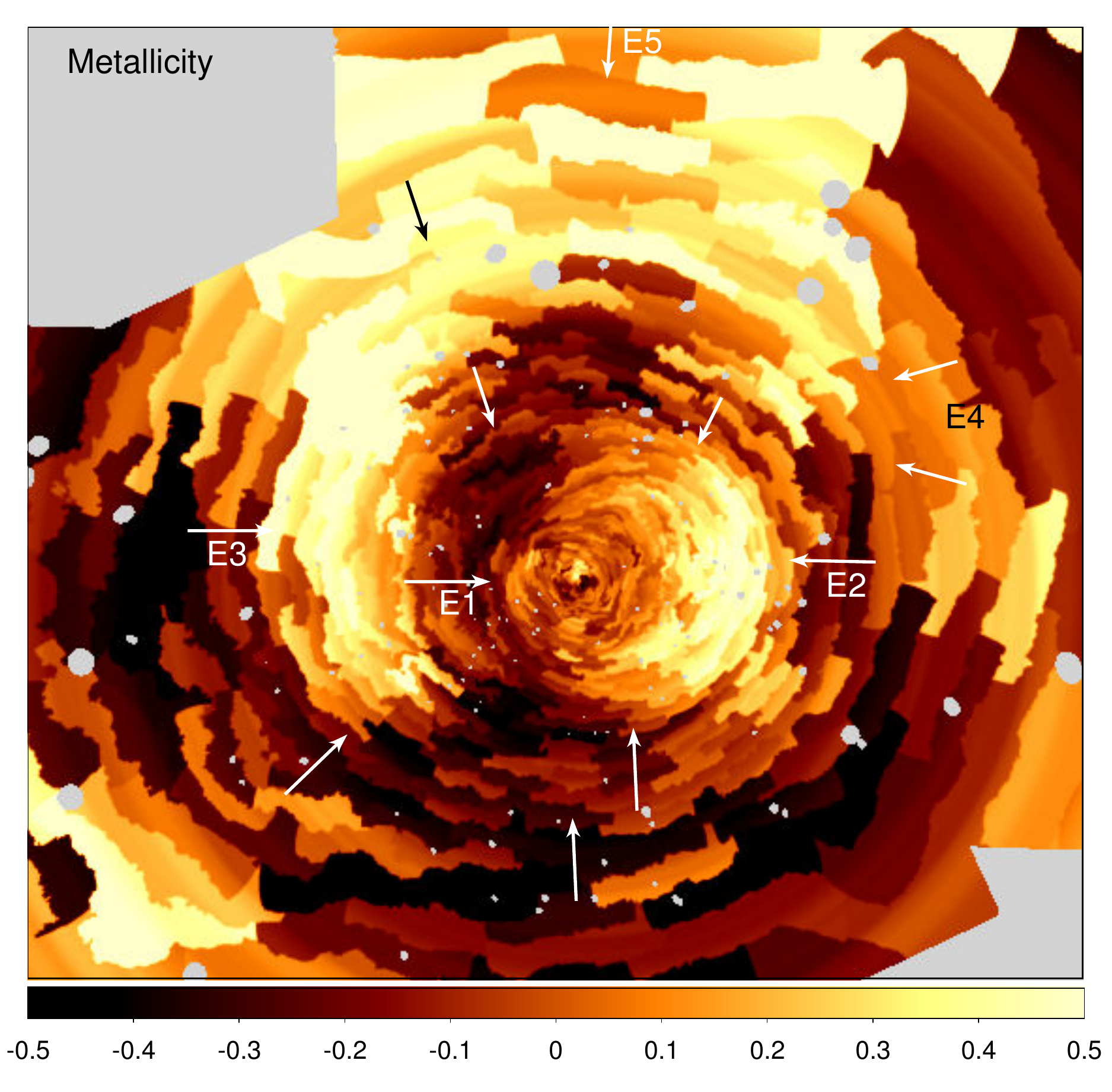}
  \caption{Fractional difference from the average at
    each radius for X-ray 0.5 to 7 keV SB, temperature and
    metallicity. The X-ray image was smoothed by a 2 arcsec
    Gaussian. The temperature and metallicity maps used the same data
    as presented in Fig.~\ref{fig:wide_T_Z}. Arrows mark edges
    (discussed in text). Also shown is the direction of the merging
    Cen\,45 subcluster.}
  \label{fig:subav}
\end{figure}

We compare the SB deviations in the cluster to the metallicity and
temperature in Fig.~\ref{fig:subav}. There is a low-density,
high-temperature material on the opposite side of the core to the
high-density, low-temperature material behind the western cold
front. This material appears to sweep from the south through the east
and towards the north.  The plot shows striking correlations between
SB, temperature and density deviations. The western cold front is
labelled E2. There are also two edges in metallicity and SB in the
opposite direction towards the eastern side of the cluster (marked E1
and E3), closer to the Cen 45 subcluster. In addition there is a
further edge outside the western cold front, going from the westernmost point of the cold front and
extending towards the north, with more metal rich gas inside
it (E4). Further to the north is the high metallicity region already
noted by \cite{WalkerCen13}. There are hints that there is a SB edge
to the north of this feature, with lower metallicity outside it
(E5). The SB enhancement towards the north-west of Fig.~\ref{fig:subav} may
be the structure found by \cite{Churazov99} (see their figure 3),
suggested to be stripped gas from NGC\,4696B.

\begin{figure}
  \centering
  \includegraphics[width=\columnwidth]{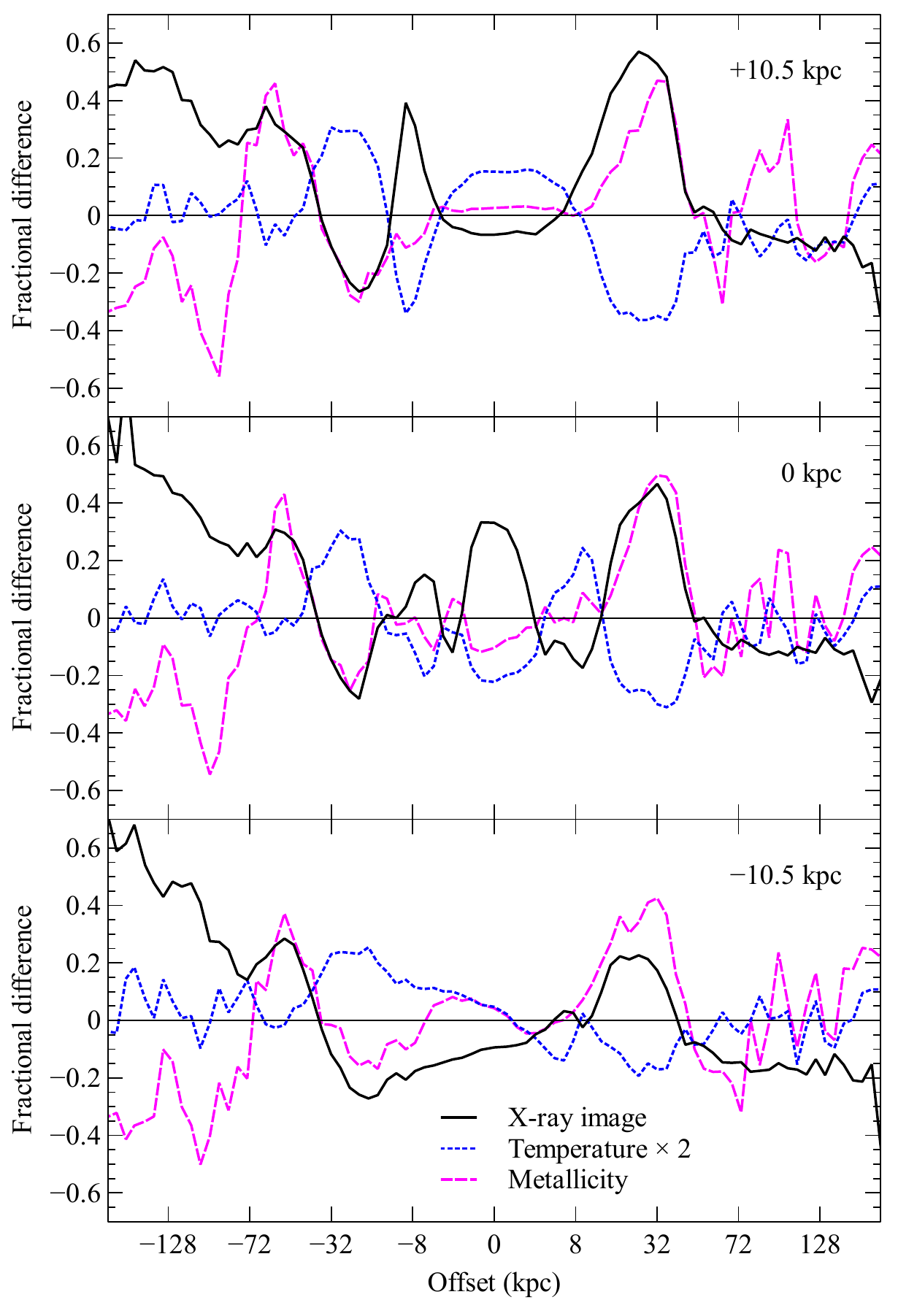}
  \caption{SB, temperature and metallicity fractional
    residual profiles in an east-west direction (west is positive)
    across the centre of the cluster with three northerly offsets
    ($-10.5$, $0$ and $10.5$~kpc), computed from
    Fig.~\ref{fig:subav}. The width of each strip is $8.4$~kpc. The
    horizontal axis has a square-root scaling. Note that the
    temperature profiles have been scaled by a factor of 2.}
  \label{fig:subav_prof}
\end{figure}

In Fig.~\ref{fig:subav_prof} are shown three rectangular profiles
across the cluster in an east-west direction. One passes through the
nucleus, while one is north of it and the other south. In some regions
there is excellent correlation (between SB and metallicity) or
anti-correlation (of SB or metallicity with temperature) between the
variables, but this is not always the case. For example, in the
central region, the bright, cool, plume is relatively low in
metallicity. There is also a fairly strong east-west SB gradient,
which makes the fluctuations difficult to see.

\begin{figure}
  \includegraphics[width=\columnwidth]{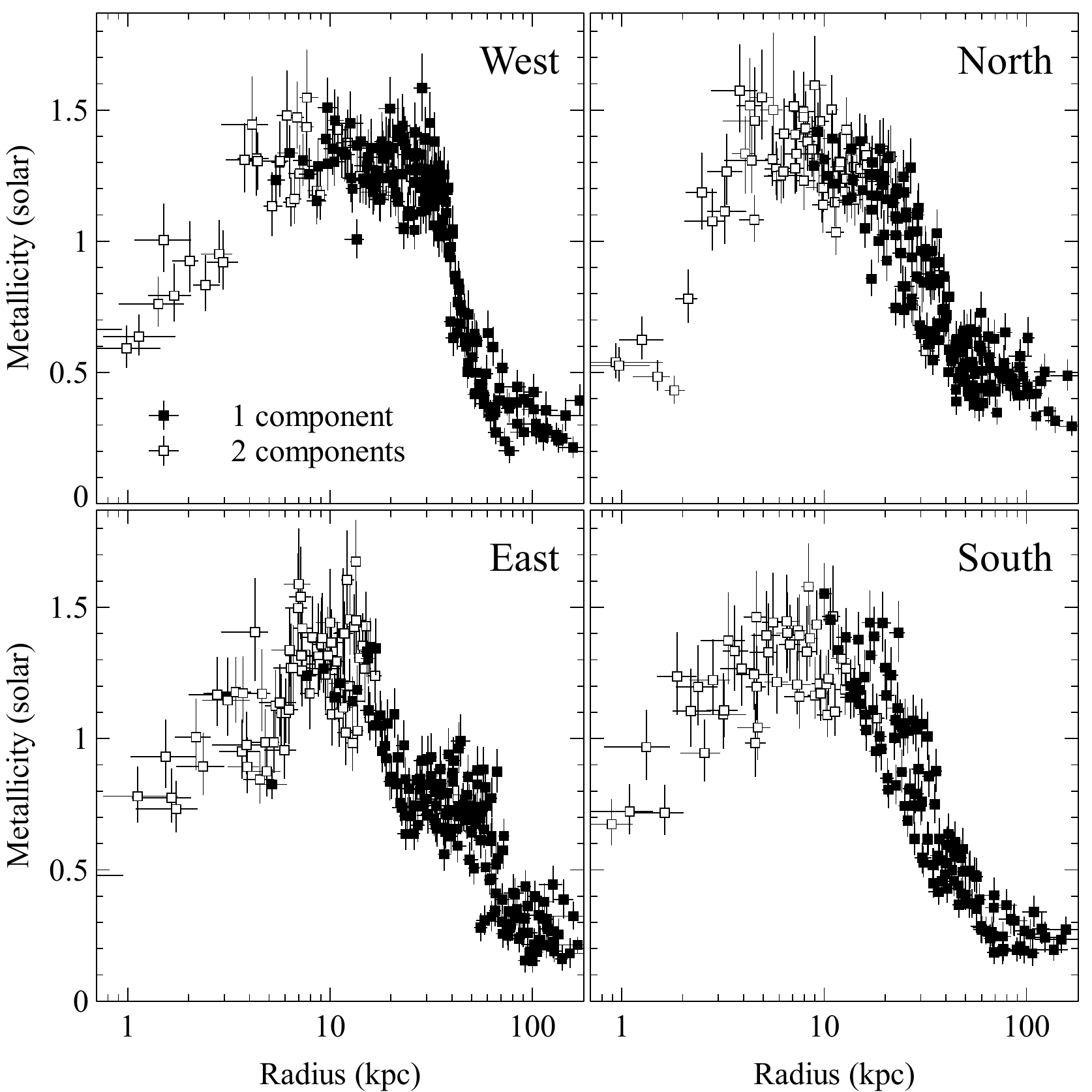}
  \caption{Radial metallicity profile of the map in
    Fig.~\ref{fig:wide_T_Z} split into $90^\circ$ sectors. The number
    of components used in the fits is indicated by open or closed
    symbols. Horizontal error bars show the radial range of each bin.}
  \label{fig:Z_prof}
\end{figure}

Fig.~\ref{fig:Z_prof} shows radial profiles in metallicity in four
quadrants to the north, south, east and west. Despite small-scale
variation, the overall metallicity profile inside the western edge
(roughly 40 kpc radius) is remarkably flat. In addition there remains
a central drop in metallicity, which we discuss further in Section
\ref{sect:elemabund}. In the eastern sector, there is a much smaller
region of flat metallicity towards the cluster core. However, the
tail-like structure has a flat metallicity profile which drops outward
of 60--70 kpc.

The spatial correlation between SB and metallicity
  enhancement, and anti-correlation with temperature is a typical
  signature of sloshing and has been found in simulations as well as
  other observations (e.g. Virgo, \citealt{Roediger11}; A\,496,
  \citealt{Roediger12}; Perseus, \citealt{Simionescu12}; A\,2029
  \citealt{PaternoMahler13}).  Similarly, sloshing in the cluster can
  distort and induce asymmetries in the distribution of metals about
  the central galaxy.

The gravitational perturbation by the merger of the Cen\,45 subcluster
may have started these sloshing motions. This is suggested by the
alignment of the larger scale SB deviations along the axis of the
merger (Fig.~\ref{fig:subav}).  After the
  interaction by the perturbing system, cold fronts move outwards with
  time \citep[e.g.][]{Ascasibar06}.  The A\,496 system,
  which has a similar temperature to Centaurus, was simulated by
  \cite{Roediger12}, who found that the radial evolution of the cold
  front mainly depends on the cluster potential.  The radii of the
  eastern and western inner Centaurus cold fronts (19 and 42 kpc,
  respectively) are similar to the fronts in their `distant' run after
  1 Gyr (see their fig. 7).  The circular symmetry of the features
  suggest that the sloshing gas motions are close to being
  perpendicular to the line of sight, as pointed out by
  \cite{Ascasibar06}. In this case, the perturbing system is not
  moving in the plane of the sky. It is not possible to tell whether
  Cen 45 is the perturber based on its position, velocity and the time
  since the interaction. Indeed, Cen 45 may not be on its first pass
  through the system and could be gravitationally bound.

\subsection{Western cold front}
\label{sect:cfront}

\begin{figure}
  \includegraphics[width=\columnwidth]{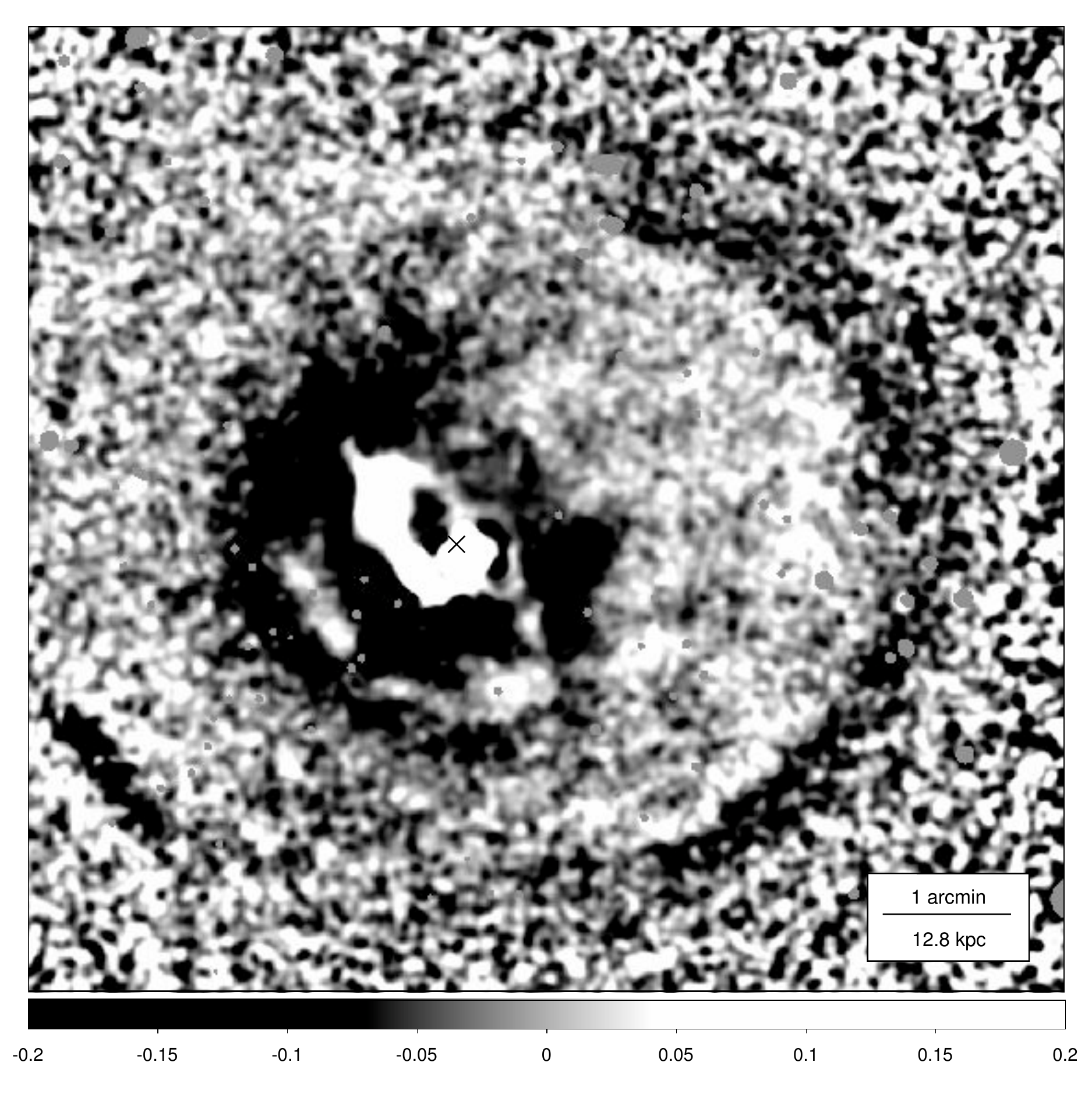}
  \includegraphics[width=\columnwidth]{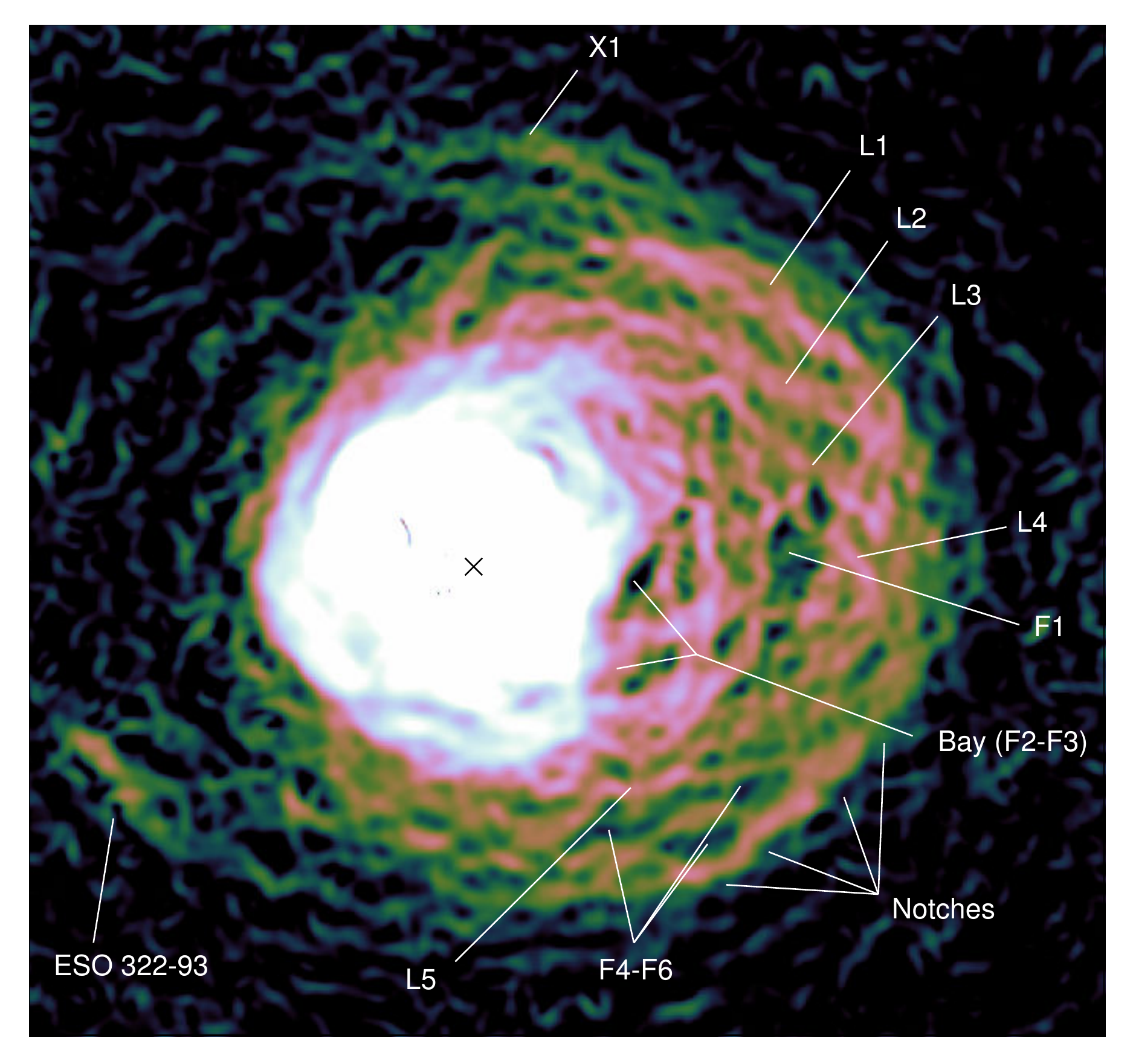}
  \caption{(Top panel) Unsharp-masked 0.5 to 7 keV image. This is the
    fractional difference between Gaussian-smoothed images with
    $\sigma=2.0$ and $15.7$ arcsec, after masking out point
    sources. `$\times$' marks the central nucleus (see Section
    \ref{sect:central}). (Bottom panel) GGM-filtered image of the same
    region, using $\sigma=7.9$ arcsec, taken from
    Fig.~\ref{fig:filtscale}. The labelled features are described in
    Section \ref{sect:sbstructure}.}
  \label{fig:unsharp}
\end{figure}

The western cold front in Centaurus is one of the clearest and nearest
examples known in a cluster, spanning nearly $180^\circ$. The pressure
across the edge is continuous, confirming it is a cold front (Appendix
\ref{sect:frontfit}).  Fig.~\ref{fig:unsharp} (top panel) shows an
unsharp-masked image of the centre and cold front region, highlighting
the SB fluctuations around the front and within the central
region. The edge is not perfectly smooth; the south-western side
appears to be where it is sharpest and directly west it appears
broader.  To the north and south of the cold front there is a rapid
transition from a sharp edge to a weak or non-existent edge. However,
some circular structure may continue on in the north-east and
south-east directions. There appear to be a number of `notches' into
the edge on length scales of $\sim 7\kpc$. The notches along the edge
can also be seen in a gradient filtered X-ray image
(Fig.~\ref{fig:unsharp} bottom panel). We discuss the SB features
inside the front in Section \ref{sect:sbstructure}.

\begin{figure}
  \includegraphics[width=\columnwidth]{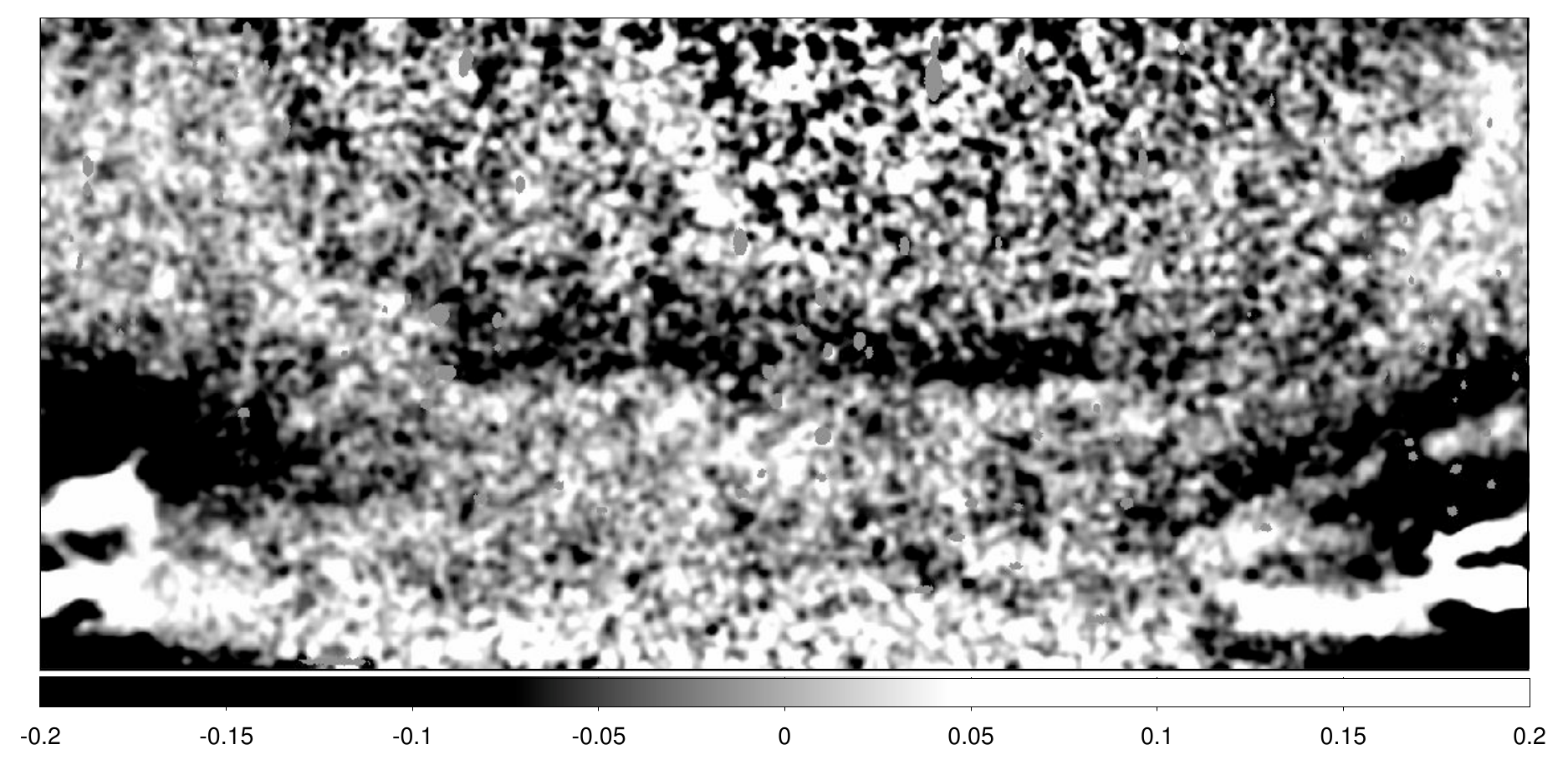}
  \caption{Unsharp-masked image of the cluster remapped to have radius
    and angle axes, with a centre chosen to make the western cold
    front lie at constant radius.  Horizontally it spans $360^\circ$
    from east (left), through north, then west (centre), then back to
    east (right). Radius runs vertically from 2 (bottom) to 68 kpc
    (top). The 0.5 to 7 keV input image was rebinned into 0.1~kpc
    by $0.25^\circ$ pixels. Unsharp masking was then applied,
    using the fractional difference between maps smoothed by 4 and 32
    pixels.}
  \label{fig:cfront_remap}
\end{figure}

The deviations from a circular cold front can be clearly seen if the
image of the cluster is remapped to radial and azimuthal coordinates
(Fig.~\ref{fig:cfront_remap}). The notches appear as wave-like
irregularities on the straight edge. The edge is best represented by a
circle with a centre at RA $192.18325$ and Dec $-41.31048$ (marked by
the sectors in Fig.~\ref{fig:unsharp}), 11~kpc to the west of the
nucleus.

\begin{figure}
  \includegraphics[width=\columnwidth]{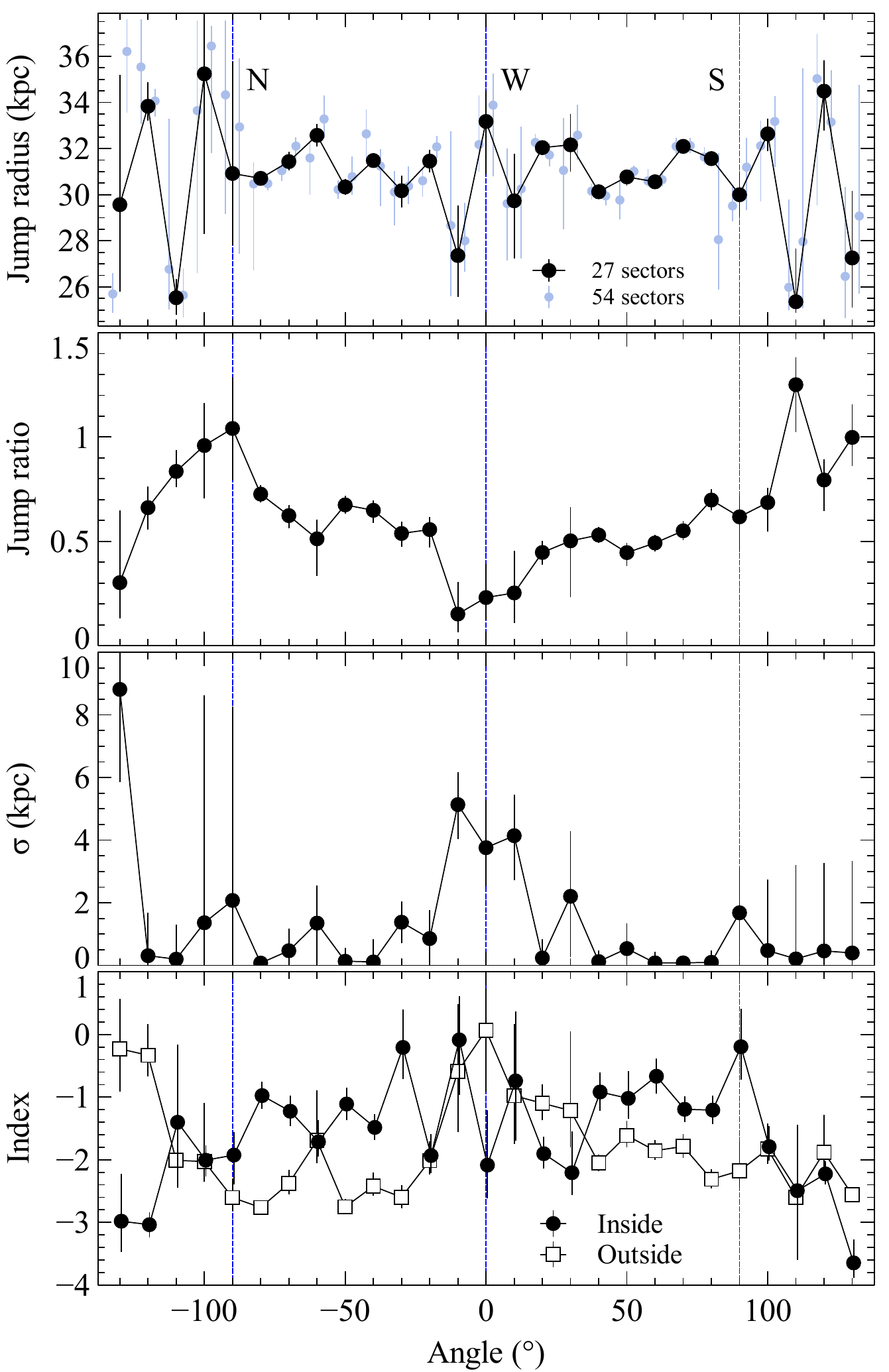}
  \caption{Fit parameters for the profiles of the 27 sectors around
    the western cold front from the north-east ($-130^\circ$) to
    south-east ($130^\circ$). These are the median and $1\sigma$
    percentiles obtained from the analysis (Appendix
    \ref{sect:frontfit}). Shown are the radius of the jump, the ratio
    in emissivity outside to inside (the jump ratio), its width
    ($\sigma$) and the powerlaw indices of the emissivity inside and
    outside. For the radius we also show the results
    using twice the number of sectors.}
  \label{fig:cfront_fit_params}
\end{figure}

To quantify the variation and width of the edge around the front we
fit models to SB profiles extracted in 27 independent $10^\circ$ sectors spanning
$270^\circ$, covering the $180^\circ$ towards the west where the front
is visible and $45^\circ$ beyond this in either direction. This
analysis is similar to that presented in \cite{Werner15}. We fit a
broken powerlaw with a broadened edge model to the profiles (see
Appendix \ref{sect:frontfit}).  Fig.~\ref{fig:cfront_fit_params} shows
for each sector the radius of the front, the jump magnitude, the width
and the powerlaw indices to either side of the edge.

Between the north and south, where the edge is well defined, the jump
radius is fairly constant at 31~kpc, with sector-to-sector variations
of about 1~kpc, likely due to the notches around the edge.
Immediately west, the front is at its broadest at 4~kpc, but it
is much narrower to the north-west and south-west. In some
locations the edge appears narrower than 1~kpc. Where the front is
narrowest towards the south-west, for example in the sectors between
$60-80^\circ$, the $1\sigma$ upper limits on the edge width are $\sim
0.3$~kpc (accounting for the PSF at the edge in
  these sectors).  If there are no magnetic fields, in the north-west
direction the density and temperature imply values for the collisional electron
mean free path of $0.18$ and $0.36$~kpc (0.8 and 1.7 arcsec), interior
and exterior to the front, respectively. In the south-west these
increase to $0.30$ and $0.48$~kpc (1.4 and 2.3 arcsec),
respectively. Where the front is narrowest to the south-east, our
upper limits are similar to the electron mean free path.

The emissivity of the ICM decreases outwards over most of the front to
around 60 per cent of the interior value, implying that the density
decreases by around 20 per cent, which is in agreement with spectral
measurements (Appendix \ref{sect:frontfit}). However, immediately to
the west the fits prefer an 80 per cent SB decrease. In the same
region the powerlaw index of the model outside the jump is flatter and
the edge is wider.  It is clear from the unsharp-masked and filtered
images in Fig.~\ref{fig:unsharp} that the edge is much less distinct
immediately to the west at the most extreme part of the front. There
may also be a SB edge which extends from the westernmost point of the
front to the north-west. This can be seen where SB edge E4 apparently
connects to the cold front E2 (Fig.~\ref{fig:subav}). This feature can
also be seen as the enhancement in the remapped cluster image
(Fig.~\ref{fig:cfront_remap}) from the centre of the front extending
upwards (to larger radius) and leftwards (north-west).

Outside the $180^\circ$ sector where the edge is obvious, the SB fits
imply that there is some sort of jump or break in SB in the radial
range examined. The allowed jump radii vary considerably, however, and
are unlikely to correspond to the same physical features in the
profiles.

The notches along the edge of the western cold front
(Figs. \ref{fig:unsharp} and \ref{fig:cfront_remap}) are consistent
with Kelvin-Helmholtz instabilities (KHIs) of length scales of $\sim
7$~kpc along the edge. KHIs in clusters were first
  seen in simulations \citep{Roediger11,ZuHone11}. A kink, which may
be a large KHI, has been observed in Abell~496
\citep{Roediger12}. Similar notches of length of a few kpc have also
been seen in the cold front in the Virgo cluster
\citep{Werner15}. These notches likely make the radius of the cold
front appear to vary by $\sim 1$ kpc
(Fig.~\ref{fig:cfront_fit_params}) between the north and south.
Comparison with simulations of cold fronts including varying levels of
viscosity or magnetic fields suggests levels of viscosity of around 10
per cent of the Spitzer value are consistent with the data
\citep{Roediger13,ZuHone15}. Larger amounts of viscosity smooth out
all instabilities, while instabilities grow very large with no
viscosity or magnetic fields.

\subsection{Linear structures inside the front}
\label{sect:sbstructure}
Using shorter observations we previously detected ripple-like SB
fluctuations using Fourier high pass filtering
\citep{SandersSound08}. These quasi-periodic fluctuations were also
seen in SB profiles in three sectors (along the north-east, south-west
and north-west directions) We interpreted these features as sound
waves generated by AGN activity.

Examining the unsharp-masked image (Fig.~\ref{fig:unsharp} top panel)
there are a number of SB fluctuations with a similar
$\sim 5-10 \kpc$ length scale to the previously found
fluctuations. They are also similar in size to the notches along the
cold front (Section \ref{sect:cfront}). However, they do not appear to
have a periodic structure.  Creating SB profiles using
the same sectors as in \cite{SandersSound08}, obtains consistent
SB profiles and residuals to $\beta$ model fits,
showing that the fluctuations are robust.

Using a GGM filter on the data instead of unsharp-masking
(Fig.~\ref{fig:unsharp} bottom panel) finds a number of quasi-linear
features (labelled L1--L5). These straight edges have corresponding
features in the unsharp-masked image, but the GGM filter appears to be
a much more powerful tool for detecting and connecting these edges in
SB. All of the labelled structures in Fig.~\ref{fig:unsharp} (bottom
panel) can be seen as SB edges in the raw data when examined
closely. The structures are also apparent in the multiple-scale
GGM-filtered image (Fig.~\ref{fig:centre} bottom panel).

We tested that these features are significant by similarly filtering a
Poisson realisation of a heavily-smoothed cluster image, finding that
the noise was at lower levels. In addition, all the labelled features
can be seen by closely inspecting the raw images. L1--L4 are
approximately parallel to each other. L1 and L4 (and perhaps L2 and
L3) also appear to bend with a sharp angle close to the edge of the
cold front. The image also highlights some flat SB
regions (F1--F6). The bay (discussed in Section \ref{sect:central})
appears as two flat regions surrounded by steep gradients. There are
also flat regions closer to the edge of the front (F1 and F4--F6). To
the north is an extension of the cluster emission extends
northwards (X1).

Several of the identified linear features L1-L6 appear to have an
approximately regular separation of $\sim 5$~kpc. L1--L3 seem to have
a sharp bend near the cold front, suggestive of waves reflecting off a
boundary. The temperature on the outside of the cold front is a factor
of $\sim 1.4$ greater than inside. Therefore the critical angle for
total internal reflection at the cold front is $58^\circ$. If the
features are sound waves, then they lie above this critical angle, and
so it is possible that the bends at the cold-front edge are caused by
total internal reflection of sound waves. Alternatively, variations in
sound speed could cause the fronts to bend if the wave passes through
a sharp temperature gradient at an angle.

The 5~kpc spacing corresponds to 6~Myr timescale at the sound speed at
2.5~keV temperature. This timescale is similar to the age of the inner
shock and cavities. It is therefore plausible that the structures
could be generated by AGN activity. The quasi-periodic nature of the
features argues against the features being turbulence generated by the
AGN. We analyse these fluctuations quantitatively in \cite{Walker15}
and examine whether they could contribute significantly to AGN
feedback heating.

Alternatively, the structures may be unrelated to
  the AGN but instead caused by sloshing. The sloshing simulations of
  \cite{Roediger13}, in particular their fig.~8, shows features
  similar in appearance. KHIs on the curved 3D surface of the cold
  front can be seen as projected features inside the cold front
  edge. The interaction of KHIs at the cold front, combined with the
  slow outwards motion of the cold front can also give rise to turbulence
  inside or underneath the cold front, which also adds fine
  structure. Linear features were also found by \cite{Werner15} inside
  the M\,87 cold front. They suggested that these could instead be
  generated by the amplification of magnetic field layers by sloshing,
  as found in the simulations of \cite{ZuHone11} and \cite{ZuHone15}.

\subsection{Metallicity blobs}
\label{sect:Zstructure}

\begin{figure}
  \includegraphics[width=\columnwidth]{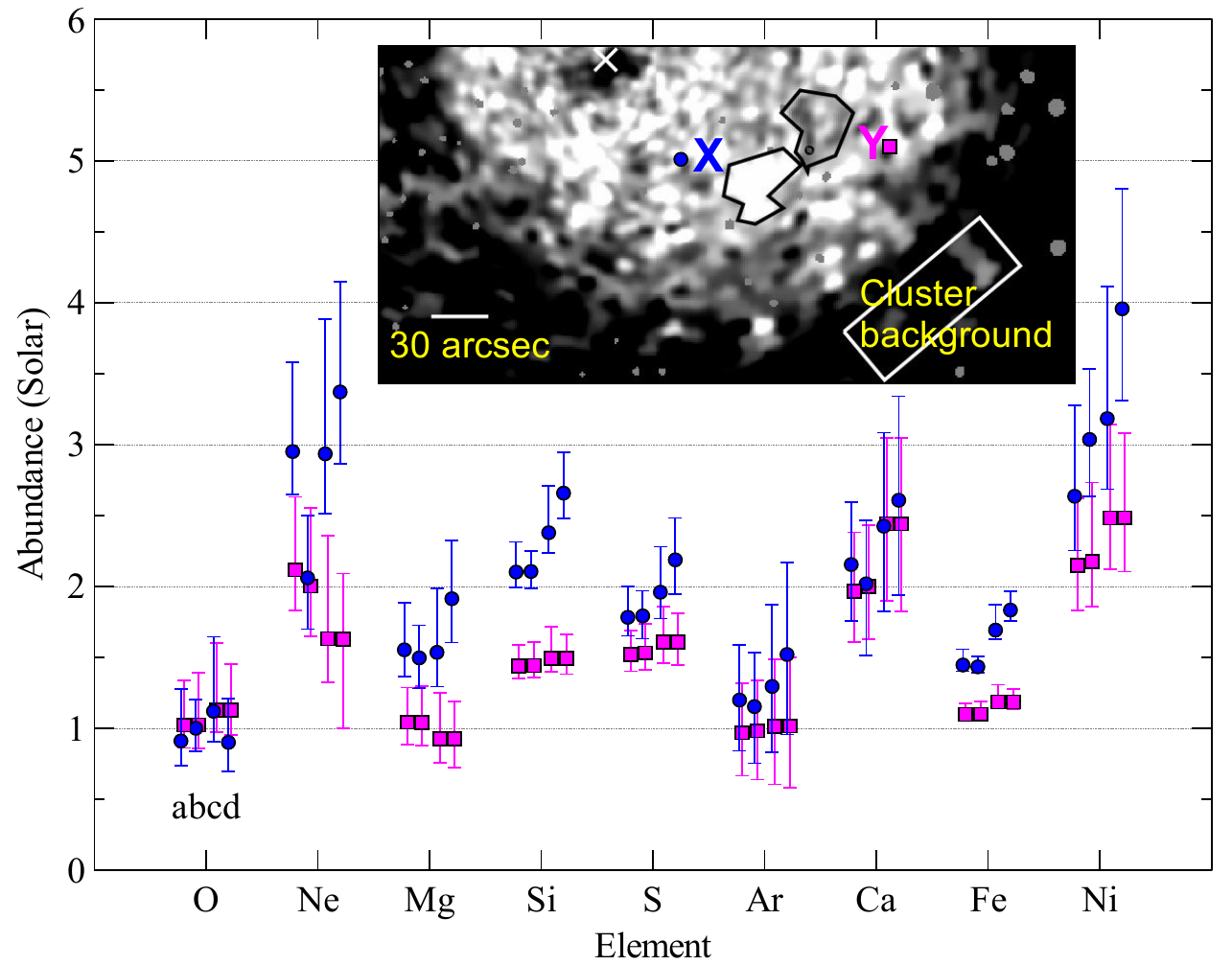}
  \caption{Comparison of the metallicity in adjacent
    high (X) and low (Y) metallicity regions, fitting (a) 1 and (b) 2
    component(s), blank-sky background, and (c) 1 and (d) 2
    component(s), cluster background (see Appendix
    \ref{sect:zfluctdetails}). The inset metallicity map was created
    from $S/N=40$ regions.}
  \label{fig:highZ}
\end{figure}

\begin{figure*}
  \centering
  \includegraphics[width=0.8\textwidth]{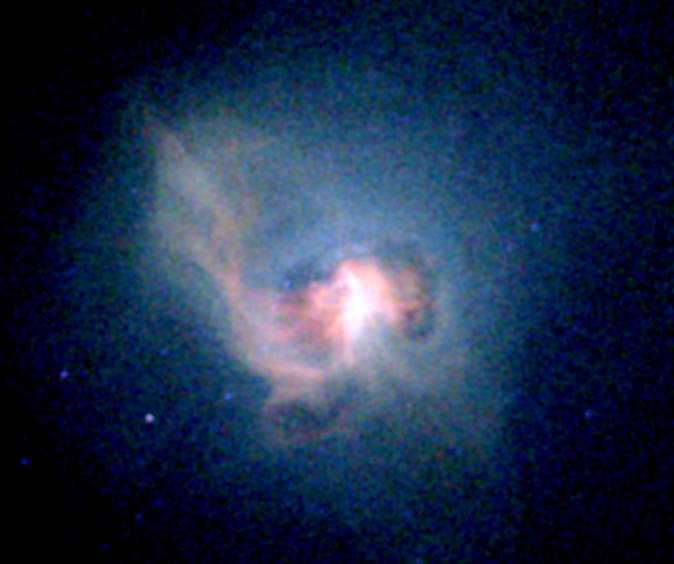} \vspace{5mm} \\
  \includegraphics[width=0.6\textwidth]{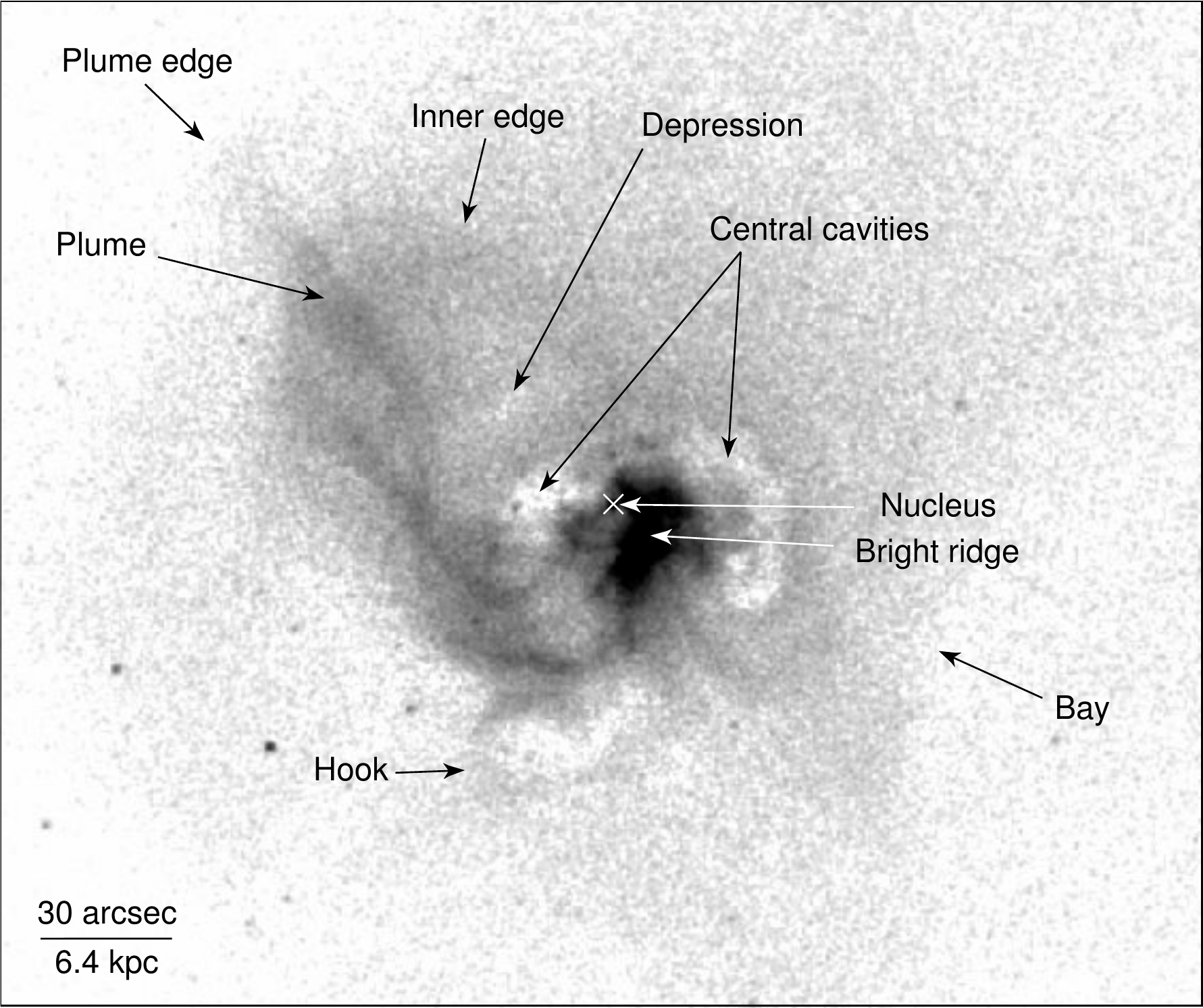}
  \caption{(Top panel) RGB image of the cluster centre. The bands
    0.5--1, 1--1.5 and 1.5--7 keV are shown as red, green and blue,
    respectively. Images were extracted using bins of 0.5 detector
    pixels (0.246 arcsec) and smoothed by a Gaussian of width 0.492
    arcsec. The image measures 2.8 by 2.3 arcmin (35 by 29
    kpc). (Bottom panel) Partially unsharp-masked 0.5-7 keV greyscale
    image showing labelled features, constructed by taking the X-ray
    data in 0.246 arcsec pixels and smoothing by a Gaussian of width 1
    pixel and subtracting the 0.5 times the image smoothed by 64
    pixels.}
  \label{fig:rgb}
\end{figure*}

There are fluctuations in metallicity interior to the western cold
front on $\sim 5-10$~kpc scales (Fig.~\ref{fig:wide_T_Z}). These
features are robust against changes to the binning scheme. We compared
two neighbouring regions offset from the nucleus with high (X) and low
(Y) metallicity (Fig.~\ref{fig:highZ}) to confirm that these
structures are significant.  Fitting the spectra separately, we
obtained the metallicities using one and two temperature component
fits, and for two different background spectra.

The Si and Fe abundances are greater in region 1 for every model. The
Fe metallicity is 30 per cent greater in region 1 using a blank-sky
background, increasing to 55 per cent if a cluster background is
used. It should be noted that if a blank-sky background is used, the
Fe metallicity calculated from the Fe-K lines is consistent ($\sim 1
\Zsun$) between X and Y. However, a cluster background makes the Fe-K
line strength consistent with the increase in metallicity seen in the
Fe-L lines. The cluster background is likely more realistic, as we
know that there is projected cluster emission. Within the
uncertainties, the enhancements in the other elements are
consistent with the Fe increase.  There is no preference for increased
amounts of Type Ia or Type II supernova products in the high
metallicity region.

We estimated the effects of projection by
  considering an $8^3 \kpc^3$ enhancement in metallicity at 25 kpc
  radius, using the profiles across the SW front (Appendix
  \ref{sect:frontfit}). For regions of this size, the measured
  enhancement is likely suppressed by a factor of $\sim 4$. Therefore
  this blob has $\sim 5$ times the metallicity of the nearby low
  metallicity material, assuming that the simple model of an enhanced
  cubic blob is correct. Region X would contain an enhancement of
  around $10^6 \Msun$ of Fe over region Y.

  The flatness of the western metallicity profile and
  sharpness of the cold front edge implies high effective central
  diffusion rates which rapidly drop with radius
  \citep{Graham06,Walker14}.  The existence of the compact high
  metallicity regions argues that diffusion by stochastic gas motions
  is low.  In contrast, sloshing of gas in the potential well does not
  lead to broadening of the distribution \citep{Roediger12}, but
  results in asymmetric and distorted distributions. If the AGN
  bubbles drag dusty clouds of high metallicity gas produced by
  stellar mass loss from the core of NGC\,4696 to larger radius, this
  will naturally flatten out the central peak and give the observed
  effective diffusion coefficient which rapidly declines with radius
  \citep{Panagoulia13}. In addition, this intriguingly raises the
  possibility that the high metallicity regions are metal-rich
  materials uplifted from the galactic centre by AGN bubbles.  High
  metallicity blobs on similar scales have appeared in other clusters
  which have good quality data (e.g. Perseus, \citealt{SandersPer07}
  and Abell 2204, \citealt{SandersA220409}) and could be produced by
  a similar process.

  If the material was injected into a much smaller region and has
  expanded to size $L$ over a timescale $t$, the diffusion coefficient
  is $D \sim L^2/(24t)$ in three dimensions. Using the sloshing
  timescale (1 Gyr; Section \ref{sect:sloshing}) and a size of 8 kpc,
  this translates into a diffusion coefficient $D \sim 10^{27} \cm^{2}
  \s^{-1}$. If the cavity age (e.g. 10~Myr; Section
  \ref{sect:central}) is a more appropriate timescale because the
  blobs are uplifted by the AGN, then the diffusion coefficient would
  be two orders of magnitude greater.

\section{Nuclear region}
\label{sect:central}
\subsection{Images and spectral maps}

Fig.~\ref{fig:rgb} (top panel) shows an RGB image of the central $\sim
30$~kpc of the cluster around the central radio source. Cool
structures in the cluster appear as red in this scheme. In
Fig.~\ref{fig:rgb} (bottom panel) is a labelled image of the same
region, with partial unsharp-masking.  Obvious is the soft cool
plume-like structure extending to the north-east
\citep{SandersCent02}. In these new, deeper, observations it can be
seen that the plume is made up of at least three separate
filaments. There are the two strong central cavities in the X-ray
either side of the nucleus which are filled with high-frequency (GHz)
radio emitting plasma \citep{Taylor02}. Connecting the plume to the
nucleus is a SB edge, labelled the inner edge. Between this edge and
the plume appears to be a SB depression. The plume appears to extend
through the edge, ending a few kpc beyond that. The inner edge may
represent the edge of an egg-shaped region of enhanced SB. To the
south-east of the core is another cool filament with a hook-like
shape. There is a SB edge to the south-west, with an unusual negative
curvature. We name it the bay, as it looks similar to the bay seen in
the core of the Perseus cluster \citep{FabianPer06}. Similar
structures are also seen in Abell 1795 and Abell 2390
\citep{Walker14}.

\begin{figure}
  \centering
  \includegraphics[width=0.85\columnwidth]{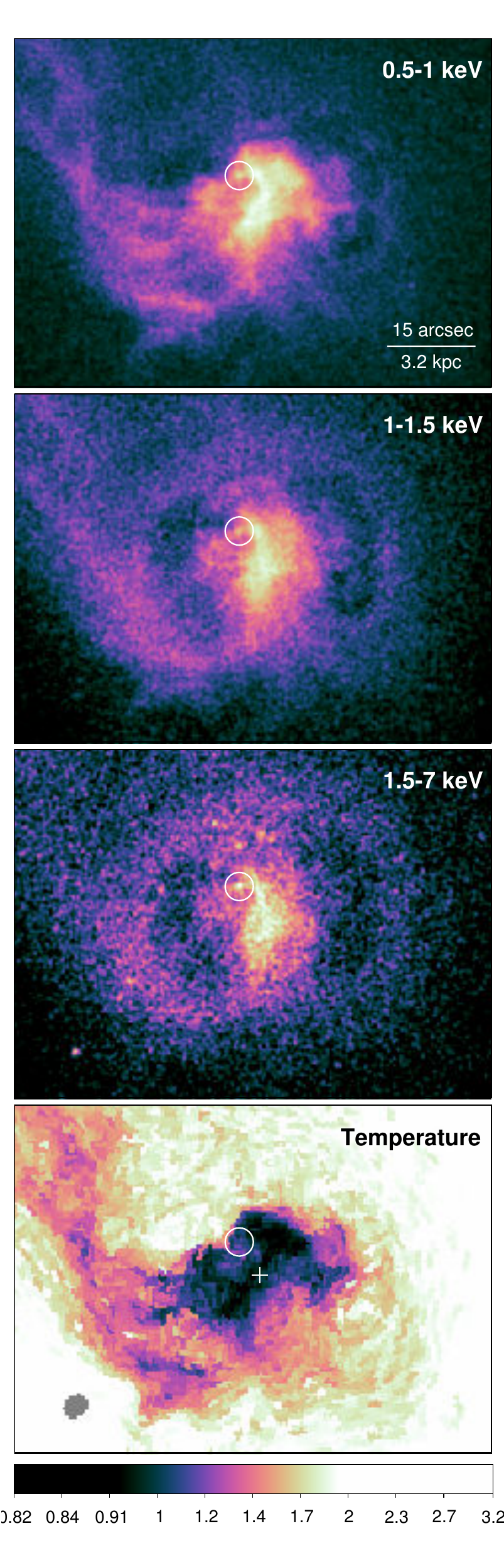}
  \caption{Detail of the central regions in three energy bands and
    comparison temperature map. X-ray images were created with 0.246
    arcsec binning with 1 pixel Gaussian smoothing. A 2.4 arcsec
    radius circle marks the radio nucleus and `+' shows the X-ray
    centroid.}
  \label{fig:vcentre}
\end{figure}

\begin{figure*}
  \centering
  \includegraphics[width=0.43\textwidth]{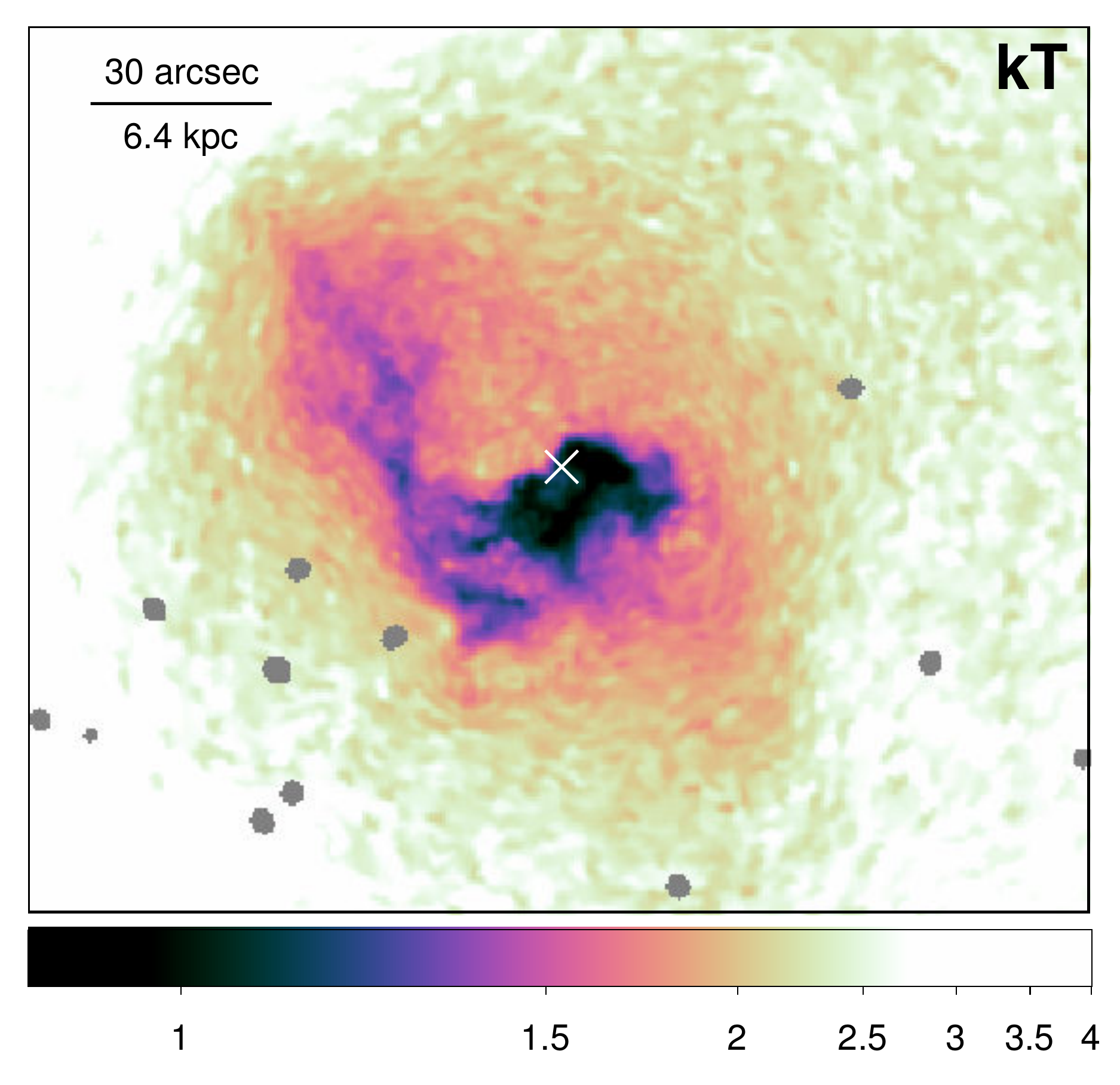}
  \includegraphics[width=0.43\textwidth]{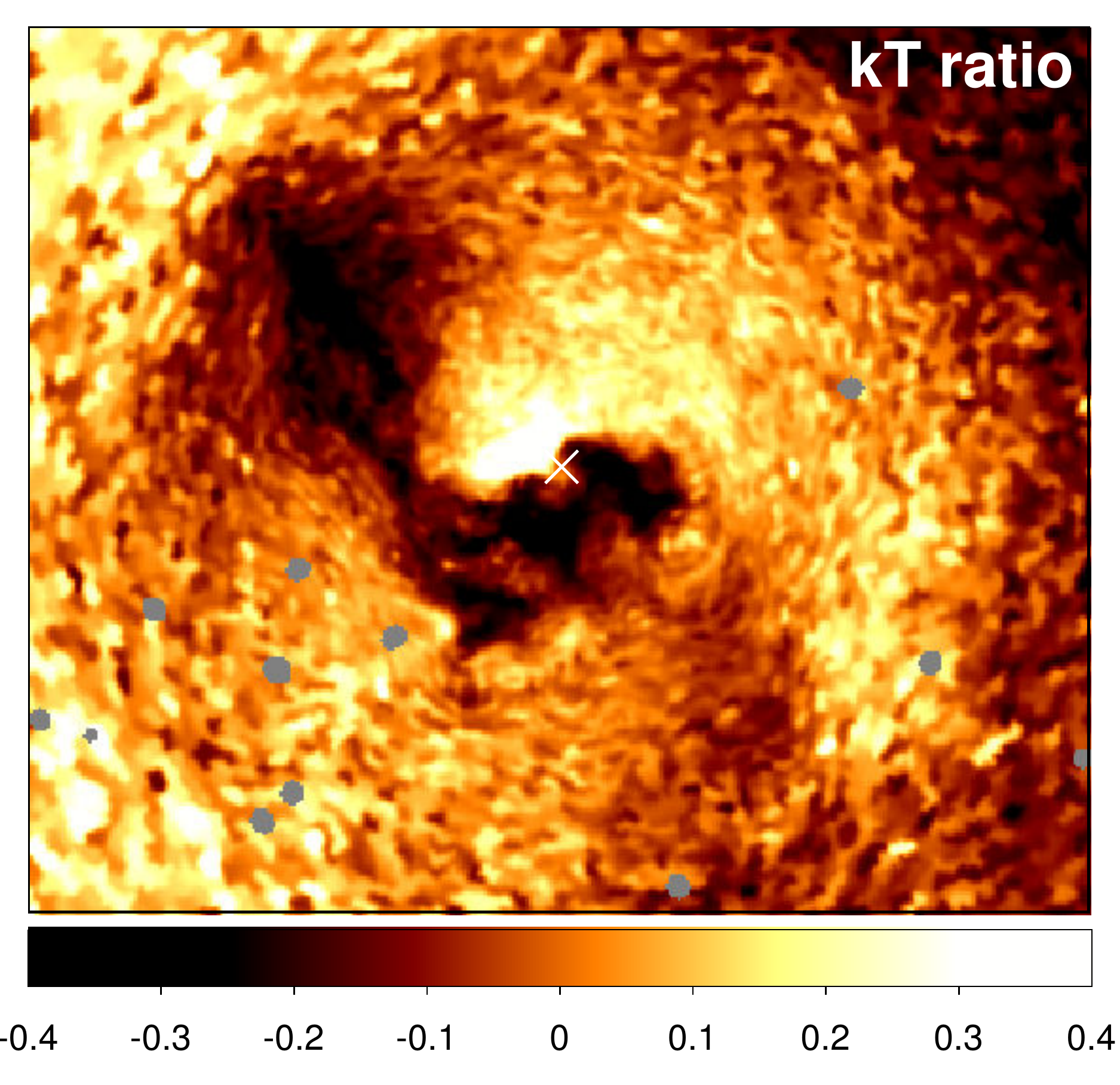} \\
  \includegraphics[width=0.43\textwidth]{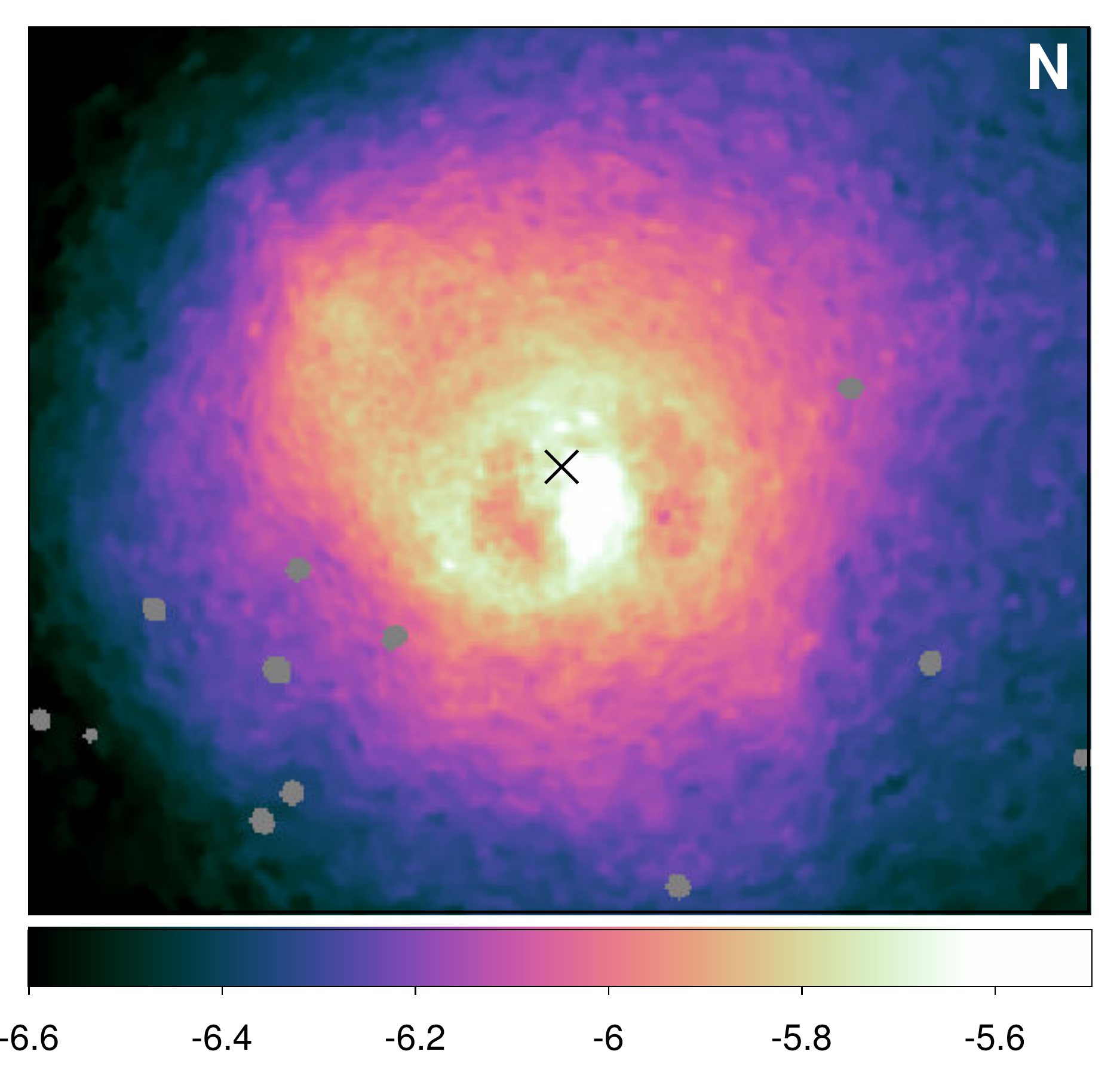}
  \includegraphics[width=0.43\textwidth]{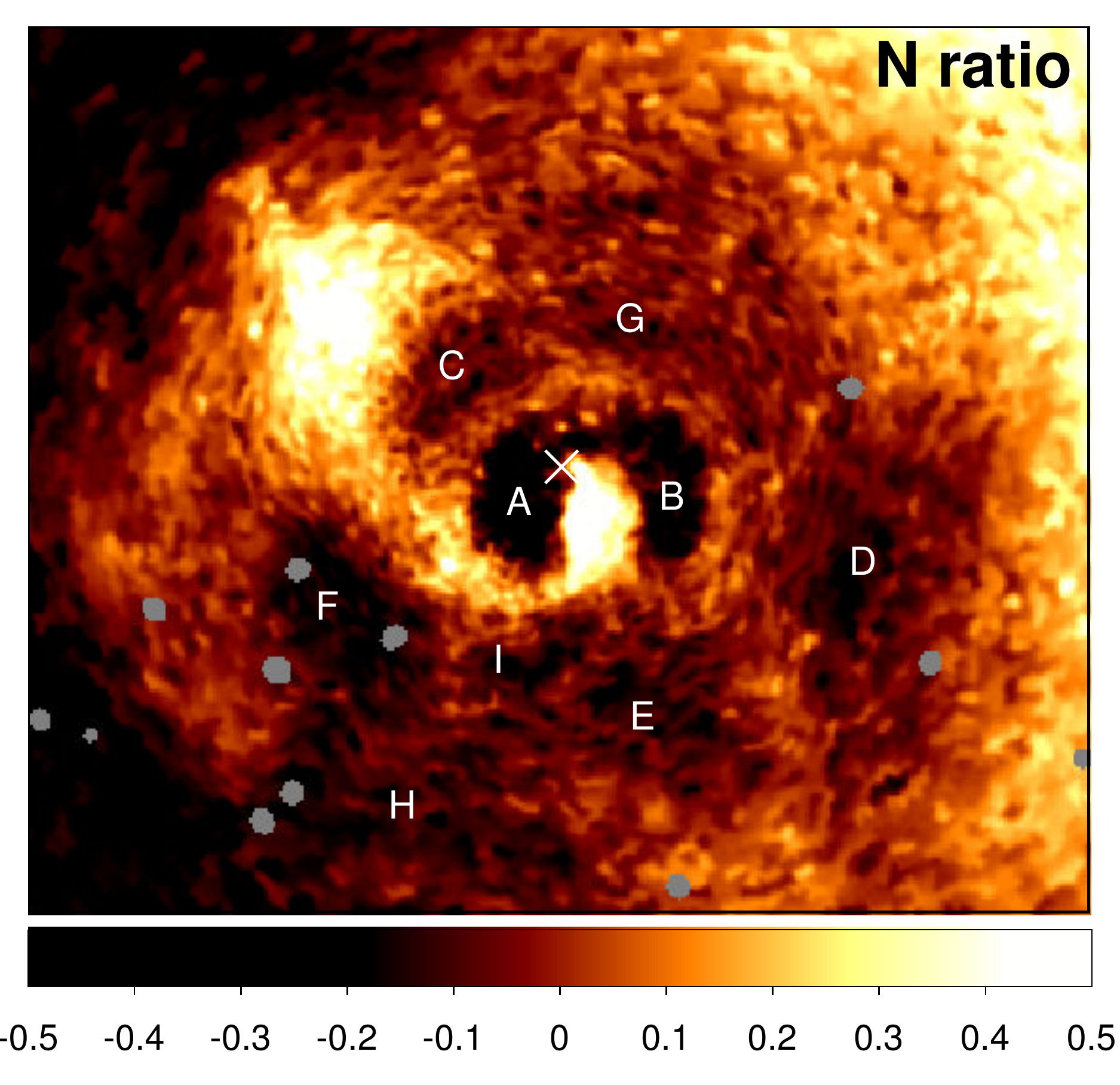} \\
  \includegraphics[width=0.43\textwidth]{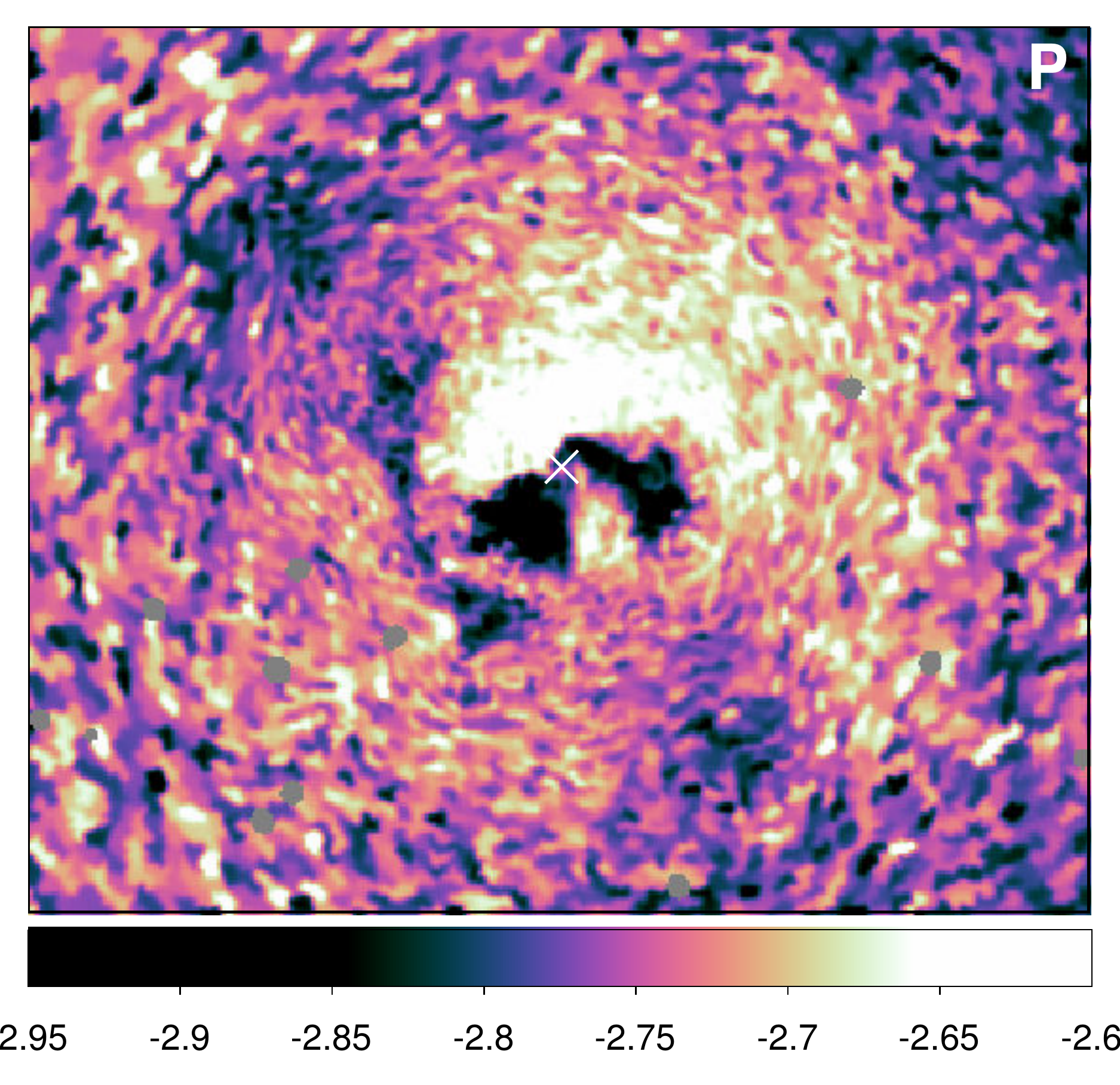}
  \includegraphics[width=0.43\textwidth]{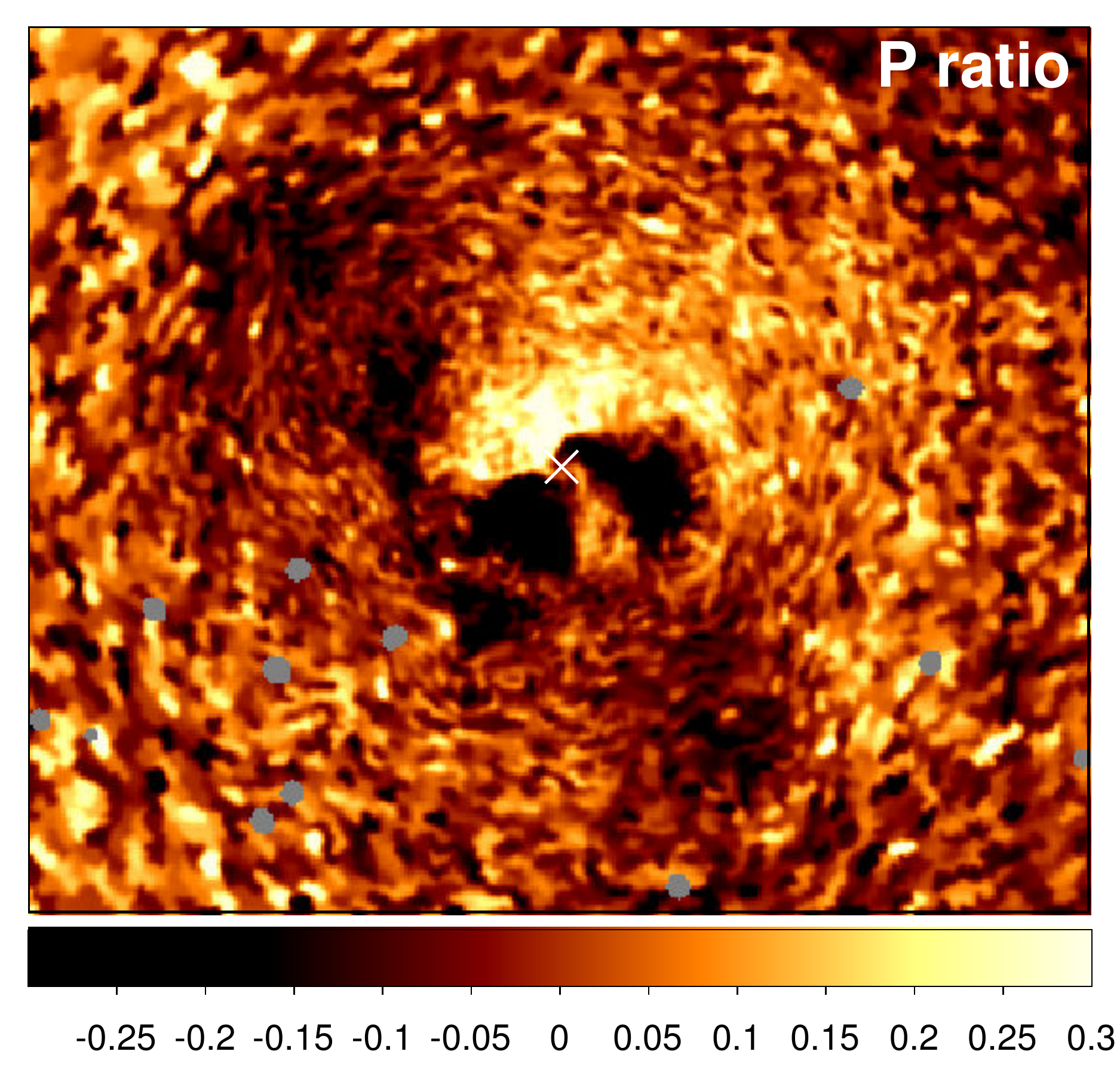}
  \caption{Central spectral maps (left) and fractional deviations from
    spherical symmetry (right). Analysis details are given in Appendix
    \ref{sect:speccentredetails}. On the left side are the (top left
    panel) temperature in keV, (centre left panel) normalisation (in
    log$_{10}$ cm$^{-5}$ arcsec$^{-2}$) and (bottom left panel)
    pseudo-pressure (in log$_{10}$ keV cm$^{-5/2}$ arcsec$^{-1}$). On
    the right side are the fractional deviations from the average at
    each radius of the respective left map.  Maps have been smoothed
    by a Gaussian of 0.492 arcsec, excluded point sources are shown as
    circles or ellipses and the central radio source is marked by
    crosses.}
  \label{fig:centre_spec_maps}
\end{figure*}

The large number of counts available in this dataset allow us to make
detailed X-ray images of the very central regions in different energy
bands (Fig.~\ref{fig:vcentre}). The 0.5--1 keV band is sensitive to
cool ($\sim 1$~keV) gas, while the 1.5--7 keV band is sensitive to
hotter material. The radio nucleus, as noted by \cite{Taylor06}, is
offset to the north-east of a ridge of bright, soft X-ray
emission. The nucleus also does not lie in the geometric centre of the
two inner cavities. Also shown in Fig.~\ref{fig:vcentre} is a
projected X-ray temperature map of the same region (taken from
Fig.~\ref{fig:centre_spec_maps}). The lowest temperature material is
strongly correlated with the soft emission, although there is not a
complete correspondence.

Maps of the central temperature, \textsc{xspec} normalisation per unit
area and pseudo-pressure are shown in Fig.~\ref{fig:centre_spec_maps}.
The normalisation per unit area scales with the density squared and
the line-of-sight depth. The pseudo-pressure is the temperature
multiplied by the normalisation per unit area, scaling as the
projected thermal pressure times the line-of-sight depth. On the left
side are the projected quantities. On the right are shown the
fractional deviations from spherical symmetry.

The temperature map shows that the soft-X-ray-emitting structures,
including the plume, seen in the X-ray images, are indeed cooler gas.
The overall temperature decreases towards the core of the cluster. The
temperature deviation map shows deviations in projected temperature
are around 30 per cent. The map shows an extension of cool material in
roughly the opposite direction to the plume from the nucleus. The
normalisation map shows the dense plume structure and the dense shells
around the two inner cavities. Examining the non-spherical structure,
there are a number of depressions visible in the residuals, which we
assign the letters A--H. The pressure maps show the thermal pressure
is lower inside the two central cavities and also towards the end of
the plume and towards the south-west, in roughly the opposite
direction.

\subsection{Shock around nucleus}
\label{sect:innershock}
\begin{figure}
  \includegraphics[width=\columnwidth]{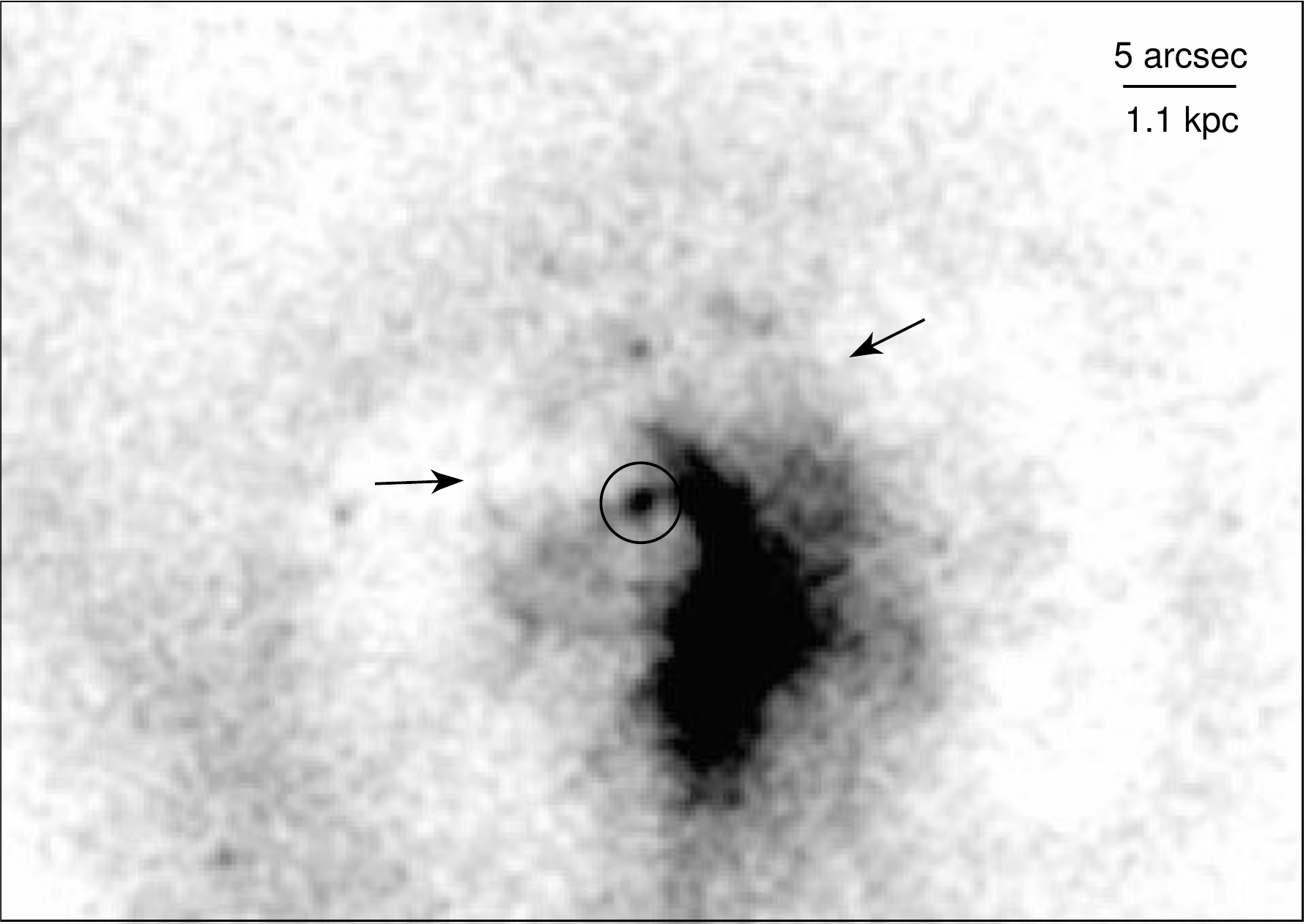}
  \caption{A 1.9-kpc-radius shell-like structure
    (marked by arrows) seen around the nucleus. The 1 to 7 keV image
    was binned in 0.246 arcsec pixels, smoothing by a 1 pixel
    Gaussian. The radio nucleus position is marked by a circle of
    radius 1.8 arcsec.}
  \label{fig:centreshell}
\end{figure}

There is a shell-like enhancement which surrounds the central nucleus
(Fig.~\ref{fig:centreshell}), not clear in the previous images.  This
shell appears to split into two separate fine edges to the south-east,
separated by 3 arcsec (0.6 kpc). The north part of the shell can be
seen in the RGB image (Fig.~\ref{fig:rgb}) as a blue rim to the north
of the red central emission. The shell has a radius of 9 arcsec (1.9
kpc). The nucleus is offset 3 arcsec (0.6 kpc) to the south-east from
the geometric centre of the shell, in the direction of the splitting
of the edge.

Its morphology suggests that it may be a shock generated by an
outburst of the central nucleus. This is consistent with the X-ray
emission from the feature being hard. However, we do not see an
obvious SB enhancement within the shock, but just its
edge, which may be due to the complex density and temperature
structure at the core of the cluster. The shock appears to be
overlapping the central cavities. If the cavities and shock were
spatially coincident, it is unclear into what the shock is shocking. A
possible explanation is that the cavities are offset along the line of
sight from each other and the shock lies between the two and around
the nucleus.

We do not know the speed of the shock, but given the sound speed for
material at 1.1~keV, an upper limit for its age would be $\sim
3.5$~Myr. The outer parts of central bubbles beyond the shock are at a
maximum projected radius of around 5.5 kpc, which would imply that
they are only a few times older than the shock, unless projected radii
are much larger than intrinsic distances, which their morphology
suggests is not the case. Estimates for the ages of the inner bubbles
in Centaurus range from $6-22$~Myr \citep{Rafferty06}. This implies
that there are repeated outbursts from the nucleus in Centaurus on
timescales of $3-19$~Myr.

A shock would explain why the X-ray-coolest gas is not the closest to
the nucleus, although this could alternatively be explained by
sloshing.  If there was a non-symmetric temperature distribution
around the nucleus before the shock was launched, the varying sound
speed might explain why the nucleus is not at the shell
centre. Indeed, the colder material seen to the west might have
reduced the shock radius in this direction. This colder material,
however, may only lie close to the nucleus in projection and could be
outside the shock region. The very central high-frequency radio
emission appears to be bounded by the shell edges to the north and
south (Section \ref{sect:radio_icm}).

\subsection{Interaction of radio source and ICM}
\label{sect:radio_icm}
\begin{figure*}
  \centering
  \includegraphics[width=\textwidth]{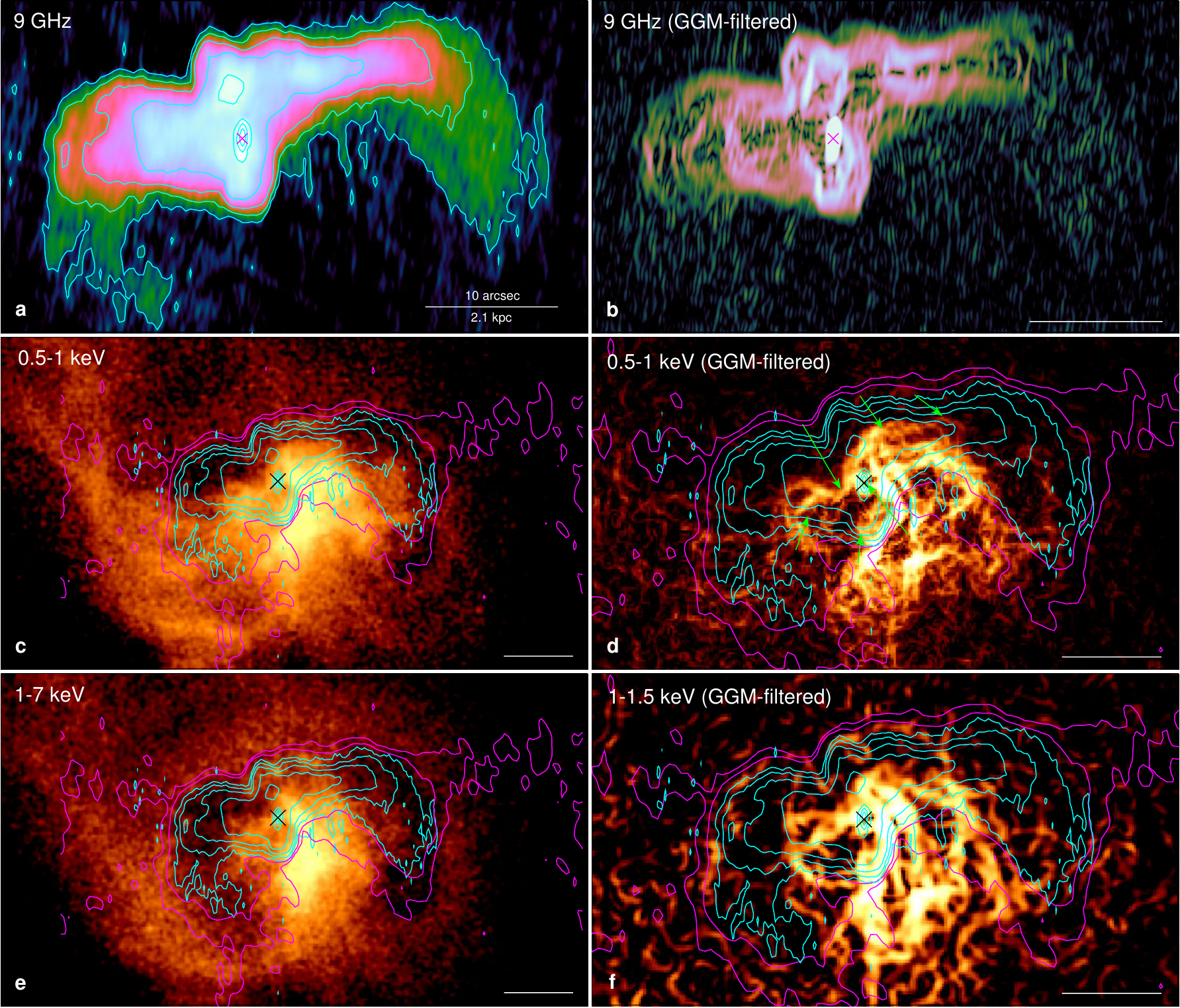}
  \caption{High frequency radio and X-ray
    comparison. (a) 9 GHz image with 8 logarithmic cyan contours at
    levels from $4\times 10^{-5}$ to 0.0607 Jy~beam$^{-1}$, with
    resolution $1.15\times 0.31$ arcsec. (b) 9 GHz image applying GGM
    filter ($\sigma=0.08$ arcsec). (c) 0.5 to 1 keV X-ray image. Also
    show are two 5 GHz contours from \protect\cite{Taylor02} at 0.1
    and 0.4 mJy~beam$^{-1}$ (magenta). (d) GGM-filtered 0.5 to 1 keV
    image, summing the GGM-filtered images with $\sigma=0.246$ and
    0.492 arcsec. Arrows mark regions of high rotation measure. (e) 1
    to 7 keV X-ray image. (f) GGM-filtered 1 to 1.5 keV image
    ($\sigma=0.492$ arcsec).}
  \label{fig:radio_xray}
\end{figure*}

The central radio source PKS\,1246-410 was observed in 2013 and 2014
using the Karl G. Jansky Very Large Array (VLA) in the A and B
configurations between 8 and 10 GHz, with a total of 8.2 hours of time
on source. Analysis details and first results are presented in
\cite{Grimes14}. The intensity map is shown in
Fig.~\ref{fig:radio_xray} (a). There are a number of edges in the
radio image, which we highlight by applying a GGM filter (b). The
filtered image suggests that the inner part of the source is extended
along the north-south direction, changing to the east-west on larger
scales. The bright nucleus is also extended in the north-south
direction, in agreement with the jet on smaller scales in VLBI
observations \citep{Taylor06}.

We compare the radio emission at 9 and 5 GHz with the soft X-rays (c),
examining the central region using a GGM-filtered soft X-ray image
(d).  It can be seen that the edges in the soft X-ray emission (as
traced by the GGM-filtered image) extend along the direction of the
radio source or around its edges. This is consistent with the skin
effects surrounding the radio source identified in the rotation measure analysis of
\cite{RudnickBlundell03}. Examining the harder X-ray emission (e) we
see that the rims of the X-ray cavity match the extent of the radio
source. In addition, edges in the inner part of the radio source
closely follow the edge of the shell-like structure, as seen in a
GGM-filtered harder image (f). The edges in the radio source can
clearly be seen in the GGM-filtered image (b).

\begin{figure*}
  \centering
  \includegraphics[width=0.9\columnwidth]{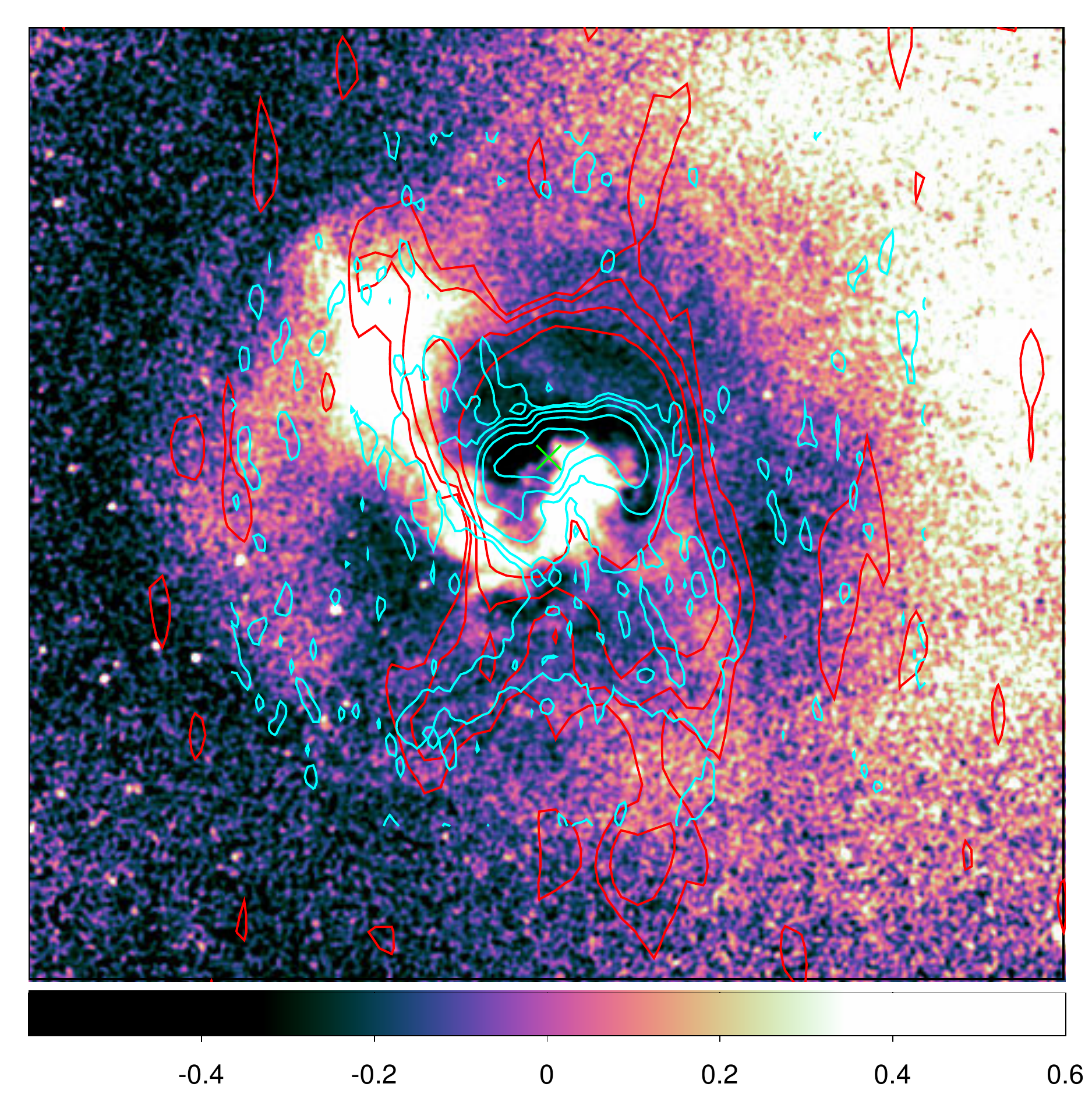}
  \includegraphics[width=0.9\columnwidth]{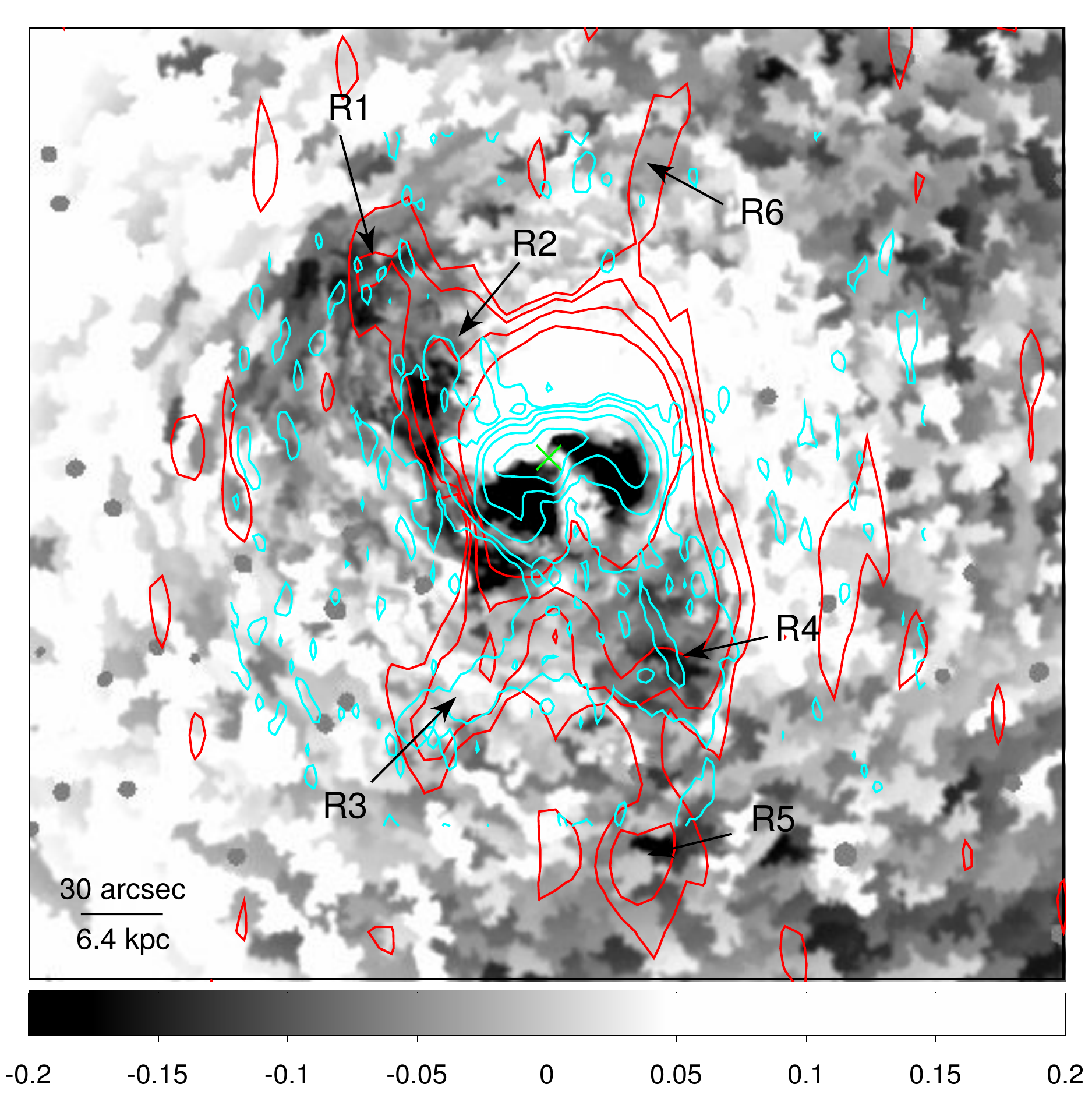}
  \caption{Lower frequency radio and X-ray/pressure comparison. (Left
    panel) 0.5 to 7 keV X-ray fractional residuals from radial
    average, showing 1.565 GHz (inner cyan) and 330 MHz (outer red)
    contours. (Right panel) Pressure azimuthal fractional residuals
    using $S/N=40$ binning. There are six logarithmic 1.565 GHz
    contours between 0.2 and 360 mJy~beam$^{-1}$ and four logarithmic
    330 MHz contours between 10 and 100 mJy~beam$^{-1}$.}
  \label{fig:radiolower}
\end{figure*}

At lower frequencies the radio emission is connected to features in
the X-ray image and pressure map (Fig.~\ref{fig:radiolower};
\citealt{Taylor02}). At 1.6~GHz, in addition to emission in the
central cavities, there are extensions to the north and south. In the
south the emission appears to bifurcate into extensions R3 and
R4. Region R4 is coincident with an area of lower pseudo-pressure and
a depression in the X-ray image (labelled E in
Fig.~\ref{fig:centre_spec_maps}). To the north there is an extension
in the radio source, R2, along the direction of the plume in a region
lower in pseudo-pressure. At 330 MHz frequencies, measured using the
VLA, R2 extends into the lower pressure region R1 and R4 continues
south along a lower pressure channel to R5. There also exists a spur
of emission into a lower pressure region R6.

The radio emission switches from a north-south direction on jet
scales, to east-west in the inner cavities to north-south again on
larger scales. The radio source and X-ray emission are connected from
sub-kpc scales to several tens of kpc. In the very inner region the
high frequency radio emission shows edges coincident with the inner
shell, or shock.  In the inner 2~kpc there are edges in the soft X-ray
emission which clearly trace along the direction of the radio source,
including an edge running along its midpoint and a sharp edge
immediately to the south of the inner radio source, south of which is
a bright ridge of X-ray emission.

\subsection{Cavities}
\label{sect:cavities}
There are a number of SB depressions in the core of the cluster, most
easily seen labelled in the residual normalisation map
(Fig.~\ref{fig:centre_spec_maps}) and the GGM-filtered image
(Fig.~\ref{fig:centre}). A and B are the central cavities coincident
with radio emission. C is the depression between the nucleus and
plume, marked by the `inner edge' in Fig.~\ref{fig:rgb}. D is the
depression bounded by the `bay' structure. E is also associated with a
depression in the thermal pressure, and lies near the `hook' structure
(which itself seems to surround another cavity, I). F, G, and H are
further depressions. All of these features could be generated by AGN
outbursts, with a large fraction showing some evidence for radio
emission (Figs. \ref{fig:radio_xray} and \ref{fig:radiolower}). These
include the central cavities (A and B), the cavity towards the plume
direction (C), the cavity in the opposite direction (E) and its
neighbouring cavity (H). There is also 330 MHz emission at the
location of the bay cavity (D). The gradient filtered image on larger
scales (Fig.~\ref{fig:unsharp} lower panel) shows an SB edge across
the bay (it is split into F2 and F3) and there may be further flat SB
areas in which could be yet-older cavities (F1 and F4-F6).

A large fraction of the volume in the inner 20 kpc
radius, or even further, may be filled by cavities and their ageing
electron populations.  There are other clusters, such as M\,87
\citep{FormanM8707} and 2A\,0335+0986 \citep{Sanders2A033509} which
show large number of possible cavities in their core. In Centaurus,
the pressure map, however, does not show reductions in thermal
pressure at the location of all these cavities, though there are
decrements in thermal pressure for A, B, C, E and I. It is possible
that displaced thermal gas may mask pressure changes. It is also the
case that in other clusters not all extended radio emission is
associated with cavities. For example, in Perseus the minihalo
emission is correlated with the X-ray substructure \citep{Fabian11}.

\subsection{Shock surrounding inner cavities}
\label{sect:centcavity}
\begin{figure}
  \centering
  \includegraphics[width=\columnwidth]{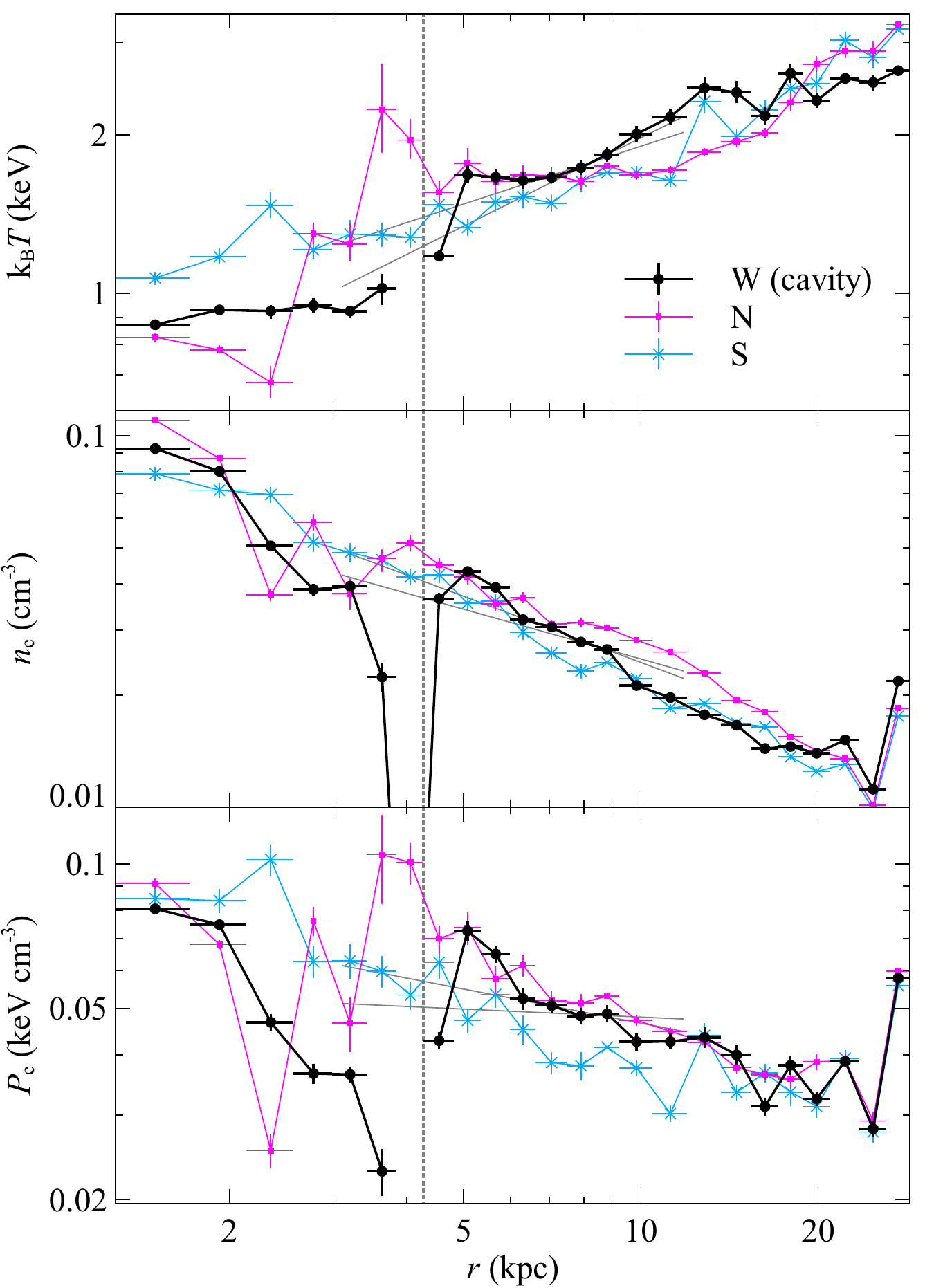}
  \caption{Deprojected temperature, density and pressure profiles
    across the western inner cavity and two neighbouring sectors (see
    Appendix \ref{sect:innercavity}). The cavity outer edge radius is
    shown by a vertical dashed line. The straight lines are two fits
    to the western points from 6--9 and 7.5--9~kpc.}
  \label{fig:cavity_prof}
\end{figure}

There are pressure and temperature enhancements outside the inner
cavities towards the north and west
(Fig.~\ref{fig:centre_spec_maps}). These could be due to shock heating
of the region due to the formation of the cavities. To examine this in
more detail we made deprojected temperature, density and pressure
profiles (Fig.~\ref{fig:cavity_prof}) across the western inner cavity
(the eastern cavity is too close to the plume), and in two comparison
regions to the north and south. We note that there
  are potential problems with deprojecting spectra in a region with a
  complex structure, although the results are qualitatively consistent
  with the projected maps.  The cavity is clearly seen in the density
profile (the 7th radial bin), where only an upper limit can be
obtained.

Immediately outside the cavity the temperature is relatively low (1.2
keV), but jumps up by 40 per cent in the next bin to 1.7 keV. It then
stays flat over $\sim 3$ kpc, then rises from 8 to 13 kpc, then
flattens again. In comparison, the northern profile appears to have
the same 1.7 keV temperature plateau from 4.2 to 12 kpc radius, but is
even hotter ($\sim 2$ keV) inside that at the radius of the cavity. To
the south there is a much more gradual rise in temperature with
radius. These trends can be seen in the deviation maps of the central
region (Fig.~\ref{fig:centre_spec_maps}).  Outside the cavity is a
1.7-kpc-thick high-density rim. The profile and images show a similar
thicker region in the northern sector. Beyond the cavity rim in the
western sector, the density drops by 15 per cent at 9 kpc radius where
the sector overlaps with the bay region.  The southern density profile
does not show the obvious jumps in density beyond 2.6 kpc radius. As
there are jumps in both temperature and density, there are
corresponding increases in pressure, with the largest pressure
increase to the north.

The magnitude of the jumps in temperature, density and pressure are
uncertain because we do not know the underlying profiles. We show fits
to two sets of points (6--9 kpc and 7.5--9 kpc) in
Fig.~\ref{fig:cavity_prof} to indicate the allowed range.  The data
are not fit beyond 9 kpc because of the density jump, likely
associated with the bay region. To the west, comparing the peak value
with the fit, the range in temperature jump is 13 to 24 per cent
(although there is a 40 per cent jump in temperature from the bin
immediately outside the cavity to the next bin), the density jumps by
18 to 27 per cent and the pressure by 32 to 45 per cent. In the
northern sector the jumps appear larger. The northern pressure jump is
likely between 60 and 130 per cent, while the temperature increases by
40 to 100 per cent and the density jump is uncertain.

The most likely explanation for these high density, temperature and
pressure rims is that they are caused by a weak shock generated by the
mechanical action of an outburst of the central AGN on the ICM. Such
shocks are seen in other clusters and groups (e.g. Perseus,
\citealt{FabianPer03}; M87, \citealt{FormanM8707}; NGC 5813,
\citealt{Randall15}). Assuming a $\gamma=5/3$ gas and using the
Rankine–Hugoniot jump conditions \citep{LandauLifshitz}, the range of
density and temperature jump values to the west would imply a Mach
number for the shock between 1.1 and 1.2, while the temperature jump
ratio implies a value between 1.2 and 1.4. The northern jump in
temperature, however, implies a much stronger shock between 1.6 and
2.2. The values are rather uncertain, however, due to the unknown
unshocked profiles. The discrepancies between the north and west may
be due to the rising cavity entraining cool material which is being
mixed into the surrounding ICM to the west, reducing the temperature
increase. It is unclear how this could affect the density in this
direction, however.

We can estimate the energy that the shock has dissipated in the ICM
(as in the Perseus cluster, \citealt{Graham08Per}). Taking the hard
X-ray band image, the western cavity is an ellipse with axes with
radii 1.8 and 2.4~kpc and the hard rim has radii 3.3 and
4.2~kpc. Assuming an electron pressure enhancement in the shock of
$0.02\keVpcmcu$ and that the shock lies completely around the western
cavity, the energy dissipated by the shock is $4\times 10^{56} \erg$
(using equation 19 in \citealt{Graham08Per}). This compares to the
adiabatic energy to expand the cavity $PV = 2 \times 10^{56} \erg$, if
the electron pressure for the cavity of $0.07\keVpcmcu$. Therefore the
shock energy is around half the energy available
($4PV$). The heating power of the shock, using the
shock width divided by its speed as a timescale, is $6 \times 10^{42}
\ergps$. This is around 75 per cent greater than the heating power
inferred per cavity \citep{Rafferty06}.

\subsection{Dust and X-ray association}
\label{sect:optical}
\begin{figure}
  \includegraphics[width=\columnwidth]{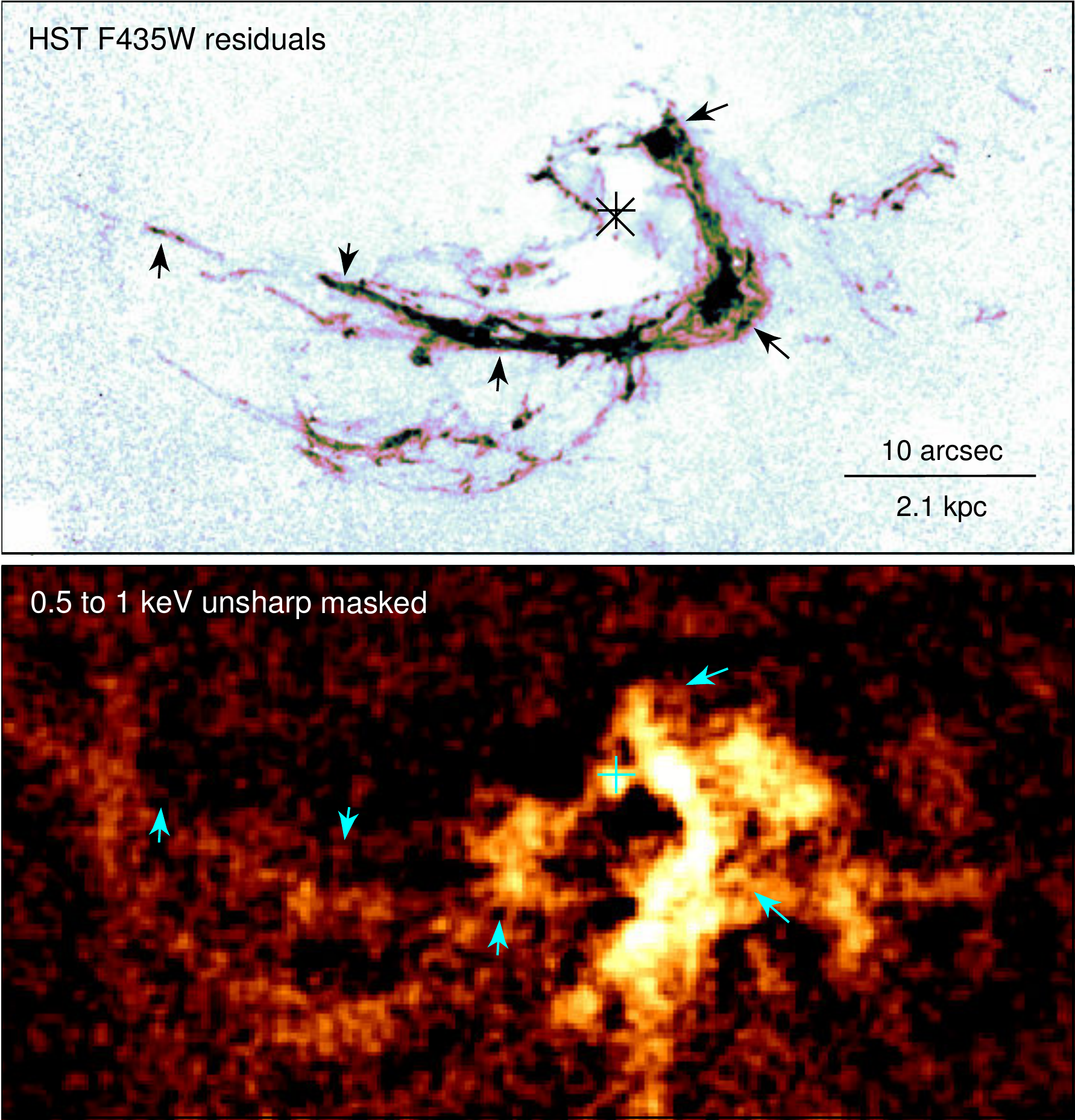}
  \caption{Central dust lanes and 
    soft X-ray comparison. (Top panel) Fractional residuals of image to a
    symmetric elliptical model fitted to an \emph{HST} F435W image of
    NGC 4696 from Hubble legacy dataset
    HST\_9427\_06\_ACS\_WFC\_F435W. (Bottom panel) Unsharp-masked
    0.5-1 keV image. The arrows are in the same location in each
    image. `$\times$' marks the radio nucleus, while `+' indicates the
    optical model centroid.}
  \label{fig:optresid}
\end{figure}

In Fig.~\ref{fig:optresid} we compare the optical morphology of the
dust lanes in NGC 4696 with the soft X-ray filaments. To remove the
smooth stellar component, we constructed an elliptical model. Given a
centre, ellipticity and rotation angle, the average at each elliptical
radius was computed. This profile was then converted to a 2D
image. The model image was fit to the data, adjusting the model
parameters. Regions containing obvious dust features and bright
sources were masked from the data and model during fitting. The ratios
of this model to the data are shown in the top
panel.

Although the morphology of the dust follows the
  soft X-ray emission, there is an offset between the regions with the
  brightest X-ray emission and the dust. We checked point source
  positions to verify that the astrometry was corrected. The
  morphological agreement indicates the dust and soft X-rays have a
  common origin. Either there is a real offset between the components,
  or there is additional bright X-ray emitting material obscured by
  dust and associated material. If the dust material is blocking
  significant X-ray emission it must lie in front of the X-ray
  emitting material or is well mixed with it. We examined the X-ray
  spectra of a region between the two rightmost arrows, along the dust
  lane and a comparison region to the east along the bright X-ray
  rim. Fitting two-temperature-component models to the spectra did not
  show any evidence for increased column density in the dust region,
  either when fixing the temperatures to be the same in the two
  regions or not. Most of the difference in flux between the regions
  appears to be around $0.9\keV$ energy, which argues against
  absorption which would primarily affect low energies.
  This would suggest that the soft X-rays and dust are
  offset, although partial covering models may account for the lack of
  obvious soft X-ray absorption.

\subsection{Plume origin}
\label{sect:plumeorigin}
The reversing SB and temperature structure in the cluster
(Fig.~\ref{fig:subav}) suggests that the plume is part of the sloshing
morphology of the cluster. Arguing against this origin is that the
plume appears to be made up of filamentary multiphase gas, connected
to the cluster core (Fig.~\ref{fig:rgb}). The southern part of the
plume is associated with cold material in other wavebands
(e.g. Fig.~\ref{fig:optresid} and \citealt{Crawford05}). The
spatial coincidence of the radio structure with the plume
(Fig.~\ref{fig:radiolower}) implies that the feature is due to a ghost
radio cavity which has risen in that direction, dragging out cold
material behind it \citep{Crawford05}.

If the plume was generated by a rising bubble, which bubble is
responsible? Around four-fifths of the way along the plume is a SB
edge (marked as Inner Edge in Fig.~\ref{fig:rgb}) which appears to
cross the plume and perhaps connects to the east-most filament. This
edge is made up of a lower temperature material, appearing to bound a
cocoon-like structure (see the temperature and normalisation ratio
maps in Fig.~\ref{fig:centre_spec_maps} and the 1 keV map in
Fig.~\ref{fig:norm_T_maps}). Inside this edge is a depression
(labelled C in Fig.~\ref{fig:centre_spec_maps}), where the majority of
the extended radio emission to the north lies
(Fig.~\ref{fig:radiolower}). This could be the responsible bubble,
although this does not explain why the plume extends beyond the edge
towards the north-east, where it rapidly ends. There is no evidence
for any other cavities or radio emission beyond the plume. It could be
that the plume is not on the plane of the sky, but has a large
component along the line of sight.

The similarity of the plume and the sloshing morphology could be due
to its shape being affected by the motions in the ICM caused by the
sloshing, giving the bend at its south. ICM motions
may have also helped displace the bubble to the north, helping to drag out
colder material behind it. This may also explain why there is no
similar plume to the south. However, cavity E, which is associated
with low frequency radio emission, is in the opposite direction from
the plume and is a plausible counterpart, albeit without the
associated plume structure. Sloshing or displacement of material by
buoyant bubbles may be the origin of the offset of the coldest X-ray
material from the nucleus (Fig.~\ref{fig:vcentre}). The coolest X-ray
emitting material is coincident with the central dust lane
(Fig.~\ref{fig:optresid}), which is offset from the X-ray, optical and
radio nuclei.

The location of the bay (D) in the hot spiral-shaped region diagonally
on the opposite side of the cool spiral-shaped plume
(Fig.~\ref{fig:centre_spec_maps}), suggests that it may be generated
by the sloshing motion. The edges of the bay, however, curve in the
opposite direction to the hot inward spiral. The bay also does not
spiral north and inward as the temperature does, but there is an edge
in SB and density to the north (Figs. \ref{fig:centre} and
\ref{fig:centre_spec_maps}). Low-frequency radio emission is also
associated with the bay region (Fig.~\ref{fig:radiolower}). Therefore
it is likely that the bay is an AGN-generated cavity and is not
generated by gas sloshing in the potential well.

\subsection{Rotation measure and magnetic fields}
\label{sect:rm}

\begin{figure}
  \centering
  \includegraphics[width=\columnwidth]{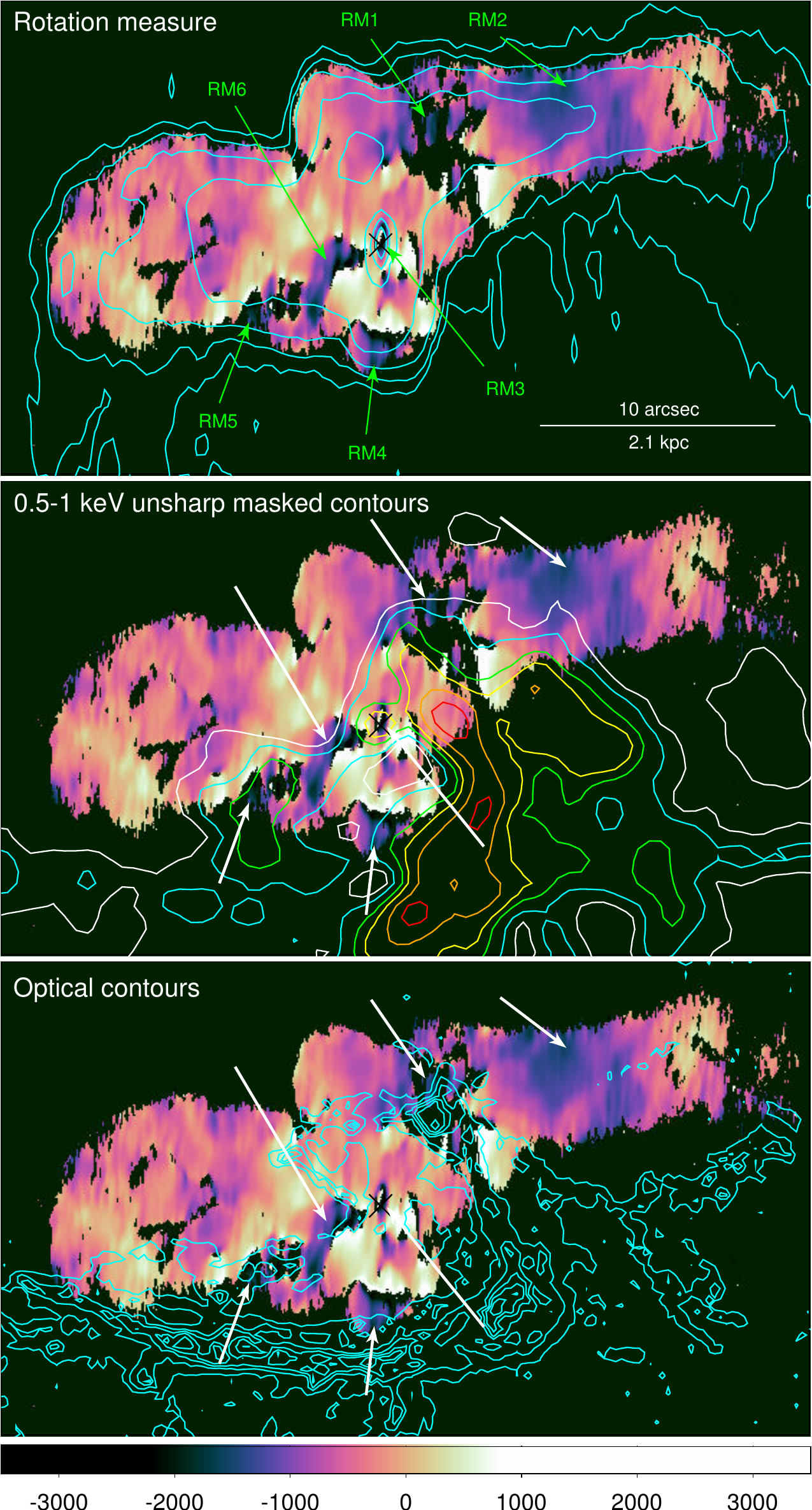}
  \caption{RM image, with radio (top), soft X-ray
    edges (centre) and optical (bottom) contours. The map has
    $1.06\times 0.26$ arcsec resolution and a scale in
    rad~m$^{-2}$. The arrows mark five regions with RMs
    $<-1500$~rad~m$^{-2}$. The RM varies between $-3500$ and
    $3500$~rad\,m$^{-2}$, with a mean of $-391$~rad\,m$^{-2}$ and a
    full width half maximum of $668$~rad\,m$^{-2}$. Regions where the
    polarisation angle uncertainty exceeds $35^\circ$ are
    excluded. The X-ray contours come from an unsharp-masked 0.5-1 keV
    image and the optical show the dust lane
    (Fig.~\ref{fig:optresid}).}
  \label{fig:RM}
\end{figure}

Fig.~\ref{fig:RM} shows the 9~GHz Rotation Measure (RM) map, plotting
contours from the intensity (top panel), soft X-rays edges (centre
panel) and optical dust lane (bottom panel).  Five regions with high
RMs (more negative than $-1500$~rad\,m$^{-2}$) are identified by
arrows (also shown in Fig.~\ref{fig:radio_xray}). These include the
nucleus (RM3), a 850-pc-long tube-like structure which extends
south-east from the nucleus (RM6) and a large gradient in RM across
the western half of the source (RM2).  The regions of high RM (marked
by the arrows) correspond to where the X-ray emission is bright,
though there is not an exact one-to-one correspondence between RM and
soft X-ray emission.

We also compare the RM map with contours showing the dust lane in the
centre of NGC 4696. The region in which the radio polarisation can be
measured is anti-coincident with the strongest dust absorption
regions. However, several of the regions with high RMs (RM1, RM4, RM5
and RM6) coincide with locations of dust absorption.

The RM shows a number of regions with large negative
($<-1500$~rad\,m$^{-2}$) values. The value of the RM is given by
\begin{equation}
\mathrm{RM} = 812 \int_0^L n_\mathrm{e} \: B_\parallel \: \mathrm{d}l \:
\mathrm{rad} \: \mathrm{m}^{-2},
\end{equation}
where $n_\mathrm{e}$ is the electron density in $\pcmcu$,
$B_\parallel$ is the component of the magnetic field along the line of
sight in $\mu$G and the integral is over length $L$ in kpc. In
PKS\,1246-410 the plausible explanations for high RM are the optical
line-emitting filaments crossing in front of the radio lobes and
filamentary soft X-ray emitting gas \citep{Taylor07} or local
interactions with the synchrotron-emitting plasma \citep{RudnickBlundell03}.

Several of the high RM locations, both in the positive and negative
directions, are coincident with filaments of soft X-ray emission. The
tube of high negative RM south and east of the nucleus (RM6) is
coincident with a soft X-ray filament, although the filament does not
run along the full length of the tube but passes into a neighbouring
region of high positive RM. There is, however, not an exact
correspondence between the soft X-ray emission and regions of high RM,
perhaps due to surface effects in the skin of the radio source \citep{RudnickBlundell03},
or because of absorption of X-rays by the dust lane
  (see Section \ref{sect:optical}). Where the X-ray emission is
brightest, along the ridge south and west of the nucleus, the
polarisation cannot be measured.

The lack of RM measurements in regions coincident with the soft X-ray
emission and dust lane may be due to depolarisation of the
source. Depolarisation is consistent with the gas surrounding the
filaments being cooler and denser and more highly magnetised. The
fields could help support the filaments, as in the Perseus cluster
\citep{Fabian08}. Depolarisation is also supported by the isolated
knots of high RM signal inside regions where the polarisation cannot
be measured (e.g. RM1 and RM5). If the regions containing large RMs
are dominated by line-emitting gas with their associated soft X-ray
filaments, and magnetic support is important as in Perseus,
the high RMs are likely more sensitive to the magnetic field
in the filaments than those in the bulk of the ICM.

\begin{figure*}
  \includegraphics[width=0.32\textwidth]{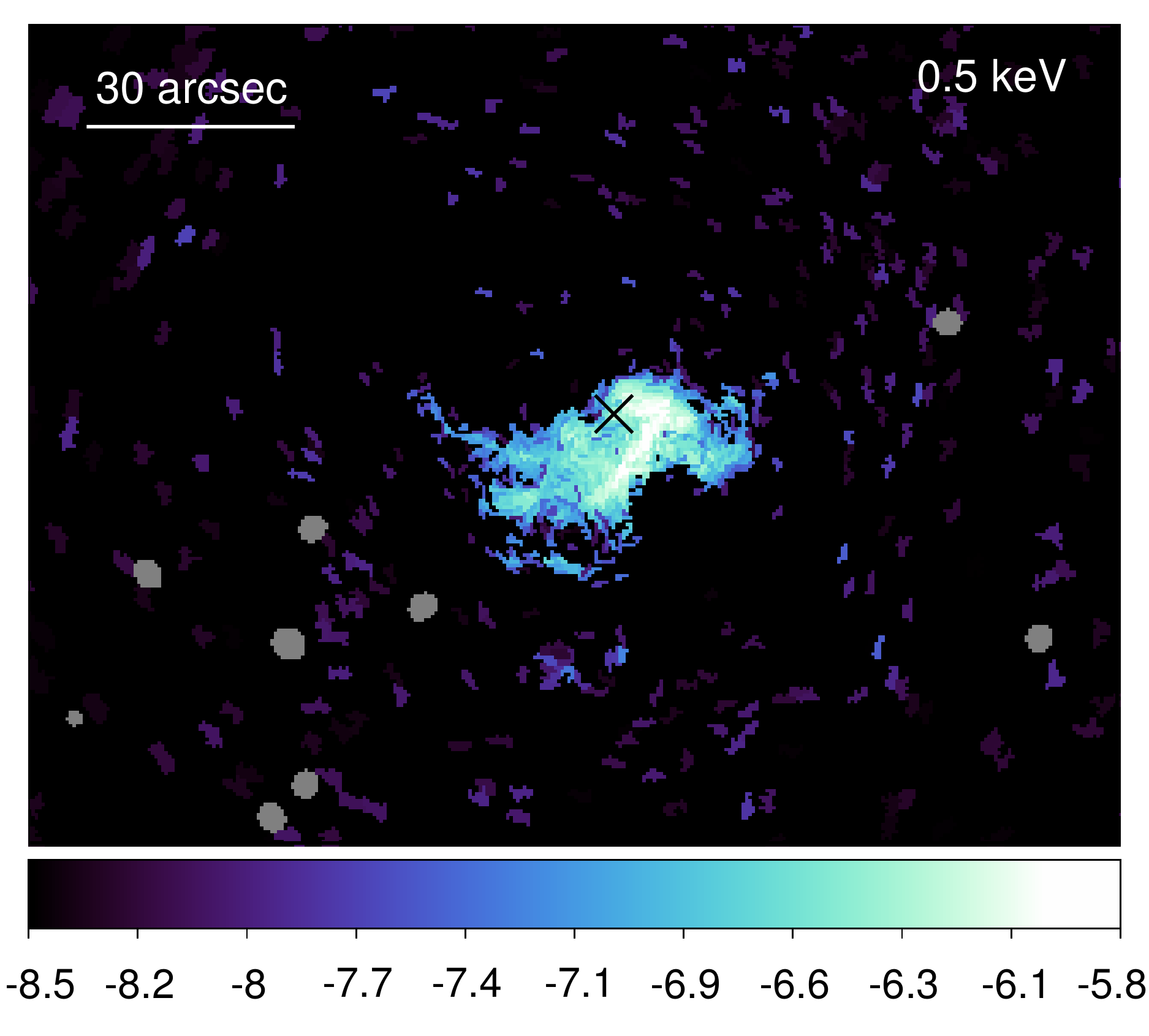}
  \includegraphics[width=0.32\textwidth]{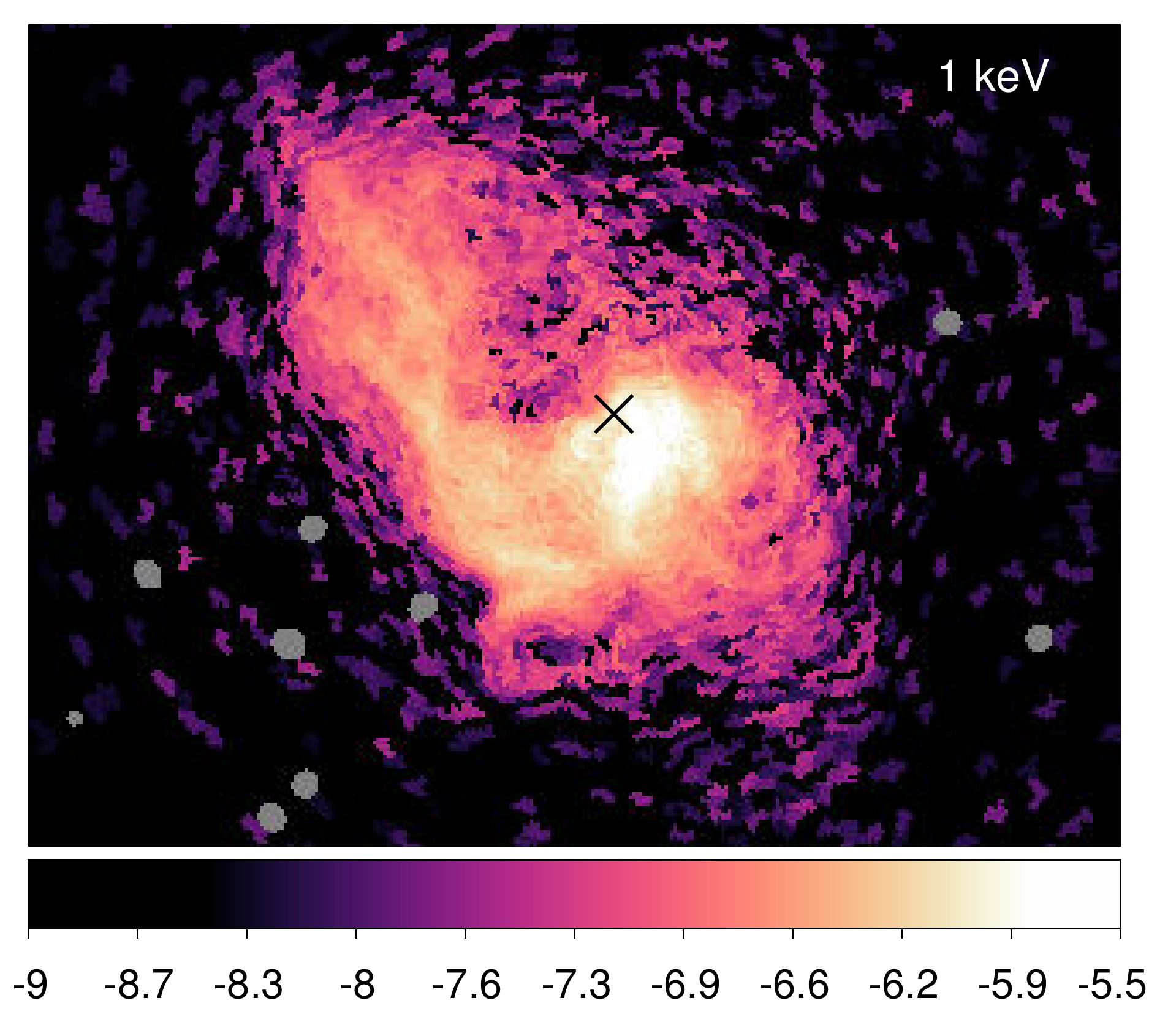}
  \includegraphics[width=0.32\textwidth]{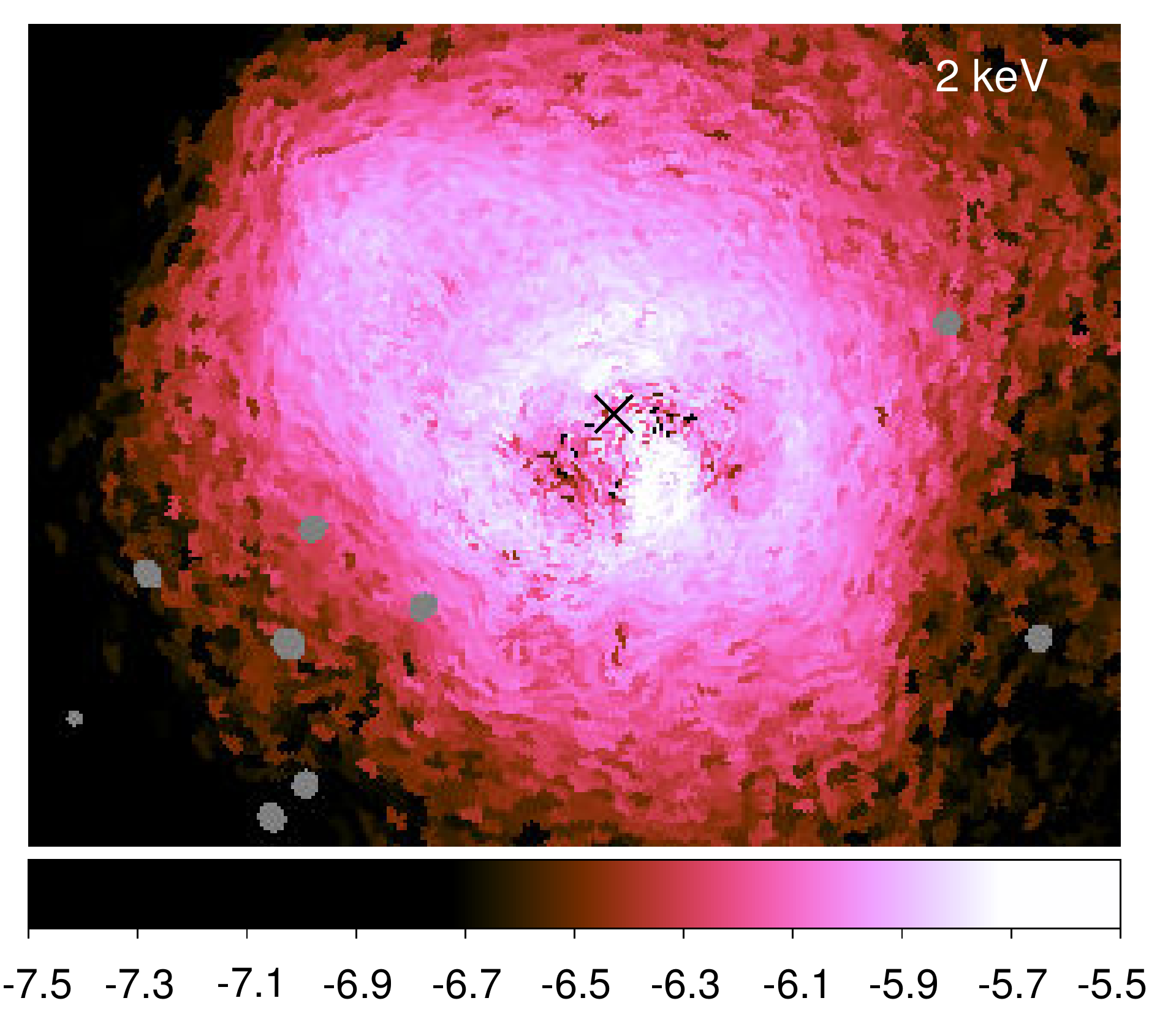}
  \caption{Normalisation per unit area for
    components with temperatures of 0.5, 1 and 2 keV. A model was
    fitted with 0.5, 1, 2 and 4 keV components with $1.2\Zsun$
    metallicity, using $S/N=20$ bins, minimising the C-statistic of
    the fit. The units are log$_{10}$ cm$^{-5}$ arcsec$^{-2}$.}
  \label{fig:norm_T_maps}
\end{figure*}

\section{Inner multiphase structure}
\label{sect:thermprof}
\subsection{Multiple temperature components}

Examining the thermodynamic state of the central region of the cluster
is important for understanding the balance between heating and cooling
and on what physical scales that is occurring. A useful technique for visualising
the location of X-ray material at different temperatures is to fit a
model made up of several fixed temperature components, allowing the
normalisation of each component to vary
\citep{FabianPer06}. Fig.~\ref{fig:norm_T_maps} shows the spatial
distribution of the three coldest components in a model made up of
0.5, 1, 2 and 4 keV temperatures. The 0.5~keV material lies along the
regions of soft X-ray emission and has a similar morphology to the
dust and H$\alpha$ emission (Fig.~\ref{fig:optresid}) and is offset
from the nucleus. The 1~keV gas traces the central ridge of
bright X-ray emission, the hook and the plume. The material closest to
2 keV appears associated with the rims of the inner cavities, the
plume and the `inner edge' (Fig.~\ref{fig:rgb}).

\begin{figure*}
  \includegraphics[width=\textwidth]{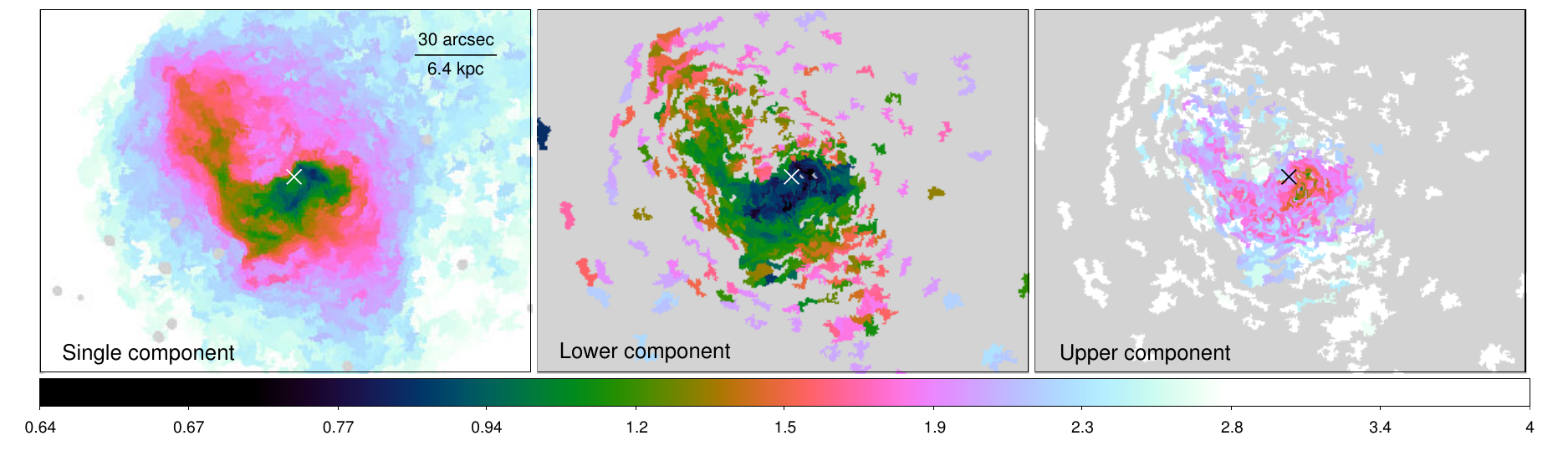}
  \caption{Comparison between the temperature (in
    keV) using a single-component model (left) and a two-component
    model (centre and right) using $S/N=40$ bins, allowing the
    metallicity to vary (assuming both components have the same
    metallicity) and minimising the C-statistic. Values for the
    two-component model are only shown when the improvement in
    C-statistic has 1 per cent probability of chance occurrence
    (determined from fits of realisations of single component
    models).}
  \label{fig:maps_2T}
\end{figure*}

The core of the cluster has phases with different temperatures which
are coincident in position. We examined these temperatures using a
two-component variable temperature model. Fig.~\ref{fig:maps_2T}
compares a single-component map against the lower and upper
temperatures of a two-temperature model, when the temperatures of the
model can be constrained. Further details and results are given in
Appendix \ref{sect:thermprofscentre}. The results show that the two
components are required over the plume region \citep[as found
previously in][]{SandersCent02} and in the opposite direction towards
cavity E (Fig.~\ref{fig:centre_spec_maps}). The lower temperature
component drops from 1.5~keV at the end of the plume to $\sim 0.7$ keV
in the regions associated with the filaments and the coldest material
from the fixed component analysis (Fig.~\ref{fig:norm_T_maps}). The
upper component has a fairly flat 2 keV temperature distribution over
much of the inner region. Its temperature, however, increases to 3 keV
at the end of the plume and drops to 1.3 keV in a few locations where
the coolest component is lowest.

\subsection{Powerlaw spectral fits}
\begin{figure}
  \includegraphics[width=\columnwidth]{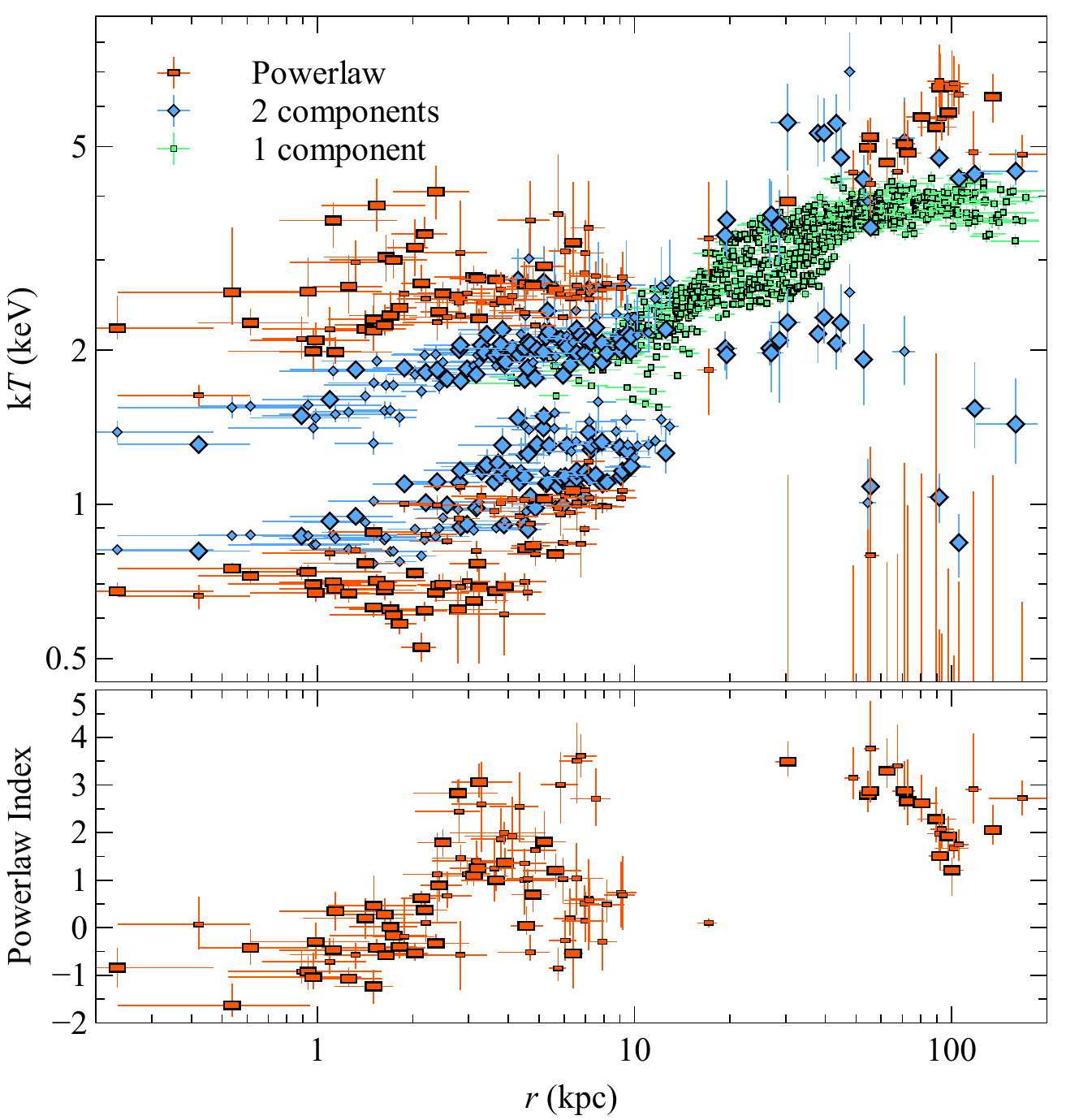}
  \caption{Powerlaw, two or single component fits to bins with
    $S/N=100$. (Top panel) Radial temperature profiles. Powerlaw upper
    and lower temperatures are shown if the upper can be determined to
    better than 1 keV and the index to at least 1. Two component
    results are shown if the temperatures can be determined to at
    least 20 per cent and the minimum temperature is $>0.4$ keV. The
    single component results are shown if neither of these criteria
    are passed. Larger markers are shown for the powerlaw or
    two-component results depending on which has the lower C-statistic
    value. (Bottom panel) The powerlaw index parameter when
    appropriate.}
  \label{fig:T_index}
\end{figure}

In order to see whether a continuous temperature distribution is more
realistic, we fitted a powerlaw-temperature-distribution model, which
assumes that the gas has a range in temperature between
$T_\mathrm{upper}$ and $T_\mathrm{lower}$. The model is constructed
using $N$ logarithmically-spaced temperature components (here 16),
where the relative normalisation of one component at temperature $T$
is given by $(T/T_\mathrm{lower})^\Gamma$. Each component is assumed
to have the same metallicity. This powerlaw model has the same number
of free parameters as a two-component model, as the free normalisation
is substituted for the powerlaw index.

A comparison between the temperature values obtained with the
two-component model and the powerlaw upper and lower temperatures is
shown in Fig.~\ref{fig:T_index}. Well constrained temperatures and powerlaw indices
for the two-component and powerlaw fits are mostly found in the central region.  The obtained
temperature range is wider when a powerlaw temperature distribution is
assumed. As the two-temperature and powerlaw models are distinct, we
cannot compare them with an F-test. However, the powerlaw model
produces better fits in the central regions compared to the
two-component model. In addition, the powerlaw index increases from
negative values (where low temperatures dominate) to positive values
(where high temperatures dominate) as a function of radius.

\subsection{Accounting for projection effects}

\begin{figure}
  \centering
  \includegraphics[width=0.95\columnwidth]{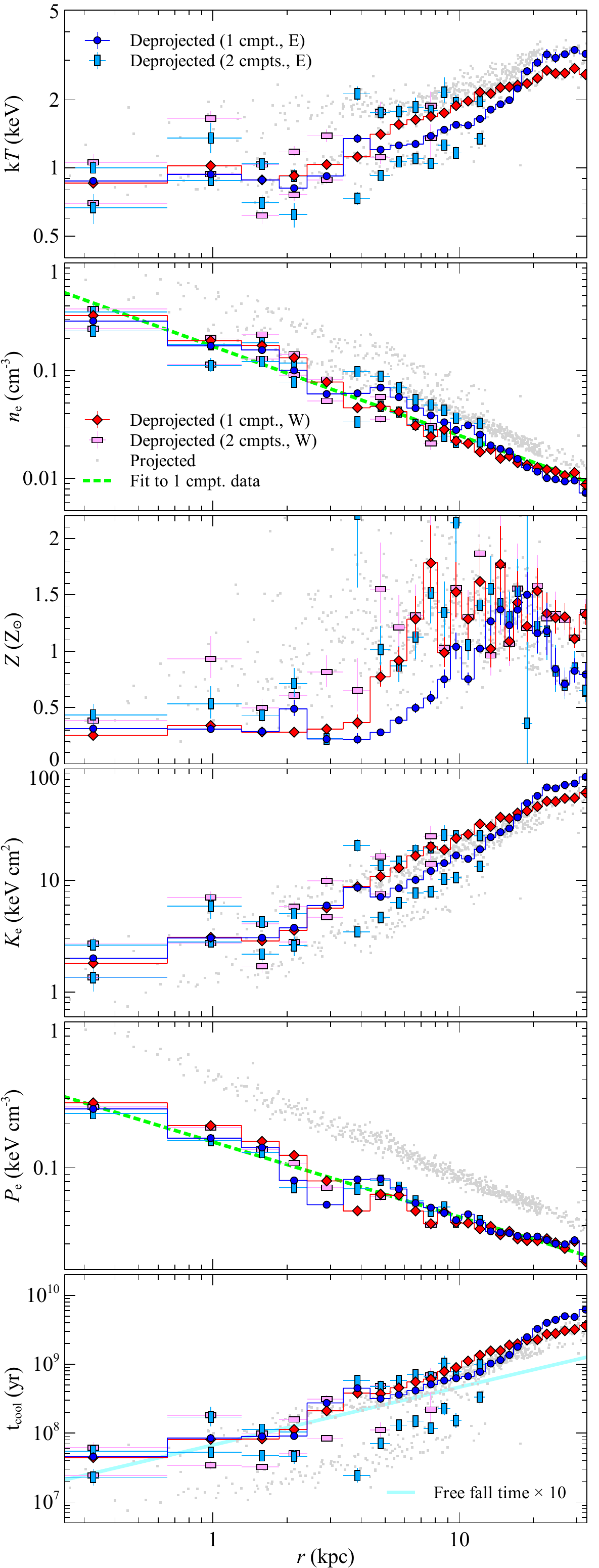}
  \caption{Deprojected profiles for the eastern (E)
    and western halves (W) of the cluster. Shown are the results for
    single- and two-component models (where the errors on the
    temperatures are less than 20 per cent). See Appendix
    \ref{append:deprojcentre} for details. We plot the results from
    projected two-temperature fits for comparison (Appendix
    \ref{sect:thermprofscentre}; plotting every 4th point). Powerlaw
    fits to the pressure and density profiles, and the estimated free
    fall time, are shown.}
  \label{fig:deproj_EW}
\end{figure}

Some of the multiple components may be due to projected spectra of
material along the line of sight. We therefore construct profiles from
deprojected spectra in east and west sectors
(Fig.~\ref{fig:deproj_EW}). The eastern side of the
  cluster centre is strongly affected by the plume which we do not
  remove.  The maximum difference in temperature between the eastern
and western sides is around 30 per cent. Excluding the cavities, the
density can differ by 40 per cent. In pressure there is generally
little difference between the two sides, except at the locations of
the cavities (between 2.4 and 4.3 kpc radius). The single-component
metallicity values show a large discrepancy between the east and west,
although when two temperature components are used, this difference
becomes much smaller.

\begin{figure}
  \includegraphics[width=\columnwidth]{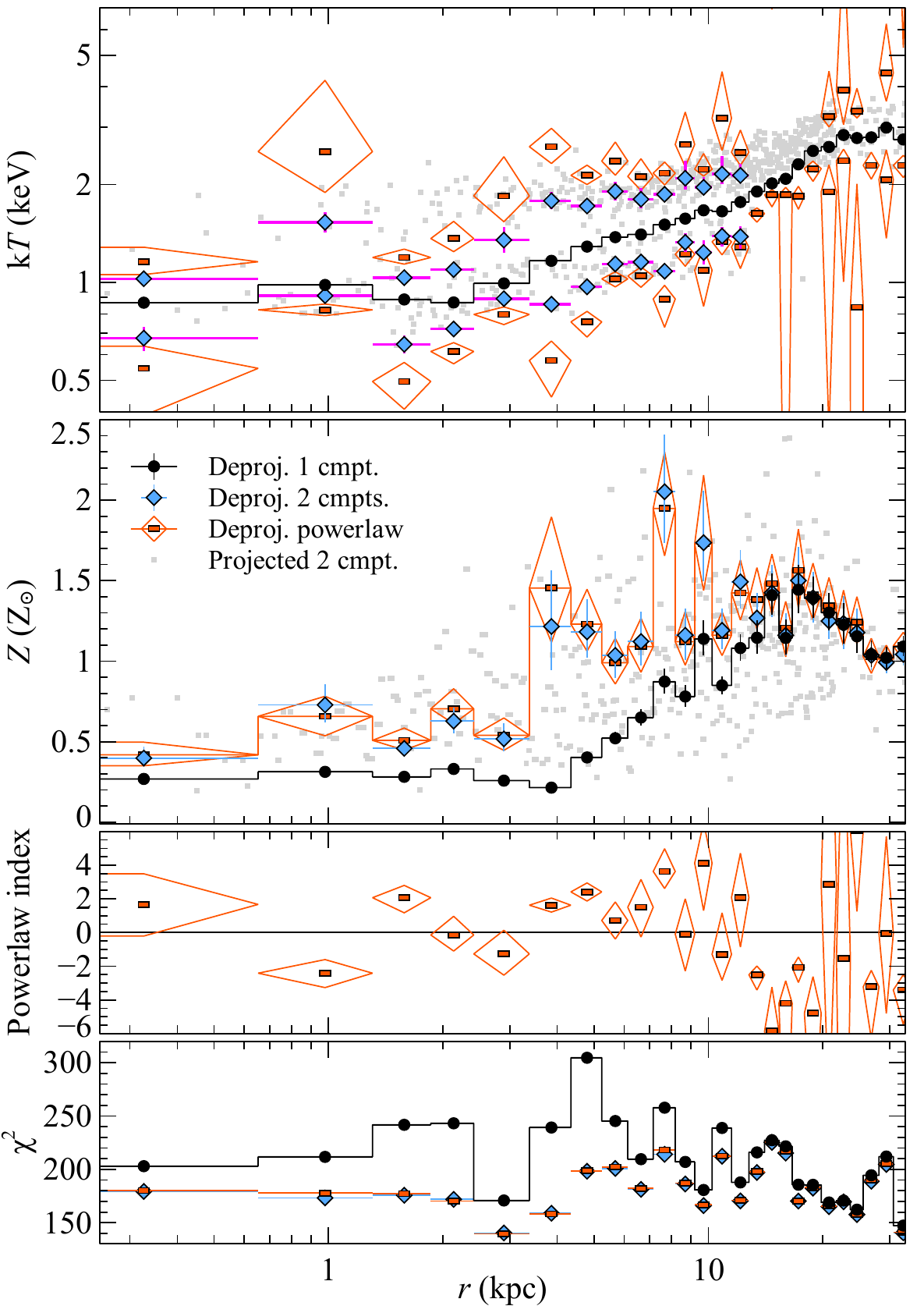}
  \caption{Comparison of powerlaw, single and two-component model
    deprojected spectral fits.  Panels are the temperature (top),
    metallicity (2nd), power law index (3rd) and fit $\chi^2$
    (bottom). The projected two-component results are also shown,
    plotting only every 4th point.}
  \label{fig:deproj_plaw}
\end{figure}

We can also fit the powerlaw distribution model to the deprojected
data. The results of these fits (Fig.~\ref{fig:deproj_plaw}) show a
similar trend to the projected results, where the powerlaw temperature
range is wider than the two temperatures of the two-component
fits. The index cannot be constrained very well, but scatters around a
value of 0. If we examine the best fitting metallicity for the
single-component, two-component and powerlaw deprojected fits and the
two-component projected fits, then we find that there is reasonable
agreement for all these models except for the single-component
deprojected results, which show a stronger drop in metallicity in the
central regions. However, all the results show a central metallicity
drop. In the figure we also plot the $\chi^2$ of the best fit to the
deprojected spectra for each of the models. The two-component and
powerlaw models fit the data better in the central regions, but there
is little difference in the fit quality between these two models
there. In the projected fits, the powerlaw model appeared to fit the
data better than the two-component model in the central regions, but
it is unclear whether this is due to projection effects or data
quality.

\subsection{Discussion}
There is projected emission down to $\sim 0.9\keV$ with a
single-component model or $0.7\keV$ with two components. This is
roughly consistent with the \emph{XMM}-RGS results
\citep{SandersRGS08} which show Fe~\textsc{xvii} emission, with
material down at $0.3-0.45\keV$. The difference may be due to
projection effects, the limited spectral resolution of ACIS and the
assumption of $1.2\times 10^{21} \pcmsq$ absorbing Hydrogen column
density in this analysis. There is dust associated with the filaments
which could contain additional absorbing material. We also note that
we always assume that the thermal components have the same
metallicity, which may not be the case.

The coolest X-ray filaments lie closest to the core of the cluster at
the location of the dust filaments, albeit with an offset
(Fig.~\ref{fig:optresid} and Fig.~\ref{fig:vcentre}). The X-ray gas is
clearly multiphase in the centre of the cluster and more than one
thermal component is required to fit the spectra
(e.g. Fig.~\ref{fig:deproj_plaw}). The multiphaseness is not due to
projection effects caused by the cluster temperature gradient and is
seen in deprojected spectra (Fig.~\ref{fig:deproj_EW}). Introducing a
two-component model into the spectral model (Fig.~\ref{fig:T_radius})
shows that there is a 0.8 to 2 keV temperature range over most of the
inner 6 kpc, rather than the more smoothly varying profile as seen in
the one-component model. Note that two components are only required in
the plume beyond this radius.  The two-component model decreases the
minimum temperature found in the region, but the change is not
dramatic (likely due to the fine spatial binning used here).  As there
is a similar range in temperatures seen using a two-component compared
to a single-component model within the inner 10 kpc, this suggests
that the multiphase filamentary structures we observe on the sky are
replicated along the line of sight in this region.

Despite the lack of a much wider range in temperatures, we do observe
slightly cooler gas down to $0.7\keV$ in the core when using two
temperature models.  The multi-temperature structure is associated with
the filaments and plume and appear suddenly within a radius of
10~kpc. There is in addition evidence for a wider range of
temperatures when fitting models with a powerlaw temperature
distribution (Fig.~\ref{fig:T_index}). Powerlaw models provide
increasingly good fits to the data compared to two component models
going to smaller radii. The index of the powerlaw also significantly
decreases at a smaller radius, indicating there is relatively more gas
at lower temperatures.  The minimum temperature in the powerlaw
distribution tends to drop down to $\sim 0.5\keV$, close to the
\emph{XMM} results.  Similar powerlaw distribution fits are also
supported when projection effects are removed
(Fig.~\ref{fig:deproj_plaw}), although no improvement in the spectral
fit over a two-component model is seen in this case.  The
two-temperature models give a range in temperature of a factor of
$\sim 2$ in the core, which increases to $\sim 4$ with powerlaw
models. The multi-component maps of the cluster core similarly give a
range of $\sim 4$ in the core of the cluster
(Fig.~\ref{fig:norm_T_maps}).

The minimum temperature of the single-component, two-component and
powerlaw models decreases roughly with radius, although the coldest
X-ray gas is offset from the nucleus and X-ray centroid
(Fig.~\ref{fig:vcentre}).  We note that the X-ray filaments in
Centaurus are coolest, densest and better defined closer towards the
nucleus (e.g. Figs. \ref{fig:rgb} and \ref{fig:vcentre},
\ref{fig:norm_T_maps}).  Dust filaments appear to extend down to the
radio nucleus, however (Fig.~\ref{fig:optresid}). If the shell at the
centre of the cluster is a shock, they must survive it. Given the
structure on small scales, it is difficult to define the centre and
account for projection.

Instead of the cool filaments being pulled out from the central galaxy
by rising bubbles, it has been suggested that cold blobs could
condense out of the ICM when the ratio of the cooling to the free-fall
time scale drops too low \citep{Gaspari12, McCourt13, Li15}. We can
estimate free-fall timescale for the cluster from the X-ray data. At a
radius $r$ the free-fall timescale is given by $t_\mathrm{ff} =
(2r/g)^{1/2}$. The gravitational acceleration, $g$, if the cluster is
in hydrostatic equilibrium is given by $g = -(1/\rho)
\mathrm{d}P/\mathrm{d}r$, where $P$ is the total pressure and $\rho$
is the mass density. The electron pressure in the cluster is
approximated by a powerlaw model (Fig.~\ref{fig:deproj_EW}),
$P_\mathrm{e} = 0.150(r/\mathrm{kpc})^{-0.51} \keVpcmcu$, while the
electron density is approximated by $n_\mathrm{e} =
0.168(r/\mathrm{kpc})^{-0.83} \pcmcu$. Therefore the gravitational
acceleration in the centre of the cluster is $\sim 1.4\times 10^{-7}
(r/\mathrm{kpc})^{-0.68} \cmpssq$. The implied free-fall time closely
tracks a 10th of the cooling time (Fig.~\ref{fig:deproj_EW}). It is
not obvious from the data for this model why there should be
multiphase gas within a certain radius given that there is not a large
variation in the ratio of cooling and free-fall times.

We note that the only multiphase gas present beyond 6 kpc radius is in
the plume and other related cool structures
(Fig.~\ref{fig:maps_2T}). There is low frequency radio emission
associated with the plume, showing it is likely material which has
been dragged out by a rising cavity. There are no cool blobs of
material outside the plume region beyond 6 kpc radius which would
indicate general condensation at larger radius where the cooling time
is 10 times the free-fall time.

\section{Central abundance drop}
\label{sect:elemabund}

\begin{figure}
  \includegraphics[width=\columnwidth]{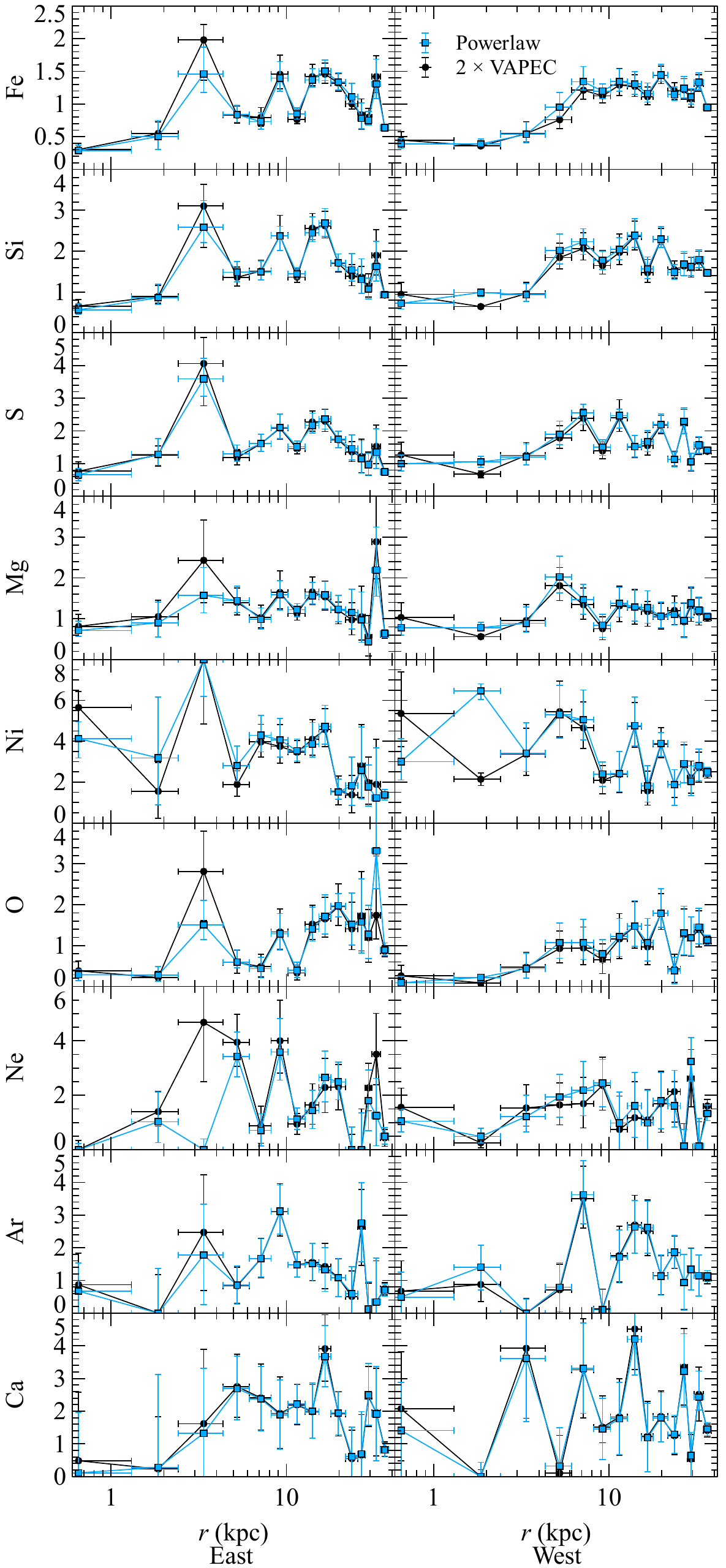}
  \caption{Elemental abundances (in Solar units) to the east and west
    of the cluster core, obtained by fitting two-component or powerlaw
    distribution \textsc{vapec} models to deprojected
    spectra. Analysis details are given in Appendix
    \ref{sect:abunddetails}.}
  \label{fig:elemental_profile}
\end{figure}

Previous work has shown a central drop in abundance in the cluster
\citep{SandersCent02,Panagoulia13}. Here we examine this on small
spatial scales using these new deeper data.  The deprojected
metallicity profiles of different elements to the east and west of the
cluster core are shown in
Fig.~\ref{fig:elemental_profile}. We note, however,
  that this analysis assumes that the phases in each radial bin have
  the same metallicity, which may not be a valid assumption. CCD-quality data are
  not good enough to test this assumption. To the eastern side of the
cluster the metallicities appear to exhibit more bin-to-bin variation
than the west. This is also seen in metallicity profiles with finer
bins (Fig.~\ref{fig:deproj_EW}). This could be real, however it may be
due to incorrect modelling of the complex thermal structure in the
plume or it may be because the assumption of spherical symmetry when
deprojecting is invalid.

The profiles are consistent with central reductions in many of the
probed elemental abundances (Fe, Si, S, Mg and O). In the other
metals, it is unclear whether the profiles are flat or declining, due
to large measurement uncertainties or differences between the two
modelling approaches. We note that several of the elemental lines are
not clearly separated in these CCD spectra (e.g. Ni and Ne lines are
close to the strong Fe-L lines) and therefore they cannot be
measured unambiguously. Furthermore, the O abundance is affected by
the choice of the absorbing column density.

We therefore confirm the central drop in metallicity in the cluster.
\cite{Panagoulia13} suggests that the metals are locked into dust in
the centre of the cluster, they are moved to larger
  radius by feedback processes, the grains are destroyed (e.g. by
  sputtering) and the metals returned to the hot phase. Such a
  scenario could also explain the metallicity blobs and flat inner
  metallicity profile (Section
  \ref{sect:Zstructure}). Fig.~\ref{fig:elemental_profile} is
consistent with this picture, confirming that Fe, Si and S abundances
drop in the centre to around a third of their peak value.
Furthermore, we also find evidence for Mg and O depletion, although
these metallicities have some modelling uncertainties. These are all
metals which can form part of dust. Unfortunately, the quality of the
data are insufficient to measure the central Ar abundance, which is
unlikely be depleted by dust formation. With the aid of detailed
models of dust composition it may be possible in the future to test
the dust deposition model in Centaurus using those elements which we
can measure.

Upcoming X-ray observatories such as \emph{Athena}
\citep{NandraAthena13} will make extremely precise measurements of the
temperature distribution and metallicity in the core of Centaurus and
those other clusters \citep{Panagoulia15} which show central drops in
metallicity. \emph{Athena} will also be able to
  measure the metallicity of different phases independently.

\section{Conclusions}
\label{sect:conclusions}
We analysed deep new \emph{Chandra} observations of the Centaurus
cluster, examining the larger-scale structure, metallicity
distribution, cold fronts and region around the nucleus. The main
conclusions of our paper are as follows:

\begin{enumerate}
\item Gradient filtering using the Gaussian gradient
    magnitude filter is able to robustly reveal previously
    hard-to-discern structures in X-ray images of galaxy clusters.
\item The correlation on larger scales of the metals with SB and
  inverse correlation with temperature implies that the asymmetry in
  the cluster is dominated by sloshing motions.
\item Enriched dusty material, uplifted by AGN activity
    would simultaneously explain the low inner metallicities, the flat
    inner metallicity profile in the core and the existence of 5-10
    kpc high metallicity blobs on the western side of the cluster.
\item We find 7 kpc `notches' around the edge of the western cold
  front which may be KHIs. There are strong point-to-point changes in
  the sharpness of the cold front on $10^\circ$ scales. In some
  sectors the width of the edge less than or equal to the electron
  mean free path.
\item We identify a 1.9-kpc-radius shell around the nucleus. This
  could be a shock generated by an AGN outburst $\lesssim 3.5$~Myr
  ago. A shock may also explain why the coldest X-ray gas is offset
  from the nucleus, though this could also be due to sloshing motions
  in the cluster. A significant fraction of energy in bubbles has already
  passed into ICM in the form of a higher pressure region.
\item Around the western cavity and to the north is a weak shock seen
  in temperature, density and pressure. To the western side the
  implied Mach numbers are between 1.1 and 1.4, while they are higher
  to the north (1.6 to 2.2).
\item There are $\sim 9$ depressions in X-ray SB
  around the core of the cluster, some of which may be associated with
  radio emission. The shock and multiple bubbles suggests that the
  nucleus is rapidly active on $5-10$~Myr timescales.
\item We confirm the existence of a quasi-periodic set of SB variations,
  which we previously claimed could be sound waves. The spacing of the
  peaks of $\sim 5$~kpc would imply a period of 6~Myr, if they are
  sound waves, which is similar to the ages of the shock and
  bubble. Alternatively, they could be generated by
    the sloshing, consisting of projected KHIs and their associated
    turbulence, or amplified magnetic field regions.
\item Around the nucleus there is filamentary multiphase X-ray gas. As
  found previously it is spatially correlated with material at lower
  temperatures. Fitting data with powerlaw temperature distributions
  gives better fits in the centre, implying a continuous temperature
  distribution over a factor of $\sim 4$ in temperature.
\end{enumerate}

\section*{Acknowledgements}
We thank the referee E.~Roediger for helpful suggestions which
improved this paper. ACF, HRR and SAW acknowledge support from the ERC
Advanced Grant FEEDBACK. Support for this work was provided by the
National Aeronautics and Space Administration through Chandra Award
Numbers G01-12156X, GO2-13149X and GO4-15121X issued by the Chandra
X-ray Observatory Center, which is operated by the Smithsonian
Astrophysical Observatory for and on behalf of the National
Aeronautics Space Administration under contract NAS8-03060.  The
scientific results reported in this article are based on observations
made by the Chandra X-ray Observatory and data obtained from the
Chandra Data Archive.  Based on observations made with the NASA/ESA
Hubble Space Telescope, and obtained from the Hubble Legacy Archive,
which is a collaboration between the Space Telescope Science Institute
(STScI/NASA), the Space Telescope European Coordinating Facility
(ST-ECF/ESA) and the Canadian Astronomy Data Centre (CADC/NRC/CSA).The
National Radio Astronomy Observatory is a facility of the National
Science Foundation operated under cooperative agreement by Associated
Universities, Inc.

\bibliographystyle{mnras}
\bibliography{refs}

\appendix
\section{Data analysis}
\label{sect:dataanalysis}

\subsection{Initial processing}

\begin{table}
  \centering
  \caption{\emph{Chandra} observation identifiers (OBSIDs) analysed.
    Listed are the observation date, the full exposure time and
    the exposure after cleaning.}
  \begin{tabular}{llrr}
    OBSID  & Date       & Full (ks) & Cleaned (ks) \\ \hline
    504    & 2000-05-22 & 31.8  & 26.0 \\
    505    & 2000-06-08 & 10.0  & 10.0 \\
    4190   & 2003-04-18 & 34.3  & 34.1 \\
    4191   & 2003-04-18 & 34.0  & 33.8 \\
    4954   & 2004-04-01 & 89.1  & 85.4 \\
    4955   & 2004-04-02 & 44.7  & 43.1 \\
    5310   & 2004-04-04 & 49.3  & 48.7 \\
    16223  & 2014-05-26 & 179.0 & 176.2 \\
    16224  & 2014-04-09 & 42.3  & 41.2 \\
    16225  & 2014-04-26 & 30.1  & 29.7 \\
    16534  & 2014-06-05 & 55.4  & 55.0 \\
    16607  & 2014-04-12 & 45.7  & 44.6 \\
    16608  & 2014-04-07 & 34.1  & 33.3 \\
    16609  & 2014-05-04 & 82.3  & 81.7 \\
    16610  & 2014-04-27 & 17.3  & 17.1 \\ \hline
    Total  &            & 779.3 & 760.0 \\
  \end{tabular}
  \label{tab:obsid}
\end{table}

The \emph{Chandra} Advanced CCD Imaging Spectrometer (ACIS-S)
observations listed in Table \ref{tab:obsid} are examined in this
paper. Dataset 504 was first presented in \cite{SandersCent02} and
observations 5310, 4954 and 4955 were first examined in
\cite{Fabian05}.
The datasets were reprocessed with \textsc{ciao} version 4.6
\citep{Fruscione06}. Bad time periods were identified by examining the
light curve on the ACIS-S1 CCD with 200s bins, clipping bins which
deviated from the standard deviation by more than $2.5\sigma$,
assuming Poisson errors. We excluded data from ACIS CCDs other than 6,
7 and 8 from our observations.  For each input dataset we created
background datasets using standard blank-sky background files. For
each CCD and observation, we took the appropriate background event
file and removed events which occurred in bad pixels of their
respective observation. The backgrounds were then reprojected to match
the observation coordinates and attitude and their exposure times were
changed to match the foreground count rate in the 9-12 keV band. In
order to make total foreground and total background spectra, it was
necessary to make the ratio of each background to the total background
exposure time be the same as the ratio of the respective foreground to
the total foreground exposure. To do this, we reduced the exposure
time keywords of background event files where this was not the case,
discarding random X-ray events in order to preserve the count rate.
Before analysis, the foreground and background event files were
reprojected to the coordinate system of the 16223 observation.

Unless otherwise specified, spectra were fit between 0.5 and 7 keV in
\textsc{xspec} \citep{ArnaudXspec} 12.8.2 using an \textsc{apec} 2.0.2
thermal model \citep{SmithApec01}.  We assume the relative solar
abundances of \cite{AndersGrevesse89}. Note that the \textsc{apec}
model fitted here was not the standard one, but was recomputed to have
temperature steps of 0.01 dex, rather than 0.1 dex, as the steps were
otherwise visible in radial plots of the temperature.  Galactic
absorption was modelled using a \textsc{phabs} model
\citep{BalucinskaChurchPhabs92}, with an equivalent hydrogen column of
$1.2\times10^{21} \pcmsq$, the average of the central region if the
column is allowed to be free (note that this is not a good choice for
the region of the edge-on disc galaxy which contains absorbing
material, leading to incorrect results there). All deprojected
profiles were calculated by fitting spectra deprojected using the
\textsc{dsdeproj} method \citep{SandersPer07,Russell08}.

Exposure maps were created using \textsc{mkexpmap}, assuming a 3.6~keV
spectrum with $0.7\Zsun$ metallicity and Galactic absorption. When
creating images of the larger regions, we masked the edges of the CCDs
before adding to remove residuals.

\subsection{Large scale spectral mapping procedure}
\label{sect:specmapping}
We used the Contour Binning algorithm \citep{SandersBin06} to select
regions with a minimum signal to noise ratio threshold ($S/N$). The
algorithm chooses bins which follow contours of SB, usually on an
adaptively smoothed image. Bins are grown pixel by pixel until the
threshold is reached. A geometric constraint factor, $C$, is also applied,
where a pixel is not added if its radius from the bin centroid
is more than $C$ times the radius of a circle with the
same area as the bin. This constraint prevents bins becoming too
elongated.

We examined spectra with a $S/N=100$ (i.e. $10^4$ counts for regions
where the background is not significant). The input smoothed image was
an adaptively smoothed image, created using a kernel with a minimum
signal to noise ratio of 60 in the 0.5 to 7 keV band. A geometric
constraint of $C=2$ was used in the binning. We masked out point
sources in the X-ray image (see Section \ref{sect:images}).

For each of the bins, each of the CCDs and each of the datasets we
extracted foreground and background spectra and created response and
ancillary response files. We added the foreground spectra from the
observations for each of the bins for each CCD. The background spectra
were similarly summed. Spectra were grouped
to have at least 8 counts per spectral bin.  The response and ancillary responses
were spatially weighted by the number of counts in the 0.5 to 7 keV
band. For each bin and each CCD we averaged the responses, weighting
by the number of counts in the respective foreground spectrum. The
spectra and responses for the different CCDs were not added to avoid
combining data with distinct responses.

When fitting we minimised the C-stat fit statistic.  If a bin had
spectra from more than one CCD, they were fit simultaneously with
additional free parameters to account for the difference in
normalisation due to geometry and detector differences.

There are regions, particularly close to the nucleus, where two
temperature components are necessary to fit the spectra. At the
temperatures seen in Centaurus, two-component modelling is usually
required if a single component is too narrow to fit the Fe-L complex,
indicating that there are comparable emitting components along the
line of sight which differ significantly in temperature. We discuss
this in more detail in Section \ref{sect:thermprof} and also examine
models with more components. We therefore also fit the spectra with a
model with two temperature components in each bin, assuming that they
both have the same metallicity (it is not possible to fit for the
metallicities of the components separately).

It is not clear what criteria should be used to decide between single-
or two-component models. We investigated using the F-test and
examining the size of the error bars on the best-fitting
temperatures. These criteria appeared to either select two components
for a number of outer regions, many with unrealistic temperatures, or
did not select the two-component models in all appropriate locations
in the centre, leading to a large scatter in the metallicity. The best
criterion appeared to be to select two components for those regions
where there was a reduction in C-stat better than 5 per cent. The
resulting combined single- and two-component metallicity map appears
similar to the two-component map, except for the level of scatter in
regions where only one component was required.

\subsection{Western cold front edge}
\label{sect:frontfit}
\begin{figure}
  \centering
  \includegraphics[width=0.9\columnwidth]{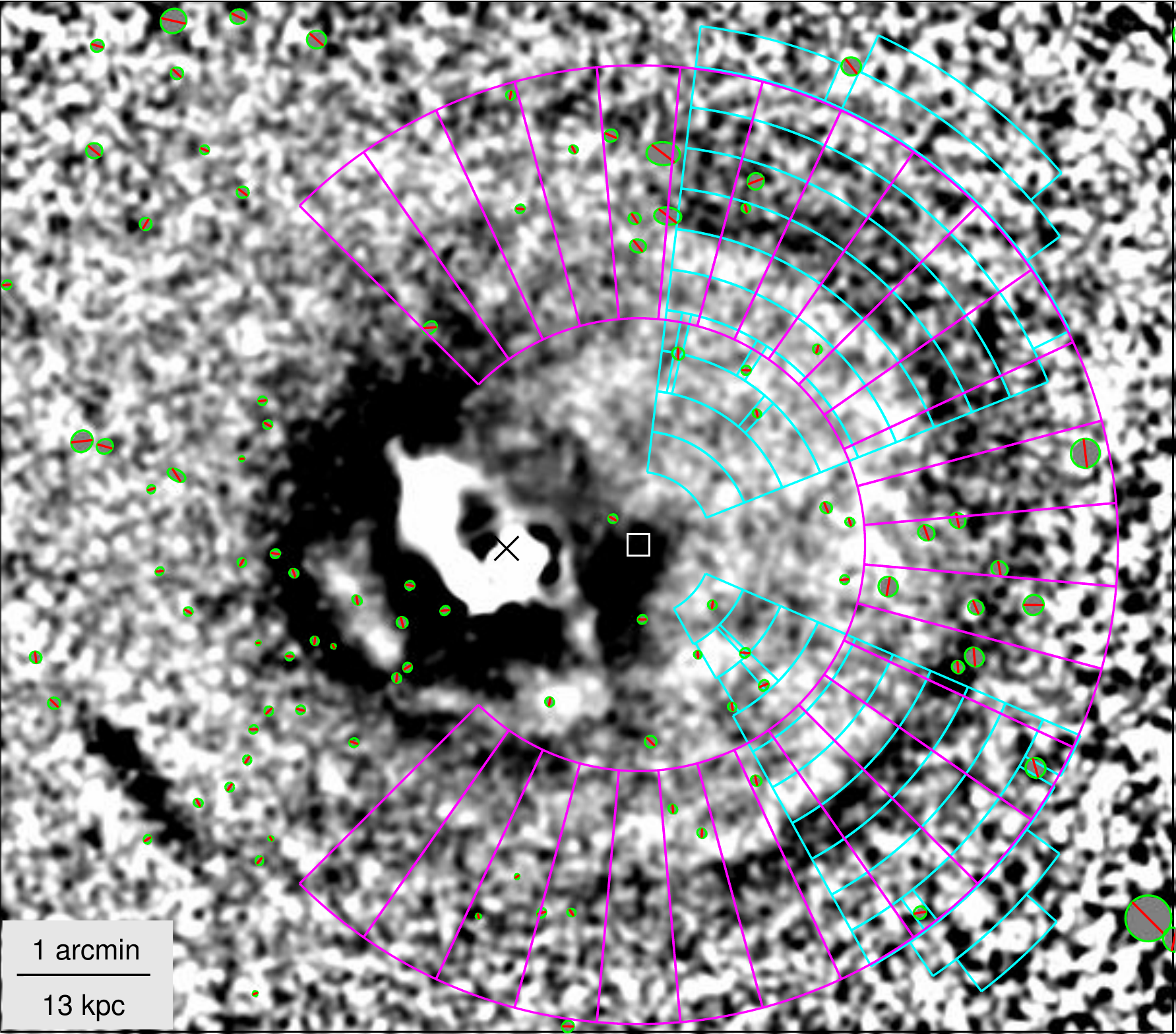}
  \caption{Regions used for examining the western cold front, shown on
    an unsharp-masked image. 27 sectors were used for SB
    extraction. Two sets of annular sectors to the north-west and
    south-west were used to examine the spectra. The cross marks the
    radio nucleus of the cluster. The box marks the centre of the
    annuli used, which is the same as used in
    Fig.~\ref{fig:cfront_remap}.}
  \label{fig:cfrontw_reg}
\end{figure}

\begin{figure}
  \centering
  \includegraphics[width=\columnwidth]{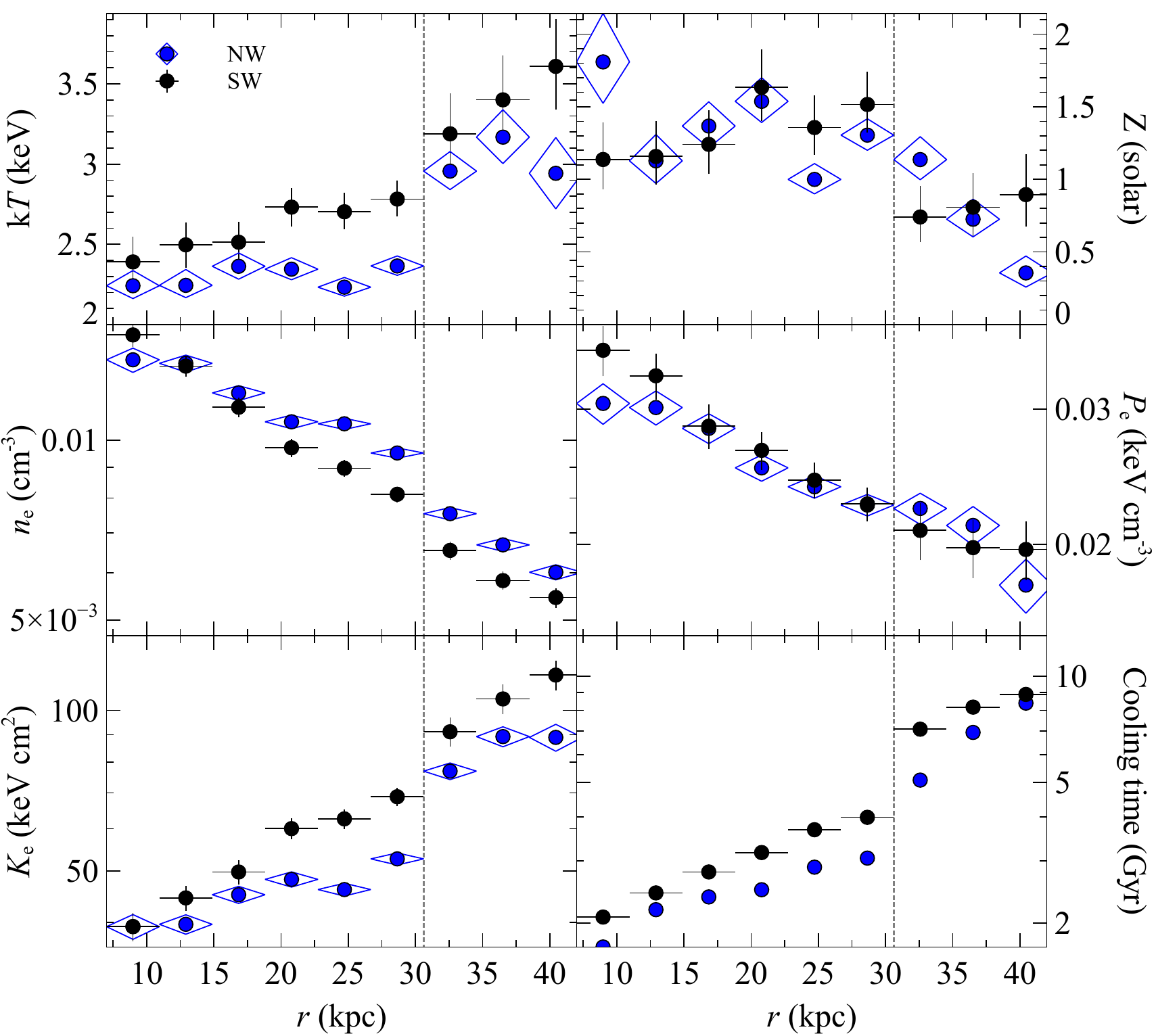}
  \caption{Deprojected thermodynamic profiles across the western cold
    front in the north-west and south-west directions
    (Fig.~\ref{fig:cfrontw_reg}). Plotted are the temperature,
    metallicity, electron density, electron pressure, electron entropy
    and mean radiative cooling time. The vertical line marks the cold
    front radius. The outer three bins are not shown due to edge
    effects giving oscillating values.}
  \label{fig:cfront_spectra}
\end{figure}

We examined deprojected spectra across the western cold front in two
different sectors (Fig.~\ref{fig:cfrontw_reg}), where the edge is
clearest towards the north- and south-west, obtaining the profiles in
Fig.~\ref{fig:cfront_spectra}. There are jumps in temperature and
density (in opposite directions) by 20 per cent at the edge, leading
to a continuous pressure distribution, confirming that the feature is
a cold front. The entropy reduces by 25--30 per cent in the inward
direction. Despite the clear correlations between the edges and
metallicity (Fig.~\ref{fig:subav_prof}), we do not see the change in
metallicity at the edge in the deprojected profiles.

To examine the SB profiles in detail around the edge,
exposure-corrected background-subtracted profiles were extracted
in the 0.5 to 4 keV band, with 0.492-arcsec binning between
radii of 1.7 and 4.5 arcmin (Fig.~\ref{fig:cfrontw_reg}). This band is
largely sensitive to the density (varying the temperature from 2.5 to
3 keV leads to a 10 per cent variation in count rate).

We took an emissivity model in which there are two powerlaws which
jump at a particular radius. The radius of the jump, the jump ratio
(the emissivity ratio outside divided by that inside), the powerlaw
indices and the overall model normalisation are free parameters.  We
allowed for broadening of the edge by interpolating between the two
powerlaw models using the integral of a Gaussian with a variable
$\sigma$. This emissivity profile was projected on the sky assuming
spherical symmetry, integrating the emissivity profile between 1 and 7
arcmin radius in 0.246~arcsec (half pixel) steps. The model was then
convolved with the PSF in the radial direction at the edge for each
sector. The PSF was calculated using \textsc{saotrace} 2.0.4 and
\textsc{marx} 5.1.0 assuming a 3~keV spectrum with $1.2\Zsun$
metallicity. The model was then resampled to have the same binning as
the data.

\begin{figure}
  \centering
  \includegraphics[width=0.8\columnwidth]{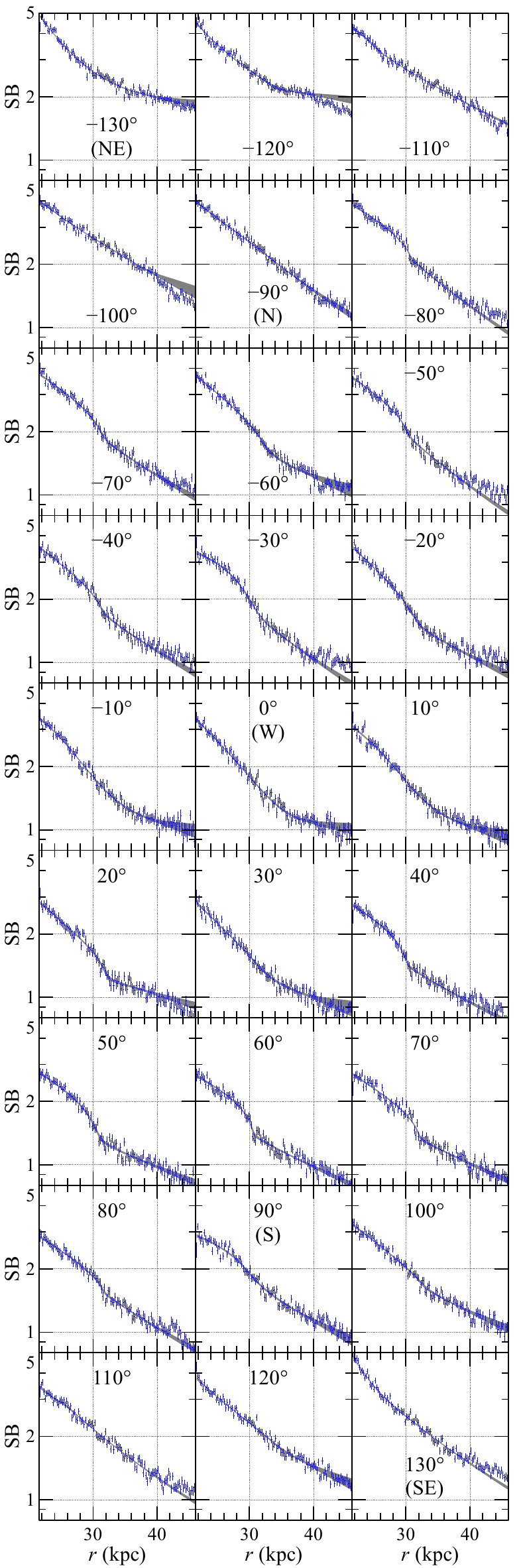}
  \caption{SB profiles in the 0.5 to 4 keV band across
    the western cold front in 27 $10^\circ$ sectors (depicted in
    Fig.~\ref{fig:cfrontw_reg}) from the north-east (top-left) through
    west to the south-east (bottom-right) in 0.984 arcsec bins. The
    0.492-arcsec-binned profiles were fitted within 40 kpc by a model
    in which the emissivity is a powerlaw which jumps to a second
    powerlaw at some radius, with the edge broadened by a
    Gaussian. The solid area shows the $1\sigma$ model uncertainties.}
  \label{fig:cfront_profiles}
\end{figure}

We applied this model to the data and examined the allowed range of
parameter values using the \textsc{emcee} \citep{ForemanMackey12}
Markov Chain Monte Carlo (MCMC) \textsc{emcee} code, which uses an
affine-invariant sampler \citep{Goodman10}.  We force the jump radius
to lie between 1.9 and 3 arcmin, the powerlaw indices to lie between
$-4$ and $0.5$, the jump ratio to lie between 0.01 and 2 (smaller
values are larger jumps) and the Gaussian $\sigma$ to be between 0.001
and 1 arcmin (the log value of $\sigma$ was varied in the MCMC). In
the MCMC analysis we used $\chi^2$ likelihoods, 400 walkers, a burn in
period of 1000 and a chain length of 1000.  The data were fit only
inside a radius of 40 kpc, as in several sectors, particularly to the
north, there is a SB break at this radius. We also investigated models
which also included this break, fitting the data out to 57 kpc
radius. The results from this more complex model are in reasonable
agreement to those those presented here. In our analysis we found that
the outer powerlaw index was affected by the outer integration
radius. Reducing the outer radius to 6 arcmin decreases the index by
0.2--0.5.  The median models and the surrounding $1\sigma$ percentiles
are shown as shaded regions in Fig.\ref{fig:cfront_profiles}. The
parameters from the fits are plotted in
Fig.~\ref{fig:cfront_fit_params}. We also did the analysis with twice
the number of sectors, finding consistent results. The radius results
are plotted in Fig.~\ref{fig:cfront_fit_params}, showing that the
variation in radius seen in the 27 bin results is real.

\subsection{Metallicity fluctuations}
\label{sect:zfluctdetails}
Spectra from regions X and Y (Fig.~\ref{fig:highZ}) were extracted and
fit with single and two-component \textsc{vapec} thermal models
(connecting the metallicity between the two components). Some of the
emission in the regions could come from projected cluster emission. To
attempt to account for this we extracted a background spectrum from an
adjacent part of the cluster outside the bright western region. We fit
the spectra using this background and secondly using a standard
blank-sky background spectrum, between 0.5 and 8 keV. As the
two-component model temperatures were not always well
constrained, we conducted a MCMC analysis to better explore the
allowed parameter space using \textsc{xspec\_emcee}
(\url{https://github.com/jeremysanders/xspec\_emcee}), applying 400
walkers, a burn-in period of 1000 and a chain length of 1000
iterations. The temperature values were assumed to lie between 0.15
and 20~keV. We allowed the O, Ne, Mg, Si, S, Ar, Ca, Fe and Ni
metallicities to be free and fixed others to Solar values. The error
bar shows the $1\sigma$ percentiles from the chain, while the point
shows the median.

\subsection{Central spectral maps}
\label{sect:speccentredetails}
We constructed detailed thermodynamic maps of the central region
(Fig.~\ref{fig:centre_spec_maps}) using $S/N=20$ spatial bins
(i.e. 400 counts where the background is not significant). The mapping
was based on an adaptively smoothed image using a kernel with a
minimum signal to noise ratio of 30 and applying a geometric
constraint of $C=3$, only examining a central $4.2 \times 4.2$ arcmin
($54\times 54$ kpc) box. We extracted spectra from
the background observations using this box. These background spectra for
each observation were also added, first artificially lowering the
exposure of background spectra (discarding events and reducing
exposure time correspondingly) so that the ratio of each background to
the total was the same as the respective foreground observation to
total. We used a central response and ancillary response matrix for
all regions examined. We note that the \textsc{xspec} normalisation is
proportional to the emission measure and defined in \textsc{xspec} to
be $10^{-14} \int n_\mathrm{e} n_\mathrm{H} \mathrm{d}V / (4\pi
D_A^2[1+z]^2)$, where the cluster lies at a redshift $z$ and angular
diameter distance of $D_A$~cm, the electron and hydrogen densities are
$n_\mathrm{e}$ and $n_\mathrm{H} \pcmcu$, respectively and are
integrated over a volume $V\cmcu$.

Note that the assumption of constant absorbing column density
where there are dust lanes in the centre of NGC 4696 where this
assumption is likely to be invalid. The metallicity was fixed to
$1.2\Zsun$, an average value for the central region excluding where
multi-component models are preferred. Fixing the metallicity allows us
to fit single temperature models to the data where multi-temperature
models may be required (see Section \ref{sect:thermprof}), obtaining
average projected temperatures and normalisations which are not biased
by unrealistic best fitting metallicities.

\subsection{Inner cavity profiles}
\label{sect:innercavity}
\begin{figure}
  \includegraphics[width=\columnwidth]{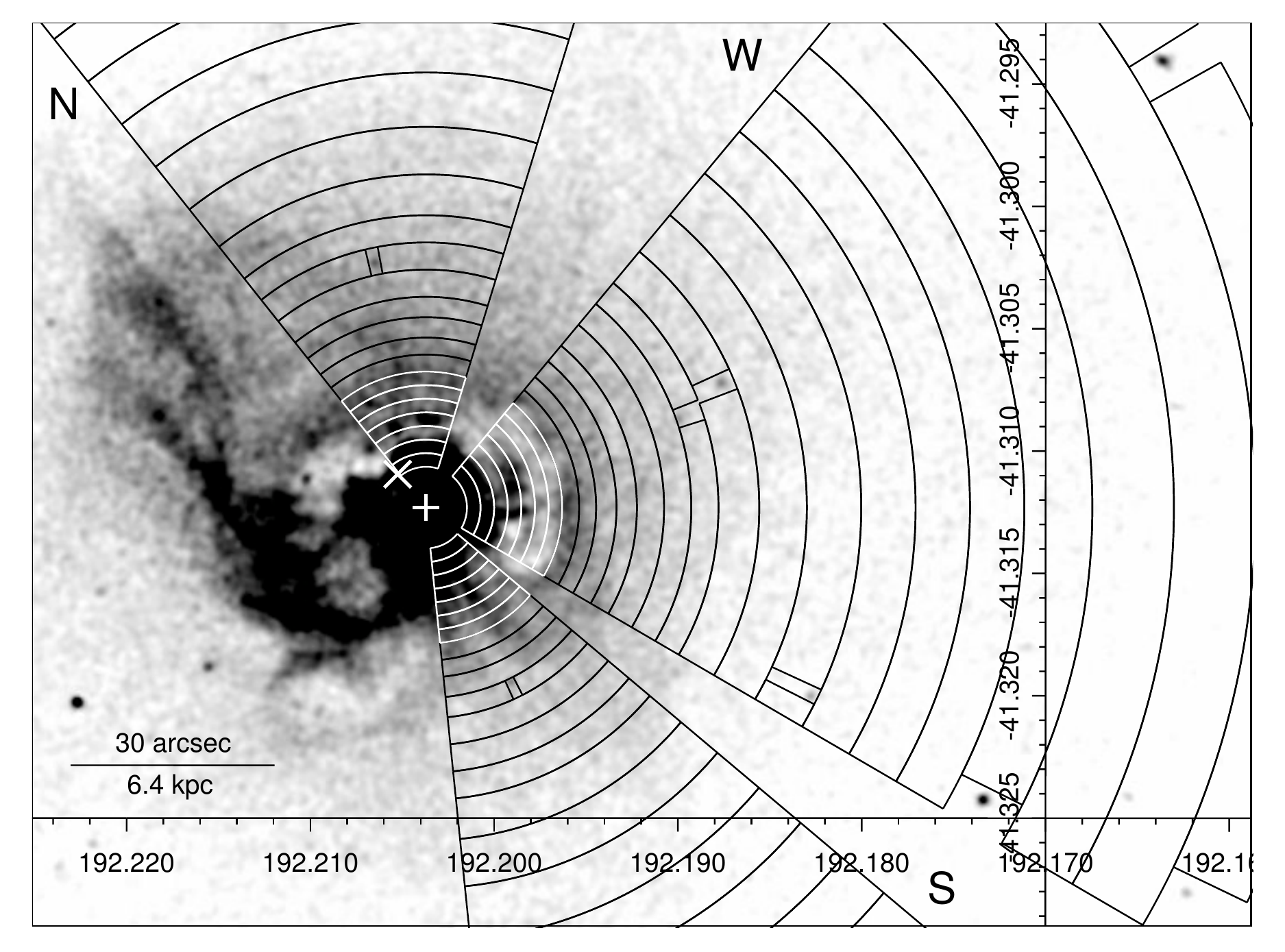}
  \caption{Regions examined to measure the ICM properties around the
    western inner cavity (W) and for comparison to the north (N) and
    south (S). The radio nucleus is marked with a `$\times$'. The
    centre of the sectors is marked by a `+'. The X-ray image is
    unsharp-masked, subtracting 0.5 of the 0.5 to 7 keV image
    smoothed by 7.9 arcsec from the image smoothed by 1 arcsec.}
  \label{fig:cavity_reg}
\end{figure}

We extracted deprojected spectra in annuli from a sector outwards from
the western inner cavity (labelled B in
Fig.~\ref{fig:centre_spec_maps}). To better match the morphology of
the inner region, we chose a centre which is that of a circle which
circumscribes the outer edges of the two inner cavities (A and B). For
comparison we also examine sectors to the north and south, avoiding
the plume to the east.  Fig.~\ref{fig:cavity_reg} shows the regions
analysed. The spectra were fitted with single component models, with
the metallicity fixed to $1.2\Zsun$.

\subsection{Deprojected central spectra}
\label{append:deprojcentre}
The spectra were extracted from $180^\circ$ east and west sectors or
complete annuli, excluding point sources. The centre used
was the X-ray centroid of the central region, as in Appendix
\ref{sect:thermprofscentre}.  We fit both single- and two-component
\textsc{apec} models to the deprojected spectra (minimising the
$\chi^2$). To fully explore the parameter space we ran a MCMC analysis
on each deprojected spectrum. The two-component results are only shown
in Fig.~\ref{fig:deproj_EW} if both temperatures can be constrained to
20 per cent. From the MCMC chains containing normalisation,
temperature and metallicity, the density, entropy, pressure and mean
radiative cooling time were computed. To compute the two-component
results the assumption that the components are in pressure equilibrium
was made. In the plot, we also compare projected thermodynamic data
points from Fig. \ref{fig:therm_r_X}.

\subsection{Elemental abundances}
\label{sect:abunddetails}
The metallicity profiles were obtained by fitting deprojected spectra,
extracted from $180^\circ$ sectors to the east and west.  To increase
the signal to noise, the radial bins used here are twice as wide in
the centre as in Fig.~\ref{fig:deproj_EW}. To properly account for the
complex temperature distribution we used two-component and powerlaw
\textsc{vapec} models. In our analysis Fe, Si, S, Mg, Ni, O, Ne, Ar
and Ca abundances were allowed to vary. C, N and Al metallicities were
tied to Fe. The same metallicities were assumed in all thermal
components in a radial bin. Input spectra were fit between 0.5 and 7
keV, minimising the $\chi^2$ statistic. Metallicities were allowed to
vary between 0 and $8\Zsun$ and temperatures were allowed to vary
between between 0.1 and 20 keV. The $1\sigma$ error bars shown were
calculated by varying the parameter until the minimum $\chi^2$
increased by 1.0. A MCMC analysis produced very similar results,
though we show the results from the simple analysis here.

\section{Central thermal structure}
\label{sect:thermprofscentre}
\begin{figure}
  \includegraphics[width=\columnwidth]{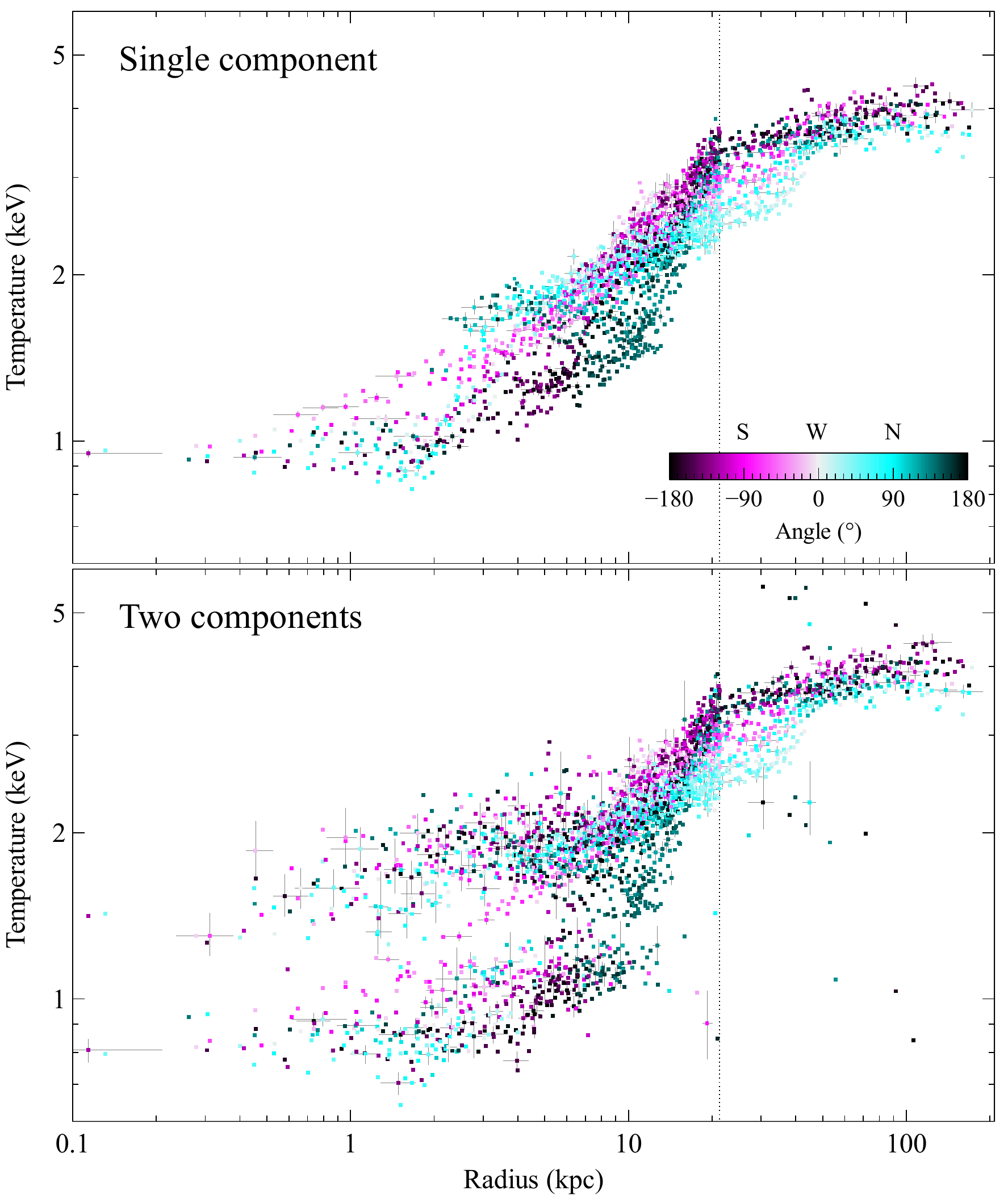}
  \caption{Radial profile of the temperatures in bins with a signal to
    noise ratio of 40 (inside the radius of $\sim 20$~kpc marked by
    the dotted line) or 100 (outside that radius). The points are
    coloured according to their angle (angles are measured from the
    west, with positive values northwards). The radius is measured
    from the central radio source. The top panel shows a single
    component fit to the spectra. The bottom panel shows the two
    temperatures for a two-component fit, if the temperatures of the
    two components can be constrained to better than 20 per cent,
    otherwise the single-component fit results are shown. Error bars
    are shown only for every 10th point.}
  \label{fig:T_radius}
\end{figure}

We constructed profiles of the temperature (Fig. \ref{fig:T_radius})
from maps with $S/N=40$ inside 10.5 kpc radius and $S/N=100$ at larger
radii (the radius was measured from the X-ray centroid, marked in
Fig.~\ref{fig:vcentre}, at 192.204 and Dec -41.3125, not from the
radio nucleus).  One aspect which is different to the procedure used in
Section \ref{sect:maps}, is that in the
two-component plot the temperatures of the two-component fit are only
shown if they can be constrained to better than 20 per cent, otherwise
the single component result is shown.  The single component fits show
that the temperatures in the centre split into hotter (at high angles)
and colder (at low angles) branches, where the colder branch is the
cool central plume.

Although we are fitting projected spectra from regions on the sky, we
can compute projected versions of the other thermodynamic properties
using geometric assumptions. Here we assume that the line of sight
depth of a region is the radius of the region.  This assumption, which
we test below, allows the normalisation to be converted to a projected
electron density, $n_\mathrm{e,proj}$. The electron pressure is
computed as $P_\mathrm{e,proj} = n_\mathrm{e,proj} \mathrm{k}T$, where
$\mathrm{k}T$ is the temperature in keV. The entropy is
$K_\mathrm{e,proj} = n_\mathrm{e,proj}^{-2/3} \mathrm{k}T$. We also
calculate a mean radiative cooling time $t_\mathrm{cool,proj}$ by
dividing the enthalpy of a unit cube of gas in the cluster, $H$, by
its emissivity calculated using an \textsc{apec} model. The enthalpy
is $H=(5/2)\: \mathrm{k}T n_\mathrm{e,proj}\: (1+1/X_\mathrm{e/H})$,
where $X_\mathrm{e/H} \sim 1.2$ is the ratio of the number of
electrons to Hydrogen nuclei in a fully ionised plasma.

For the two-component fits, we assume that both components come from
the same volume but they are in pressure equilibrium with each
other. In this case, the electron density for component 1, in
$\pcmcu$, is given by
\begin{equation}
  n_\mathrm{e,proj,1} = \left(
    \left[ N_1 + N_2 \frac{T_2^2}{T_1^2} \right]
    \frac{4\pi D_A^2[1+z]^2}{10^{-14} \: V} X_\mathrm{e/H}
  \right) ^{1/2},
\end{equation}
where $N_1$ and $N_2$ are the \textsc{xspec} normalisations of the two
components, respectively, $T_1$ and $T_2$ are the respective
temperatures, $V$ is the assumed volume (radius times area of bin on
the sky, in $\cmcu$) and $D_A$ is the angular diameter of the source
in cm and $z$ its redshift. Monte Carlo realisations of the input data
are used to derive the error bars in the computed quantities.

\begin{figure*}
  \centering
  \includegraphics[width=\textwidth]{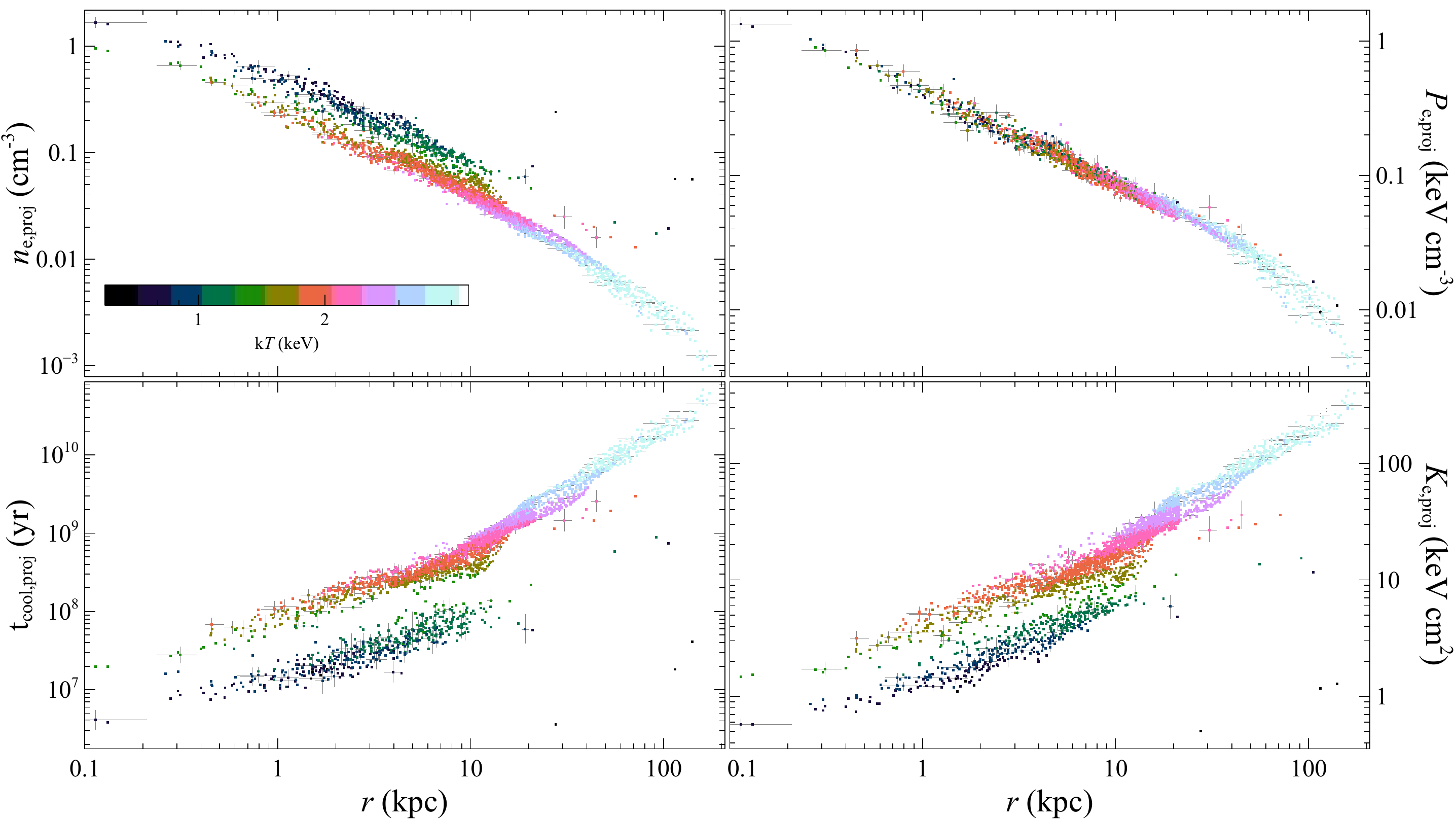}
  \caption{Radial plots of the projected electron density, electron
    entropy, electron pressure and mean radiative cooling time. The
    results are shown for two thermal components (with the number of
    components selected as in Fig. \ref{fig:T_radius}. Error bars are
    shown only for every 10th point. Points are coloured according to
    temperature. Note that for the pressure profile the assumption of
    pressure equilibrium forces the two-component results to have the
    same pressure.}
  \label{fig:therm_r_X}
\end{figure*}

In Fig.~\ref{fig:therm_r_X} we examine the radial thermodynamic
quantities computed from the two-component fits (showing the
two-component result if both temperatures can be constrained,
otherwise the single-component result). These plots show that there
are distinct low-temperature branches in terms of density, cooling
time and entropy, seen most easily in terms of cooling time. This
material has high density, short cooling times and low entropy.  This
component appears to only be strong inside 12 kpc radius. If the
two-component error threshold is increased, then additional points on
these tracks are seen at larger radius, but not in great numbers.

\begin{figure*}
  \centering
  \includegraphics[width=\textwidth]{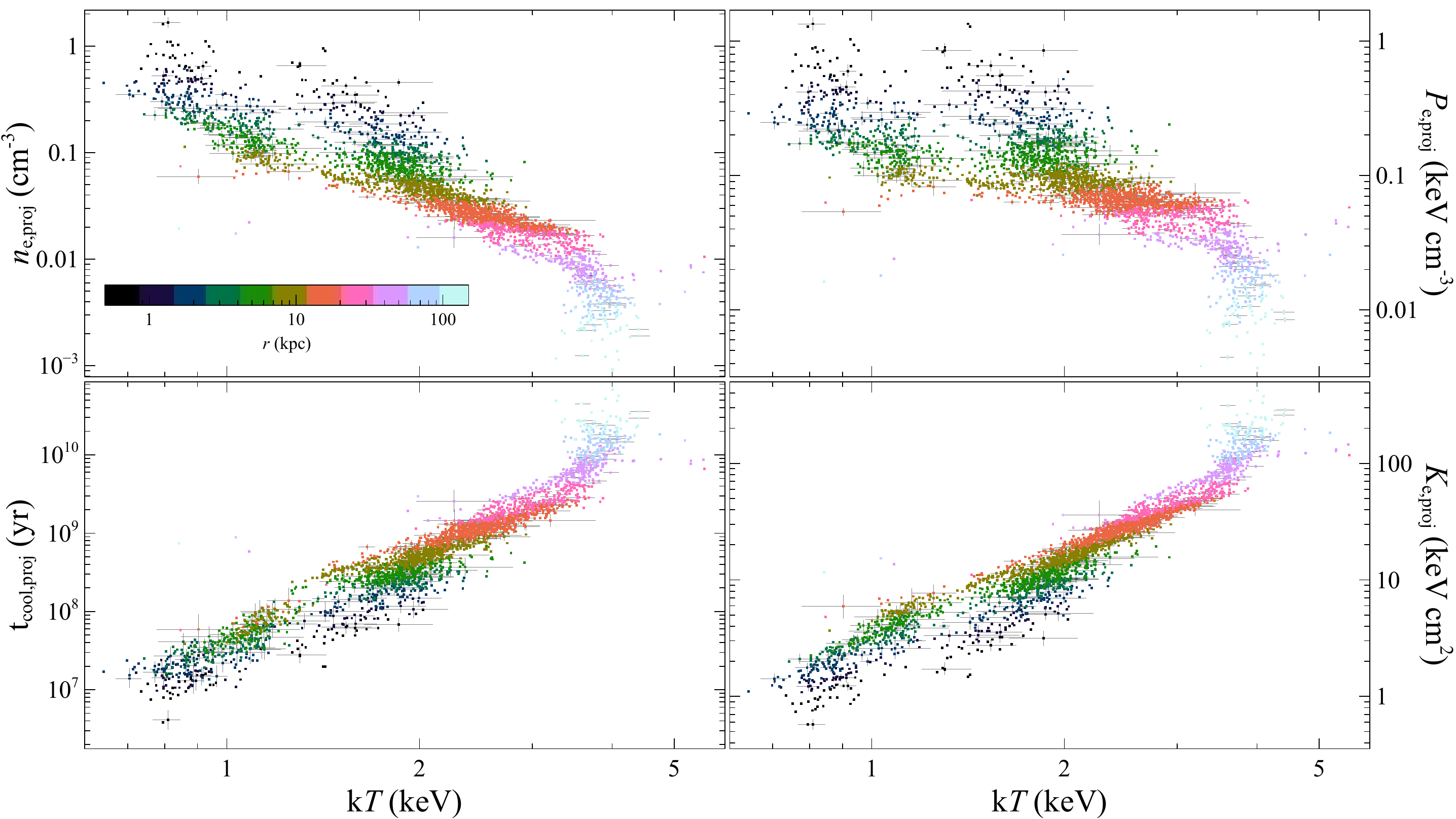}
  \caption{Plots of the two-component projected electron density,
    electron entropy, pressure and mean radiative cooling time as a
    function of temperature. Error bars are only shown for every 10th
    value. Points are coloured according to radius.}
  \label{fig:therm_T_X}
\end{figure*}

These two branches are also seen in a plot of the temperature against
density or pressure (Fig.~\ref{fig:therm_T_X}). However, if the
cooling time or entropy is plotted against the temperature, there
appears to be a continuous distribution, although there may be a lack
of values with temperatures around 1.3 keV.

When these projected values are compared to deprojected spectra
(Fig.~\ref{fig:deproj_EW}) we find reasonable agreement in some
quantities, while others are offset.  Projection effects appear to
only weakly affect the temperature in Centaurus. Both the high- and
low-temperature components are apparent in the deprojected results
over the same radial range. The radius-as-depth assumption appears to
over-predict the densities by a factor of 60 per cent at larger radius
and 100 per cent in the centre. There is a similar over-estimation of
the pressure and under-estimation of the cooling time and entropy.

\end{document}

%% file: defn.tex
\newcommand{\Mpc}{\rm\thinspace Mpc}
\newcommand{\kpc}{\rm\thinspace kpc}

\newcommand{\km}{\rm\thinspace km}

\newcommand{\cm}{\rm\thinspace cm}

\newcommand{\cmpssq}{\hbox{$\cm\s^{-2}\,$}}

\newcommand{\cmcu}{\hbox{$\cm^3\,$}}
\newcommand{\pcmcu}{\hbox{$\cm^{-3}\,$}}

\newcommand{\yr}{\rm\thinspace yr}

\newcommand{\s}{\rm\thinspace s}

\newcommand{\keVpcmcu}{\hbox{$\keV\cm^{-3}\,$}}

\newcommand{\Msun}{\hbox{$\rm\thinspace M_{\odot}$}}

\newcommand{\Msunpyr}{\hbox{$\Msun\yr^{-1}\,$}}

\newcommand{\keV}{\rm\thinspace keV}

\newcommand{\erg}{\rm\thinspace erg}

\newcommand{\ergpcmsqps}{\hbox{$\erg\cm^{-2}\s^{-1}\,$}}

\newcommand{\ergps}{\hbox{$\erg\s^{-1}\,$}}

\newcommand{\kmps}{\hbox{$\km\s^{-1}\,$}}

\newcommand{\kmpspMpc}{\hbox{$\kmps\Mpc^{-1}$}}

\newcommand{\Zsun}{\hbox{$\thinspace \mathrm{Z}_{\odot}$}}

\newcommand{\pcmsq}{\hbox{$\cm^{-2}\,$}}

